\documentclass[a4paper,UKenglish,cleveref, autoref, thm-restaten,nolineno ]{socg-lipics-v2019}

\title{On Guillotine Separable Packings for the Two-dimensional Geometric Knapsack Problem}

\titlerunning{On Guillotine Separable Packings for the Two-dimensional Geometric Knapsack Problem}

\author{Arindam Khan}{Indian Institute of Science, Bengaluru, India}{arindamkhan@iisc.ac.in}{}{}

\author{Arnab Maiti}{Indian Institute of Technology, Kharagpur, India}{arnabmaiti@iitkgp.ac.in}{}{}

\author{Amatya Sharma}{Indian Institute of Technology, Kharagpur, India}{amatya65555@iitkgp.ac.in}{}{}

\author{Andreas Wiese}{Universidad de Chile, Santiago, Chile}{awiese@dii.uchile.cl}{}{}
\authorrunning{A. Khan, A. Maiti, A. Sharma, and A. Wiese}
\Copyright{Authors} 



\category{} 

\relatedversion{} 

\supplement{}

\EventEditors{John Q. Open and Joan R. Access}
\EventNoEds{2}
\EventLongTitle{42nd Conference on Very Important Topics (CVIT 2016)}
\EventShortTitle{CVIT 2016}
\EventAcronym{CVIT}
\EventYear{2016}
\EventDate{December 24--27, 2016}
\EventLocation{Little Whinging, United Kingdom}
\EventLogo{}
\SeriesVolume{42}
\ArticleNo{23}

\usepackage{dsfont}
\usepackage[lined,boxed,commentsnumbered]{algorithm2e}
\usepackage{booktabs}
\usepackage[nocompress]{cite} 

	\usepackage[dvipsnames]{xcolor}
	\usepackage{tikz,pgfplots}

	\usetikzlibrary{patterns}

	\usetikzlibrary{arrows.meta}
	\usepackage{forest}

	\usepackage{algorithmic}
	\newtheorem{thm}{Theorem}
	\newtheorem{lem}[thm]{Lemma}
	
	\newcommand{\eps}{\varepsilon}
	
	\newcommand{\epss}{\varepsilon_{{small}}}
	\newcommand{\epsl}{\varepsilon_{{large}}}

	\newcommand{\epsau}{\varepsilon_{{ra}}}

	\newcommand{\opt}{OPT}
	
	\newcommand{\optla}{OPT_{{large}}}
	\newcommand{\optsm}{OPT_{{small}}}
	\newcommand{\optsk}{OPT_{{skew}}}
	\newcommand{\optho}{OPT_{{hor}}}
	\newcommand{\optve}{OPT_{{ver}}}

	\newcommand{\Rsm}{I_{{small}}}
	\newcommand{\Rla}{I_{{large}}}
	\newcommand{\Rho}{I_{{hor}}}
	\newcommand{\Rve}{I_{{ver}}}
	\newcommand{\Rsk}{I_{{skew}}}

	\newcommand{\R}{I}
	
	\newcommand{\bottomc}{bottom}
	\newcommand{\topc}{top}
	\newcommand{\leftc}{left}
	\newcommand{\rightc}{right}
	\newcommand{\height}{h}
	\newcommand{\width}{w}
	\newcommand{\profit}{p}
	\newcommand{\area}{a}

	\newcommand{\tgk}{2GK}
	\newcommand{\tgkr}{2GK(R)}

	\newcommand{\tggk}{2GGK}
	\newcommand{\tggkr}{2GGK(R)}
	\newcommand{\tggkc}{2GGK-C}

	\newcommand{\Lc}{\boldsymbol{\mathsf{L}}}

	\newcommand{\Bc}{\boldsymbol{\mathsf{B}}}

    \global\long\def\Rlp{I_{\mathrm{low,ps}}}%
    \global\long\def\Rlph{I_{\mathrm{low,ps,hor}}}%
    \global\long\def\Rlpv{I_{\mathrm{low,ps,ver}}}%
	
	\global\long\def\C{\mathcal{C}}%
    \global\long\def\B{\mathcal{B}}%
    \global\long\def\Rh{I_{\mathrm{high}}}%
    \global\long\def\Rl{I_{\mathrm{low}}}%

	\ifdefined\DEBUG
	\newcommand{\ari}[1]{\textcolor{Red}{#1}}
	\newcommand{\ama}[1]{\textcolor{Green}{#1}}
	 
	 \newcommand{\arn}[1]{\textcolor{Violet}{#1}}
	 \newcommand{\aw}[1]{\textcolor{NavyBlue}{#1}}
	  \def\rem#1{{\marginpar{\raggedright\scriptsize #1}}}
	  \newcommand{\arir}[1]{\rem{\textcolor{Red}{$\bullet$ #1}}}
	  \newcommand{\amar}[1]{\rem{\textcolor{Green}{$\bullet$ #1}}}
	  \newcommand{\madr}[1]{\rem{\textcolor{Maroon}{$\bullet$ #1}}}
	  \newcommand{\arnr}[1]{\rem{\textcolor{Violet}{$\bullet$ #1}}}
	  \newcommand{\awr}[1]{\rem{\textcolor{NavyBlue}{$\bullet$ #1}}}
	\else
	  \newcommand{\ari}[1]{#1}
	  
	  \newcommand{\arn}[1]{#1}
	  \newcommand{\ama}[1]{#1}
	 \newcommand{\aw}[1]{#1}
	  \newcommand{\arir}[1]{}
	  \newcommand{\madr}[1]{}
	  \newcommand{\arnr}[1]{}
	  \newcommand{\amar}[1]{}
	  \newcommand{\awr}[1]{}

	\fi

\begin{document}
\maketitle

\thispagestyle{plain}

\begin{abstract}
In two-dimensional geometric knapsack problem, we are given a set of $n$ axis-aligned rectangular items and an axis-aligned square-shaped knapsack. Each item has integral width, integral height and an associated integral profit. The goal is to find a (non-overlapping axis-aligned) packing of a maximum profit subset of rectangles into the knapsack. 
A well-studied and frequently used constraint in practice is to  allow only packings that \aw{are guillotine separable,} 
i.e., every rectangle in the packing can be obtained by recursively applying a sequence of edge-to-edge axis-parallel cuts that do not intersect any item of the solution. In this paper we study approximation algorithms for the geometric knapsack problem under guillotine cut constraints. 
We present polynomial time $(1+\eps)$-approximation algorithms for the cases with and without allowing rotations by 90 degrees, assuming that all input numeric data are polynomially bounded in $n$. In comparison, the best-known approximation factor for this setting is $3+\eps$ [Jansen-Zhang, SODA 2004], even in the cardinality case where all items have the same profit. 

Our main technical contribution is a structural lemma  which shows that any guillotine packing can be converted into another {\em structured} guillotine packing with almost the same profit. In this packing, each item is completely contained in one of a constant number of boxes and $\Lc$-shaped regions, inside which the items are placed by a simple greedy routine. In particular, we provide a clean sufficient condition when such a packing obeys the guillotine cut constraints which might be useful for other settings where these constraints are imposed.

\end{abstract}


\section{Introduction}\label{sec:intro} 
Geometric packing problems have many important
applications in cutting stock \cite{gilmore1965multistage}, VLSI
design \cite{Hochbaum1985}, logistics \cite{CKPT17}, smart-grids
\cite{GGIK16}, etc. Two-dimensional geometric knapsack (\tgk),
a multidimensional generalization of the classical knapsack problem,
is one of the central problems in this area. We are given a set of
$n$ axis-aligned (open) rectangles (also called {\em items}) $I:=\{1,2,\dots,n\}$,
where rectangle $i$ has integral width $\width\aw{(i)}$, integral height 
$\height\aw{(i)}$ and an associated integral profit $\profit(i)$. We
are also given an axis-aligned square knapsack $K:=[0,N]\times[0,N]$,
where $N\in\mathbb{N}$. The goal is to select a subset of items $I'\subseteq I$
of maximum total profit $\profit(I'):=\sum_{i\in I'}\profit(i)$ so
that they can be packed in the knapsack. The packing needs to be {\em
axis-parallel} and {\em non-overlapping}, i.e., such packing maps
each rectangle $i\in I'$ to a new translated open rectangle $R(i):=(\leftc\aw{(i)},\rightc\aw{(i)})\times(\bottomc\aw{(i)},\topc\aw{(i)})$
where $\rightc\aw{(i)}=\leftc\aw{(i)}+\width\aw{(i)}$, $\topc\aw{(i)}=\bottomc\aw{(i)}+\height\aw{(i)}$,
$\leftc\aw{(i)}\ge0,\bottomc\aw{(i)}\ge0$, $\rightc\aw{(i)}\le N,\topc\aw{(i)}\le N$
and for any $i,j\in I'$, we must have $R(i)\cap R(j)=\emptyset$.
In \tgk, items are not allowed to be rotated. There is another variant
with rotations \aw{that we denote by \tgkr}, where items are allowed to be rotated by $90$
degrees.\awr{removed definitions of cardinality versions, not relevant for this paper I believe}

\tgk~has rich connections with many important problems, such as
maximum independent set of rectangles (MISR) \cite{AW2013}, 2-D bin packing
\cite{bansal2014binpacking}, strip packing \cite{HarrenJPS14, Galvez0AJ0R20}, storage
allocation \cite{MomkeW20}, unsplittable flow \cite{GMW018}, mixed packing \cite{KSS21}, fair resource allocation \cite{PKL21}, etc.
Leung et al.~\cite{leung1990packing} showed that the problem is
strongly NP-hard. Jansen and Zhang \cite{Jansen2004} gave $(2+\eps)$-approximation
algorithms for both \tgk~and \tgkr, where $\eps>0$ is an arbitrarily small constant. 
Finally, G{á}lvez et al.~\cite{GalvezGHI0W17}
broke the barrier of $2$ by giving a $1.89$-approximation algorithm
for \tgk~and $(3/2+\eps)$-approximation algorithm for \tgkr. 
Furthermore, if the input data is quasi-polynomially
bounded (i.e., $N \aw{\le} n^{(\log n)^{c}}$ \aw{for some $c>0$}~) then there exists a quasi-polynomial
time approximation scheme (QPTAS) \aw{for both problems} \cite{adamaszek2015knapsack}. Polynomial
time approximation schemes (PTASs) are known for many special cases:
if all items are small \cite{FishkinGJ05}, if all items are squares
\cite{jansen2008ipco,HeydrichWiese2017}, if the profit of each item equals
its area \cite{bansal2009structural}, and if we allow resource augmentation
(i.e., the size of the knapsack can be slightly increased) \cite{fishkin2005packing,jansen2007new}.
However, it is an open problem to construct a PTAS, \aw{even with pseudo-polynomial running time}.
%

One can view geometric packing as a cutting problem where we are given
a large sheet or stock unit (maybe metal, glass, wood, rubber, or
cloth), which should be cut into pieces out of the given input set.
Cutting technology often only allows axis-parallel end-to-end
cuts called {\em guillotine cuts}. See \cite{sweeney1992cutting,BansalLS05}
for practical applications and software related to guillotine packing.
\aw{In this setting, we seek for solutions in which we can cut out
the individual objects by a recursive sequence of guillotine cuts that 
do not intersect any item of the solution.}
The related notion of {\em $k$-stage packing} was originally introduced
by Gilmore and Gomory \cite{gilmore1965multistage}. Here each stage
consists of either vertical or horizontal guillotine cuts (but not
both). On each stage, each of the sub-regions obtained on the previous
stage is considered separately and can be cut again by using horizontal
or vertical guillotine cuts. In $k$-stage packing, the number of
cuts to obtain each rectangle from the initial packing is at most
$k$, plus an additional cut to trim (i.e., separate the rectangles
itself from a waste area). Intuitively, this means that in the
cutting process we change the orientation of the cuts $k-1$ times.
The case where $k=2$, usually referred to as {\em shelf packing}, has
been studied extensively.

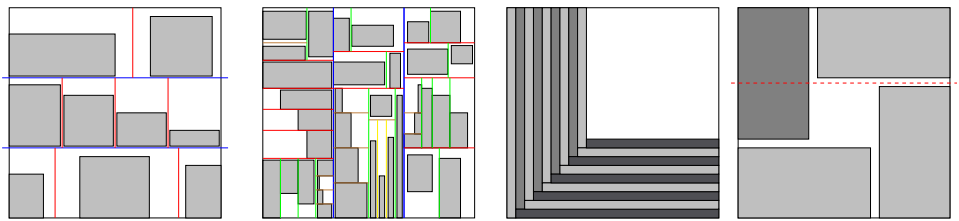
\begin{figure}[!tbh]
\captionsetup[subfigure]{justification=centering} \hspace{20pt}
\begin{subfigure}[t]{.23\textwidth} \centering \resizebox{!}{2.8 cm}{
\begin{tikzpicture}
				\draw[thick] (0,0) rectangle (6.0,6.0);
				\draw[solid, fill = lightgray] (0,0) rectangle (.97,1.25);
				\draw[solid, fill = lightgray] (2,0) rectangle (3.97,1.75);
				\draw[solid, fill = lightgray] (5,0) rectangle (6,1.5);

				\draw[solid, fill = lightgray] (0,2.05) rectangle (1.45,3.8);
				\draw[solid, fill = lightgray] (1.55,2.05) rectangle (2.95,3.5);
				\draw[solid, fill = lightgray] (3.05,2.05) rectangle (4.45,3);
				\draw[solid, fill = lightgray] (4.55,2.05) rectangle (5.95,2.5);

				\draw[solid, fill = lightgray] (0,4.05) rectangle (3,5.25);
				\draw[solid, fill = lightgray] (4,4.05) rectangle (5.75,5.75);
				
				\draw[thick, blue] (-0.2,2.0) -- ( 6.2, 2.0);
				\draw[thick, blue] (-0.2,4.0) -- ( 6.2, 4.0);

				\draw[thick, red] (1.3,0) -- ( 1.3, 2.0);
				\draw[thick, red] (4.8, 0) -- ( 4.8, 2.0);

				\draw[thick, red] (1.5,4.0) -- ( 1.5, 2.0);
				\draw[thick, red] (4.5, 4.0) -- ( 4.5, 2.0);
				\draw[thick, red] (3,4.0) -- ( 3, 2.0);

				\draw[thick, red] (3.5,4.0) -- ( 3.5, 6.0);
			\end{tikzpicture}} 
\label{figa} \end{subfigure} \begin{subfigure}[t]{.23\textwidth}
\centering \resizebox{!}{2.8cm}{ \begin{tikzpicture}

				\draw[solid, fill = lightgray] (4, 2) rectangle (4.5, 2.4);
				\draw[solid, fill = lightgray] (4.3, 2.4) rectangle (4.5, 3);
				\draw[solid, fill = lightgray] (4.4, 3) rectangle (4.5, 3.8);
				\draw[solid, fill = lightgray] (4.5, 2) rectangle (4.8, 3.7);
				\draw[solid, fill = lightgray] (4.8, 2) rectangle (5.3, 3.5);
				\draw[solid, fill = lightgray] (5.3, 2) rectangle (5.8, 3);
				\draw[solid, fill = lightgray] (4.75, 5) rectangle (5.6, 5.9);
				\draw[solid, fill = lightgray] (4.1, 5) rectangle (4.7, 5.6);
				\draw[solid, fill = lightgray] (5, 0) rectangle (5.6,1.7);
				\draw[solid, fill = lightgray] (4.1, .75) rectangle (4.8, 1.8);
				\draw[solid, fill = lightgray] (4.1, 4.1) rectangle (5.24,4.82);
				\draw[solid, fill = lightgray] (5.34, 4.4) rectangle (5.94,4.92);

				\draw[solid, fill = lightgray] (2.05, 3) rectangle (2.25,3.7);
				\draw[solid, fill = lightgray] (2.05, 2) rectangle (2.5,3);
				\draw[solid, fill = lightgray] (2.05, 1) rectangle (2.7,2);
				\draw[solid, fill = lightgray] (2.05, 0) rectangle (2.95,1);

				\draw[solid, fill = lightgray] (3.05,0) rectangle (3.2, 2.2);
				\draw[solid, fill = lightgray] (3.3,0 ) rectangle (3.45, 1.2);
				\draw[solid, fill = lightgray] (3.55,0) rectangle (3.7, 2.3);
				\draw[solid, fill = lightgray] (3.8,0) rectangle (3.95, 3.5);

				\draw[solid, fill = lightgray] (3.05,2.9) rectangle (3.65, 3.5);

				\draw[solid, fill = lightgray] (2, 4.75) rectangle (2.45,5.7);
				\draw[solid, fill = lightgray] (2.52, 4.9) rectangle (3.7,5.5);

				\draw[solid, fill = lightgray] (2, 3.8) rectangle (3.45,4.45);
				\draw[solid, fill = lightgray] (3.6, 3.7) rectangle (3.9,4.7);

				\draw[solid, fill = lightgray] (0, 4.5) rectangle (1.2,4.9);
				\draw[solid, fill = lightgray] (1.3, 4.6) rectangle (2,5.9);
				\draw[solid, fill = lightgray] (0, 5.1) rectangle ( 1.23, 5.9);

				\draw[solid, fill = lightgray] (1.25, 1.7) rectangle (1.95,2.5);
				\draw[solid, fill = lightgray] (1, 2.5) rectangle (1.95,3.1);
				\draw[solid, fill = lightgray] (.5, 3.1) rectangle (1.95,3.65);
				\draw[solid, fill = lightgray] (0, 3.7) rectangle (1.95,4.45);

				\draw[solid, fill = lightgray] (0, 1.65) rectangle (.5,0);
				\draw[solid, fill = lightgray] (.5, 1.65) rectangle (1,.7);
				\draw[solid, fill = lightgray] (1, 1.65) rectangle (1.45,0.4);

				\draw[solid, fill = lightgray] (1.55, 0) rectangle (1.95,.4);
				\draw[solid, fill = lightgray] (1.55, .4) rectangle (1.7,.8);
				\draw[solid, fill = lightgray] (1.55, .8) rectangle (1.6,1.2);
				\draw[solid, fill = lightgray] (1.55, 1.2) rectangle (2,1.65);
				
					\draw[thick] (0,0) rectangle (6.0,6.0);

				\draw[thick, blue] (2.0,0) -- ( 2.0, 6);
				\draw[thick, blue] (2.0,0) -- ( 2.0, 6);
				\draw[thick, blue] (4.0,0) -- ( 4.0, 6);
				\draw[thick, blue] (4.0,0) -- ( 4.0, 6);	
				
				\draw[thick, red] (4.0,2.0) -- ( 6.0, 2.0);
				\draw[thick, red] (4.0,4.0) -- ( 6.0, 4.0);	
				\draw[thick, red] (4.0,5.0) -- ( 6.0, 5.0);

				\draw[thick, red] (0.0,4.5) -- ( 2.0, 4.5);	
				\draw[thick, red] (0.0,1.7) -- ( 2.0, 1.7);
				\draw[thick, red] (0,2.5) -- ( 2.0, 2.5);	
				\draw[thick, red] (0, 3.1) -- ( 2.0, 3.1);
				\draw[thick, red] (0,3.7) -- ( 2.0, 3.7);

				\draw[thick, red] (2.0,4.75) -- ( 4.0, 4.75);	
				\draw[thick, red] (2.0,3.7) -- ( 4.0, 3.7);

				\draw[thick, green] (1.25,4.5) -- (1.25, 6);	

				\draw[thick, green] (1.5,0) -- ( 1.5, 1.7);
				\draw[thick, green] (1,0) -- ( 1, 1.7);
				\draw[thick, green] (.5,0) -- ( .5, 1.7);				

				\draw[thick, green] (3.,0) -- (3, 3.7);	

				\draw[thick, green] (3.5,3.7) -- ( 3.5, 4.75);

				\draw[thick, green] (2.5,4.75) -- ( 2.5, 6);

				\draw[thick, green] (4.5, 2) -- ( 4.5, 4);
				\draw[thick, green] (4.8, 2) -- ( 4.8, 4);
				\draw[thick, green] (5.3, 2) -- ( 5.3, 4);

				\draw[thick, green] (4.75, 5) -- (  4.75, 6);
				\draw[thick, green] (5, 0) -- (  5, 2);
				\draw[thick, green] (5.25, 4) -- (5.25, 5);

				\draw[thick, green] (3.75, 0) -- (3.75, 3.7);

				\draw[thick, brown] (0,5) -- ( 1.25, 5);

				\draw[thick, brown] (1.5,0.4) -- ( 2, .4);
				\draw[thick, brown] (1.5,0.8) -- ( 2, .8);
				\draw[thick, brown] (1.5,1.2) -- ( 2, 1.2);	

				\draw[thick, brown] (3,1) -- ( 2, 1);
				\draw[thick, brown] (3,2) -- ( 2, 2);
				\draw[thick, brown] (3,3) -- ( 2, 3);			

				\draw[thick, brown] (4,2.4) -- ( 4.5, 2.4);
				\draw[thick, brown] (4,3) -- ( 4.5, 3);

				\draw[thick, brown] (3, 2.8) -- (3.75, 2.8);

				\draw[thick, yellow] (3.5, 0) -- (3.5, 2.8);
				\draw[thick, yellow] (3.25, 0) -- (3.25, 2.8);
				
		\end{tikzpicture}} 
\label{figb} \end{subfigure} \begin{subfigure}[t]{.2\textwidth}
\centering \resizebox{!}{2.8cm}{ \begin{tikzpicture}
					\draw[thick](0,0) rectangle (6,6);
					\draw[solid, fill = lightgray] (0,0) rectangle (.25,6);
					\draw[solid, fill = gray] (0.25,0.25) rectangle (.5,6);
					\draw[solid, fill = lightgray] (0.5,0.5) rectangle (.75,6);
					\draw[solid, fill = gray] (0.75,0.75) rectangle (1,6);					
					\draw[solid, fill = lightgray] (1,1) rectangle (1.25,6);
					\draw[solid, fill = gray] (1.25,1.25) rectangle (1.5,6);
					\draw[solid, fill = lightgray] (1.5,1.5) rectangle (1.75,6);
					\draw[solid, fill = gray] (1.75,1.75) rectangle (2,6);
					\draw[solid, fill = lightgray] (2,2) rectangle (2.25,6);
	
					\draw[solid, fill = darkgray] (0.25,0) rectangle (6,.25);
					\draw[solid, fill = lightgray] (0.5,0.25) rectangle (6,.5);
					\draw[solid, fill = darkgray] (0.75,0.5) rectangle (6,.75);
					\draw[solid, fill = lightgray] (1,0.75) rectangle (6,1);
					\draw[solid, fill = darkgray] (1.25,1) rectangle (6,1.25);
					\draw[solid, fill = lightgray] (1.5,1.25) rectangle (6,1.5);
					\draw[solid, fill = darkgray] (1.75,1.5) rectangle (6,1.75);
					\draw[solid, fill = lightgray] (2,1.75) rectangle (6,2);
					\draw[solid, fill = darkgray] (2.25,2) rectangle (6,2.25);
				\end{tikzpicture}} 
\label{figc} \end{subfigure} \begin{subfigure}[t]{.23\textwidth}
\centering \resizebox{!}{2.8cm}{ \begin{tikzpicture}
				\draw[thick] (0,0) rectangle (6.0,6.0);

				\draw[solid, fill = lightgray] (0,0) rectangle (3.75,2);
				\draw[solid, fill = gray] (0,6) rectangle (2,2.25);
				\draw[solid, fill = lightgray] (6,6) rectangle (2.25,4);
				\draw[solid, fill = lightgray] (6,0) rectangle (4,3.75);
				
				\draw[thick, dashed, red] (-0.2,3.85) -- (6.2, 3.85 );			
			\end{tikzpicture}} 
\label{figd} \end{subfigure} \caption{The first three packing are guillotine separable packings of 2-stages, 5-stages, and many stages, respectively. The last packing is not a guillotine packing as any end-to-end cut in the knapsack intersects at least one of the packed rectangles.}
\label{fig:guillotinestages1}
\end{figure}

In this paper, we study the two-dimensional knapsack problem under
guillotine cuts (\tggk). The input is the same as for \tgk,
but we require additionally that the items in the solution can be
separated by a sequence of guillotine cuts, and we say that then they
are \emph{guillotine separable}. 
NP-hardness of \tggk~follows from a reduction from the \aw{(one-dimensional) knapsack problem}.
Christofides et al.~\cite{christofides1977algorithm} studied the
problem in 1970s. Since then many heuristics have been developed to
efficiently solve benchmark instances, based on tree-search \cite{viswanathan1988exact},
branch-and-bound \cite{hadjiconstantinou1995exact}, dynamic optimization
\cite{beasley1985exact}, tabu search \cite{alvarez2002tabu}, genetic
algorithms \cite{parada1995hybrid}, etc. Despite a staggering number
of recent experimental papers \cite{dolatabadi2012exact,wei2015bidirectional,borgulya2019eda,lodi2017partial,di2013algorithms,clautiaux2018combining,furini2016modeling, clautiaux2019pattern},
there was little theoretical progress for \tggk, due to limitations
of past techniques. Since 2004, the $(3+\eps)$-approximation for
\tgk~by Jansen and Zhang~\cite{Jansen2004} has been the best-known
approximation algorithm for \tggk. Recently, Abed et al.~\cite{AbedCCKPSW15}
have studied approximation algorithms for the cardinality cases of \tggk~and \tggkr~and 
have given a QPTASs, assuming the input data to be quasi-polynomially
bounded.

Most algorithms for \tgk~utilize a {\em container packing} (see
Section \ref{sec:prelim}) which arranges the items in the knapsack
such that they are packed inside a constant number of axis-aligned
boxes (containers). The best sizes and locations of these containers 
can be guessed efficiently since there are only a constant number
of them. Then inside each container the items are packed either
\aw{in} one-stage packings or \aw{in} two-stage packings~ (if items are small). However, G{á}lvez
et al.~\cite{GalvezGHI0W17} show that \aw{one cannot obtain a better approximation
ratio than $2$ with container-based
packings with only $O(1)$ many containers,} due to interaction between \emph{horizontal} (wide and thin)
and \emph{vertical} (tall and narrow) items. To break this barrier,
they use a {\em corridor-decomposition} where the knapsack is divided
into axis-parallel polygonal regions called {\em corridors} with
constant number of  regions called {\em subcorridors}.
Vertical (resp. horizontal) items are packed in only vertical (resp.
horizontal) subcorridors. After simplifying the interaction between
vertical and horizontal items, they define two types of packings.
In one packing, they {\em process} the subcorridors to obtain a
container-based packing. In the other, a profitable subset of long
horizontal and long vertical items are packed in an $\Lc$-shaped
region. They prove that the best of these two packings achieves
a better approximation ratio than $2$. However, it is not clear how
to use this approach for \tggk: even if we start with an optimal
guillotine packing, the rearrangements of items may not preserve guillotine
separability, and hence they might not lead to a feasible solution
to \tggk.


\subsection{Our contribution}

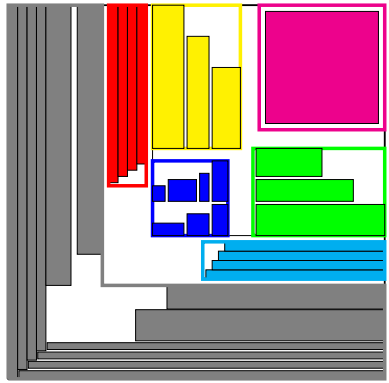
\begin{figure}[h]
		\centering
		\resizebox{!}{5cm}{
		\begin{tikzpicture}		
			\draw[thick] (0,0) rectangle (6,6);	
			\draw[solid] (0,0)--(0,6)--(1.5,6)--(1.5,1.5)--(6,1.5)--(6,0)--(0,0);

			\draw[solid, fill=gray] (0,0)rectangle(.15,6);
			\draw[solid, fill=gray] (.15,.15)rectangle(.3,6);
			\draw[solid, fill=gray] (.3,.3)rectangle(.45,6);
			\draw[solid, fill=gray] (.45,.45)rectangle(.6,6);
			\draw[solid, fill=gray] (.6,1.5)rectangle(1,6);
			\draw[solid, fill=gray] (1.1,2)rectangle(1.5,6);

			\draw[solid, fill=gray] (0.17,0)rectangle(6,.13);
			\draw[solid, fill=gray] (0.32,.17)rectangle(6,.28);
			\draw[solid, fill=gray] (0.47,0.32)rectangle(6,.43);
			\draw[solid, fill=gray] (0.62,0.47)rectangle(6,.58);
			\draw[solid, fill=gray] (2.03,.61)rectangle(6,1.11);
			\draw[solid, fill=gray] (2.53,1.12)rectangle(6, 1.5);

			\draw [solid] (6,6)rectangle(2.3,2.3);

			\draw [solid] (1.6,6)rectangle(2.2,3.1);
			\draw [solid] (6,1.6)rectangle(3.1,2.2);

			\draw[solid, fill=red] (1.6,3.15)rectangle(1.75,6);
			\draw[solid, fill=red] (1.75,3.25)rectangle(1.9,6);
			\draw[solid, fill=red] (1.9,3.35)rectangle(2.05,6);
			\draw[solid, fill=red] (2.05,3.45)rectangle(2.2,6);
			\draw[solid, fill=cyan] (3.15,1.6)rectangle(6,1.75);
			\draw[solid, fill=cyan] (3.25,1.75)rectangle(6,1.9);
			\draw[solid, fill=cyan] (3.35,1.9)rectangle(6,2.05);
			\draw[solid, fill=cyan] (3.45,2.05)rectangle(6,2.2);

			\draw [ultra thick, magenta](6,6)rectangle(4,4);
			\draw[solid, fill=magenta] (5.9,5.9)rectangle(4.1,4.1);

			\draw [ultra thick, yellow](2.3,3.7) rectangle (3.7,6);
			\draw[solid, fill=yellow] (2.3,3.7)rectangle(2.8,6);
			\draw[solid, fill=yellow] (2.85,3.7)rectangle(3.2,5.5);
			\draw[solid, fill=yellow] (3.25,3.7)rectangle(3.7,5);

			\draw[ultra thick, green] (3.9,2.3) rectangle (6,3.7);
			\draw[solid, fill=green] (3.95,2.3)rectangle(6,2.8);
			\draw[solid, fill=green] (3.95,2.85)rectangle(5.5,3.2);
			\draw[solid, fill=green] (3.95,3.25)rectangle(5,3.7);

			\draw [ultra thick, blue] (2.3,2.3) rectangle (3.5,3.5);
			\draw[solid, fill=blue] (2.3,2.3)rectangle(2.8,2.5);
			\draw[solid, fill=blue] (2.85,2.3)rectangle(3.2,2.65);
			\draw[solid, fill=blue] (3.25,2.3)rectangle(3.5,2.8);

			\draw[solid, fill=blue] (2.3,2.85)rectangle(2.5,3.1);
			\draw[solid, fill=blue] (2.55,2.85)rectangle(3,3.2);
			\draw[solid, fill=blue] (3.05,2.85)rectangle(3.2,3.3);
			\draw[solid, fill=blue] (3.25,2.85)rectangle(3.5,3.5);

			\draw [ultra thick, red] (1.6,6)rectangle(2.2,3.1);
			\draw [ultra thick, cyan] (6,1.6)rectangle(3.1,2.2);
			\draw[ultra thick, gray] (0,0)--(0,6)--(1.5,6)--(1.5,1.5)--(6,1.5)--(6,0)--(0,0);
		\end{tikzpicture}}
	\caption{\aw{A structured packing of items into compartments that each have the shape of an $\Lc$- or a rectangular box.}} 
	\label{fig:packing}
\end{figure}

In this paper, we obtain $(1+\eps)$-approximation algorithms \aw{with pseudo-polynomial running time for both \tggk~and \tggkr, i.e., the running time is a polynomial if 
the (integral) input numbers are all polynomially bounded in $n$}. 
The key idea is to show that there
are $(1+\eps)$-approximate solutions in which the knapsack is
divided into simple {\em compartments} that each have the shape of a rectangular
box or an $\Lc$, see Figure \ref{fig:packing}. Inside each compartment, the items are
placed in a very simple way, e.g., all horizontal items are simply
stacked on top of each other, all vertical items are placed side
by side, and all small items are packed greedily with the Next-Fit-Decreasing-Height
algorithm \cite{coffman1980performance}, see Figure \ref{fig:packing}. 
To establish this structure,
we crucially exploit that the optimal solution is guillotine separable;
in particular, in \tgk~(where the optimal solution might not be
guillotine separable) more complicated compartments may be necessary
for near-optimal solutions, e.g., with the form of a ring.

While the items in our structured solution are guillotine separable,
we cannot separate the compartments by guillotine cuts since we cannot
cut out an $\Lc$-shaped compartment with such cuts. This makes it difficult
to compute a solution of this type since it is not sufficient to ensure
that (locally) within each compartment the items are guillotine separable
(which is immediately guaranteed by our simple packings inside them).
Therefore, our compartments have an important additional property:
they can be separated by a \emph{pseudo guillotine cutting sequence.}
This is a cutting sequence in which each step is either a guillotine
cut, or a cut along two line segments that separates a rectangular
area into an $\Lc$-shaped compartment and a smaller rectangular area,
see Figure \ref{fig:pseudonice}.
We prove a strong property for compartments that admit
such a pseudo guillotine cutting sequence: we show that 
\aw{if we pack items into such compartments in the simple way mentioned above,}
\aw{this will \emph{always}} yield a solution that is \emph{globally }guillotine separable.
This property and our structural result might have applications also
in other settings where we are interested in solutions that are guillotine
separable.


Our strong structural result allows us to construct algorithms (for
the cases with and without rotations) that are relatively simple:
we first guess the constantly many compartments in the structured
solution mentioned above. Then we compute up to a factor $1+\eps$, 
the most profitable set of items that can be placed nicely into them,
using a simplified version of a recent algorithm in \cite{GalSocg21}. 
The resulting
solutions use \aw{up to $\Theta(\log (nN))$} stages (unlike e.g.,
solutions of the Next-Fit-Decreasing-Height algorithm \cite{coffman1980performance} that
need only two stages). We \aw{prove a lower bound, showing} 
that there is a family of instances of \tggk~that does not admit $(2-\eps)$-approximate
solutions with only $o(\log N)$-stages. 

\amar{if this fig hasn't been referred for color-coding, we can probably remove the colors, make all items gray?}

\awr{changed the caption of Figure 2, was not understandable at this point IMO.}

\subsection{Other related work}
\aw{There are many well-studied geometric packing problems.} 
In the 2D bin packing problem~(2BP), we are given a set of rectangular items and unit square bins, and the goal is to pack all the items into a minimum number of bins. The problem is APX-hard \cite{bansal2006bin} and the \aw{currently} best   
\aw{known} 
approximation ratio is $1.405$ \cite{bansal2014binpacking}. In the 2D strip packing problem (2SP), we are given a set of rectangular items and a fixed-width unbounded-height strip, and the goal is to pack all the items into the strip such that the height of the strip is minimized. Kenyon and R{\'{e}}mila gave an APTAS for the problem \cite{kenyon2000near} using a 3-stage packing. 

Both 2BP and 2SP are well-studied \aw{in the} guillotine \aw{setting} \cite{pietrobuoni2015two}. 
Caprara \cite{caprara2002packing} gave a 2-stage $T_{\infty} (\approx 1.691)$-approximation for 2BP.
Afterwards, Caprara et al.~\cite{caprara2005fast} gave an APTAS for 2-stage 2BP and 2-stage 2SP. 
Later, Bansal et al.~\cite{BansalLS05} showed an APTAS for guillotine 2BP. 
Bansal et al.~\cite{bansal2014binpacking} conjectured that the worst-case ratio between the best guillotine 2BP and the best general 2BP is $4/3$. If true, this would  imply a $(\frac43+\eps)$-approximation algorithm for 2BP.  
Seiden et al.~\cite{seiden2005two} gave an APTAS for guillotine 2SP. Both the APTAS for guillotine 2BP and guillotine 2SP are based on the fact that general guillotine 2BP or guillotine 2SP can be approximated arbitrarily well by $O(1)$-stage packings, and such $O(1)$-stage packings can be found efficiently. Interestingly, we showed that this property is not true for \tggk. 

Pach and Tardos \cite{pach2000cutting}  conjectured that, for any set of $n$ non-overlapping axis-parallel rectangles, there is a guillotine cutting sequence separating $\Omega(n)$ of them. 
Recently, the problem has received attention in \cite{AbedCCKPSW15, KMR20} \aw{since} a proof of this conjecture \aw{would} imply a $O(1)$-approximation for \aw{the Maximum Independent Set of Rectangles problem, which is
a long-standing open problem.} 



\section{Methodology\label{sec:prelim}}
\awr{moved this part}
We first present our algorithm for the cardinality case, i.e.,
assuming that $\profit(i)=1$ for each item $i\in\R$. In Appendix \ref{apx:algo-weighted}
we \aw{will present a generalization} to arbitrary item profits. 
For each $n\in \mathbb{N}$ we define $[n]:=\{1,2,\dots,n\}$. 

We classify the input
items according to their heights and widths. For two constants $1\geq\epsl>\epss>0$
to be defined later, we classify each item $i\in \R$ as: 
\begin{itemize}
\item {\em Large}: $\width_{i}\ >\epsl N$ and $\height_{i}\ >\epsl N$; 
\item {\em Small}: $\width_{i}\leq\epss N$ and $\height_{i}\leq\epss N$; 
\item {\em Horizontal}: $\width_{i}\ >\epsl N$ and $\height_{i}\leq\epss N$; 
\item {\em Vertical}: $\height_{i}\ >\epsl N$ and $\width_{i}\leq\epss N$; 
\item {\em Intermediate:} Either $\epsl N\geq\height_{i}\ >\epss N$
or $\epsl N\geq\width_{i}\ >\epss N$. 
\end{itemize}
Using standard shifting arguments, one can show that we can ignore
intermediate items.

\begin{lem} \label{class} \cite{GalvezGHI0W17} Let $\eps>0$ and
$f(.)$ be any positive increasing function such that $f(x)<x$ $\forall x\in(0,1]$.
Then we can efficiently find $\epsl,\epss\in\Omega_{\eps}(1)$, with
$\eps\ge f(\eps)\ge\epsl\ge f(\epsl)\ge\epss$ so that the total profit
of intermediate rectangles is at most $\eps\profit(\opt)$. \end{lem}

\noindent We define {\em skewed} items to be items that are horizontal
or vertical. Let $\Rla$, $\Rsm$, $\Rho,\Rve,\Rsk$ be the set of
large, small, horizontal, vertical, and skewed rectangles, respectively.
The corresponding intersections with $\opt$ (the optimal guillotine
packing) defines the sets $\optla$, $\optsm$, $\optho,$ $\optve,\optsk$, 
respectively.


\subsection{Compartments}

Our goal is to partition the knapsack into compartments, such that there
is an $(1+\eps)$-approximate solution whose items are placed
in a structured way inside these compartments. We will use two types
of compartments: \emph{box-compartments} and \emph{$\Lc$-compartments}. 
\begin{definition}
[Box-compartment] A \emph{box-compartment} $B$ is an axis-aligned rectangle
that satisfies $B\subseteq K$. \amar{we may have used $\Bc$ instead of box- at some places, if $\Bc$ is no longer needed we can change the command itself $\backslash Bc := 'box'$}
\end{definition}
\begin{definition}
[$\Lc$-compartment] An $\Lc$-compartment $L$ is a subregion of $K$ bounded
by a simple rectilinear polygon with six edges $e_{0},e_{1},\dots,e_{5}$
such that for each pair of horizontal (resp. vertical) edges
$e_{i},e_{6-i}$ with $i\in\{1,2\}$ there exists a vertical (resp.
horizontal) line segment $\ell_{i}$ of length less than $\ari{\epsl N/2}$ 
such that both $e_{i}$ and $e_{6-i}$ intersect $\ell_{i}$ but no
other edges intersect $\ell_{i}$. 
\end{definition}

Since the length of the line segments $\ell_{i}$ is less
than $\epsl N/2$, this implies that inside an $\Lc$-compartment $L$ we cannot
place large items, inside the horizontal part of $L$ we cannot place
 vertical items, and inside the vertical part of $L$ we cannot
place horizontal items.

We seek for a structured packing inside of these compartments according
to the following definitions. Inside box-compartments, we want only
one type of items and we want that the skewed items are placed in
a very simple way, see Figure \ref{fig:packing}.
\begin{definition}
\label{def:structured-boxes}Let $B$ be a box-compartment and let $I_{B}\subseteq I$
be a set of items that are placed non-overlappingly inside $B$. We
say that the placement of $I_{B}$ is \emph{nice} if the items in
$I_{B}$ are guillotine separable and additionally
\begin{itemize}
\item $I_{B}$ contains only one item, or
\item $I_{B}\subseteq I_{hor}$ and the items in $I_{B}$ are stacked on
top of each other inside $B$, or
\item $I_{B}\subseteq I_{ver}$ and the items in $I_{B}$ are placed side
by side inside $B$, or
\item $I_{B}\subseteq I_{small}$ and for each item $i\in I_{B}$ it holds
that $w_{i}\le\eps \cdot w(B)$ and $h_{i}\le\eps \cdot h(B)$
\end{itemize}
\end{definition}

\begin{figure}[h]
	\captionsetup[subfigure]{justification=centering}
	\hspace{-10pt}
	\begin{subfigure}[b]{.32\textwidth}
		\centering
		\resizebox{!}{4.5cm}{
		\begin{tikzpicture}			
			\draw[ultra thick](1.5,6)--(1.5,1.5)--(6,1.5)--(6,0)--(0,0)--(0,6)--(1.5,6);
	
			\draw[solid, fill=lightgray] (.26,0) rectangle (6,.06); 
			\draw[solid, fill=lightgray] (.28,.06) rectangle (6,.12); 
			\draw[solid, fill=lightgray] (.34,.12) rectangle (6,.18); 
			\draw[solid, fill=lightgray] (.40,.18) rectangle (6,.24);
			\draw[solid, fill=lightgray] (.51,.26) rectangle (6,.35);
			\draw[solid, fill=lightgray] (.6,.35) rectangle (6,.43); 
			\draw[solid, fill=lightgray] (.67,.43) rectangle (6,.49);
			\draw[solid, fill=lightgray] (.76,.51) rectangle (6,.63); 
			\draw[solid, fill=lightgray] (.88,.63) rectangle (6,.73);
			\draw[solid, fill=lightgray] (1.01,.76) rectangle (6,.87); 
			\draw[solid, fill=lightgray] (1.25,.87) rectangle (6,.98);
			\draw[solid, fill=lightgray] (1.51,1.01) rectangle (6,1.19); 
			\draw[solid, fill=lightgray] (1.75,1.19) rectangle (6,1.4);
			\draw[solid, fill=lightgray] (.15,.6) rectangle (.35,6);
			\draw[solid, fill=lightgray] (.4,.8) rectangle (.60,6);
			\draw[solid, fill=lightgray] (.65,1) rectangle (.85,6);
			\draw[solid, fill=lightgray] (.9,1.2) rectangle (1.1,6);
			\draw[solid, fill=lightgray] (1.15,1.45) rectangle (1.35,6);

			\draw[dashed, thick] (0.35, 6.2)--(0.35,-0.2);

			
			\draw [thick,decorate,decoration={brace,amplitude=8pt}] (6,.18) -- (6,0);

			\draw (6.7,.09) node {\large \textbf{$I_{hor}'$}}; 
%

			\draw (0.15,6.2) node {\large \textbf{$i_v$}};

			\draw (0.35,6.5) node {\large \textbf{$l_v$}};
		\end{tikzpicture}}
		\caption{}
		\label{fig:lguil1}
	\end{subfigure}
	\begin{subfigure}[b]{.32\textwidth}
		\centering
		\resizebox{!}{4.5cm}{
		\begin{tikzpicture}						
			\draw[ultra thick](1.5,6)--(1.5,1.5)--(6,1.5)--(6,0)--(0,0)--(0,6)--(1.5,6);
	
			\draw[solid, fill=lightgray] (.26,0) rectangle (6,.06); 
			\draw[solid, fill=lightgray] (.28,.06) rectangle (6,.12); 
			\draw[solid, fill=lightgray] (.34,.12) rectangle (6,.18); 
			\draw[solid, fill=lightgray] (.40,.18) rectangle (6,.24);
			\draw[solid, fill=lightgray] (.51,.26) rectangle (6,.35);
			\draw[solid, fill=lightgray] (.6,.35) rectangle (6,.43); 
			\draw[solid, fill=lightgray] (.67,.43) rectangle (6,.49);
			\draw[solid, fill=lightgray] (.76,.51) rectangle (6,.63); 
			\draw[solid, fill=lightgray] (.88,.63) rectangle (6,.73);
			\draw[solid, fill=lightgray] (1.01,.76) rectangle (6,.87); 
			\draw[solid, fill=lightgray] (1.25,.87) rectangle (6,.98);
			\draw[solid, fill=lightgray] (1.51,1.01) rectangle (6,1.19); 
			\draw[solid, fill=lightgray] (1.75,1.19) rectangle (6,1.4);
			\draw[solid, fill=lightgray] (.15,.6) rectangle (.35,6);
			\draw[solid, fill=lightgray] (.4,.8) rectangle (.60,6);
			\draw[solid, fill=lightgray] (.65,1) rectangle (.85,6);
			\draw[solid, fill=lightgray] (.9,1.2) rectangle (1.1,6);
			\draw[solid, fill=lightgray] (1.15,1.45) rectangle (1.35,6);

			\draw[dashed, thick] (0.35, 6.2)--(0.35,-0.2);

			\draw[thick] (-0.2, .18) -- (6.2, .18);
			
			\draw [thick,decorate,decoration={brace,amplitude=8pt}] (6,.18) -- (6,0);

			\draw (6.7,.09) node {\large \textbf{$I_{hor}'$}}; 

			\draw (-.5,.18) node {\large \textbf{$l_h$}};

			\draw (0.15,6.2) node {\large \textbf{$i_v$}};

			\draw (0.35,6.5) node {\large \textbf{$l_v$}};
		
		\end{tikzpicture}}
		\caption{}
		\label{fig:lguil2}
	\end{subfigure}
	\begin{subfigure}[b]{.32\textwidth}
		\centering
		\resizebox{!}{4.5cm}{
		\begin{tikzpicture}
			\draw[ultra thick](1.5,6)--(1.5,1.5)--(6,1.5)--(6,0.18)--(0,0.18)--(0,6)--(1.5,6);
	
			\draw[solid, fill=lightgray] (.40,.18) rectangle (6,.24);
			\draw[solid, fill=lightgray] (.51,.26) rectangle (6,.35);
			\draw[solid, fill=lightgray] (.6,.35) rectangle (6,.43); 
			\draw[solid, fill=lightgray] (.67,.43) rectangle (6,.49);
			\draw[solid, fill=lightgray] (.76,.51) rectangle (6,.63); 
			\draw[solid, fill=lightgray] (.88,.63) rectangle (6,.73);
			\draw[solid, fill=lightgray] (1.01,.76) rectangle (6,.87); 
			\draw[solid, fill=lightgray] (1.25,.87) rectangle (6,.98);
			\draw[solid, fill=lightgray] (1.51,1.01) rectangle (6,1.19); 
			\draw[solid, fill=lightgray] (1.75,1.19) rectangle (6,1.4);
			\draw[solid, fill=lightgray] (.15,.6) rectangle (.35,6);
			\draw[solid, fill=lightgray] (.4,.8) rectangle (.60,6);
			\draw[solid, fill=lightgray] (.65,1) rectangle (.85,6);
			\draw[solid, fill=lightgray] (.9,1.2) rectangle (1.1,6);
			\draw[solid, fill=lightgray] (1.15,1.45) rectangle (1.35,6);


			\draw[thick] (-0.2, .18) -- (6.2, .18);
			
			\draw [thick,decorate,decoration={brace,amplitude=8pt}] (6,-.22) -- (6,-.4);

			\draw (6.7,-.31) node {\large \textbf{$I_{hor}'$}}; 

			\draw (-.5,.18) node {\large \textbf{$l_h$}};



			\draw[ultra thick] (0,-.4) rectangle (6,-.22);
			\draw[solid, fill=lightgray] (.26,-.4) rectangle (6,-.34); 
			\draw[solid, fill=lightgray] (.28,-.34) rectangle (6,-.28); 
			\draw[solid, fill=lightgray] (.34,-.28) rectangle (6,-.22);

		\end{tikzpicture}}
		\caption{}
		\label{fig:lguil3}
	\end{subfigure}
	\caption{(a) \aw{A nicely packed set of skewed items inside an $\Lc$-compartment. The vertical cut $l_v$ 
	separates the leftmost vertical item $i_v$ from the other vertical items but it intersects the horizontal items in $I_{hor}'$. (b) However, then the horizontal cut $l_h$ separates the items in $I_{hor}'$ from the other horizontal items without intersecting any vertical item. (c) The corresponding guillotine cut that partitions the   
	$\Lc$-compartment  into a box-compartment and a smaller $\Lc$-compartment. 
	}
	}
	\label{fig:lguil}
\end{figure}
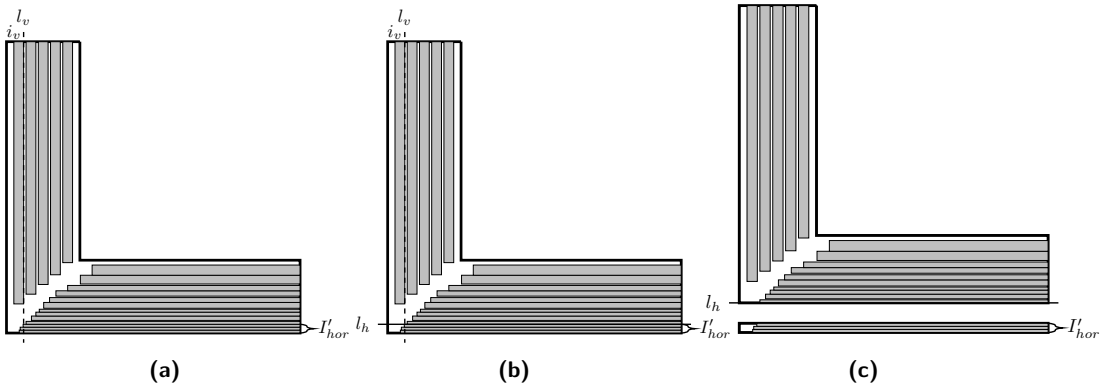

Inside $\Lc$-compartments we allow only skewed items and we want them to be placed in a similar
way as in the boxes, see Figure \ref{fig:packing} and \ref{fig:lguil}. 
\begin{definition}
Let $L$ be an $\Lc$-compartment and let $I_{L}\subseteq I$ be a set of
items that are placed non-overlappingly inside $L$. We say that the
placement of $I_{L}$ is \emph{nice }if 
\begin{itemize}
\item $I_{L}\subseteq\Rsk$, and
\item the items in $I_{L}\cap I_{hor}$ are stacked on top of each other
inside $L$, and
\item the items in $I_{L}\cap I_{ver}$ are stacked side by side inside
$L$.
\end{itemize}
\end{definition}

A nice placement inside an $\Lc$-compartment yields
a guillotine separable packing.
\begin{lem}\label{Lguillotine}
Consider a set of items $I_{L}\subseteq I$ that is placed nicely
inside an $\Lc$-compartment $L$. Then $I_{L}$ is guillotine separable.
\end{lem}

\begin{proof}[Proof sketch]
\aw{One can show that there always exists a guillotine cut that separates one or more horizontal
or vertical items in $~I_{L}$ from the other items in $I_{L}$, see Figure \ref{fig:lguil}. Then this argument
is applied recursively.} \ama{See Appendix \ref{sec:omit} for full proof.}
\end{proof}

\subsection{Pseudo-guillotine separable compartments}

We seek to partition the knapsack into box- and $\Lc$-compartments and
then place items into these compartments. We  also want to
ensure that the resulting solution is guillotine separable. We could
guarantee this if there was a guillotine cutting sequence that separates
all compartments and require that the items inside the compartments are
placed nicely. Then, we could first separate all compartments by the
mentioned cutting sequence and then separate the items inside
of each compartment by guillotine cuts (as they are packed nicely).

However, there is no guillotine cutting sequence that cuts out an
$\Lc$-compartment from the knapsack since no guillotine cut can separate
the $\Lc$-compartment from the area at the ``inner'' part of the $\Lc$-compartment. Therefore, we require for the compartments in our knapsack
only that there is a \emph{pseudo-guillotine cutting sequence}. A
pseudo-guillotine cutting sequence has the following two operations
(see Figure \ama{\ref{fig:pseudonice}}): given a rectangle $R\subseteq K$ it 
\begin{itemize}
\item applies a horizontal or vertical guillotine cut that separates $R$
into two disjoint rectangles $R_{1},R_{2}$ and then continues recursively
with $R_{1}$ and $R_{2}$, or
\item for an $\Lc$-compartment $L\subseteq R$ such that $R\setminus L$ is a
rectangle, it partitions $R$ into $L$ and $R\setminus L$ and then
continues recursively with $R\setminus L$ (but not with $L$). Note
that we cannot do this operation with every $\Lc$-compartment $L'\subseteq R$
since possibly $R\setminus L'$ is not a rectangle.
\end{itemize}
We formalize this in the following definition.
\begin{definition}
\label{def:pseudoGc}
A \emph{pseudo-guillotine cutting sequence }(for compartments)\emph{
}for a set of compartments $\C$ is a binary tree $T=(V,E)$ where for
each vertex $v\in V$ there is an associated shape $S_{v}\subseteq K$
such that 
\begin{itemize}
\item for the root $r\in V$ of $T$ it holds that $S_{r}=K$,
\item for each internal vertex $v$ with children $u,w$ it holds that 
\begin{itemize}
\item $S_{v}$ is a rectangle with $S_{v}=S_{u}\dot{\cup}S_{w}$ (so in
particular $S_{u}$ and $S_{w}$ are disjoint),
\item either $S_{u}$ and $S_{w}$ are both rectangles or one of them is
an $\Lc$-compartment and the other is a rectangle,
\end{itemize}
\item for each compartment $C\in\C$ there is a leaf $v\in V$ such that $S_{v}=C$.
\end{itemize}
\end{definition}

Observe that each $\Lc$-compartment corresponds to a leaf node in $T$. \amar{doesn't the line above tell the same thing?}

Now the important insight is that if a set of compartments $\C$ admits
a pseudo-guillotine cutting sequence, then \emph{any} \underline{nice}
placement of items inside these compartments is guillotine separable
(globally). In particular, given such compartments $\C$, we can place
items inside the compartments in $\C$ without needing to worry whether
the resulting packing will be guillotine separable globally, as long
as we place these items nicely. Intuitively, this is true since we
can use the cuts of the pseudo-guillotine cutting sequence as a template
for a global cutting sequence for the items: whenever the former sequence 
\begin{itemize}
\item makes a guillotine cut we simply do the same cut,
\item when it separates an $\Lc$-compartment $L$ from a rectangular region $R$,
we separate the items inside $L$ by a sequence of guillotine cuts;
it turns out that we can do this since all items inside $L$ are placed
nicely and skewed.
\end{itemize}
Finally, we separate the items inside each box-compartment $B$ by guillotine
cuts, using the fact that the items inside $B$ are placed nicely.
\begin{lem}
\label{lem:pseudo-enough}Let $\C$ be a set of compartments inside
$K$ that admit a pseudo-guillotine cutting sequence. Let $\R'\subseteq\R$
be a set of items that are placed nicely inside the compartments in
$\C$. Then there is a guillotine cutting sequence for $\R'$.
\end{lem}
\begin{proof}[Proof sketch.]
\aw{Let $P$ denote the pseudo-guillotine cutting sequence. 
We construct a guillotine cutting sequence for $\R'$ based on $P$. We follow the cuts of $P$.
Whenever $P$ 
makes a guillotine cut, then we also do this guillotine cut. When $P$ separates
an $\Lc$-compartment $L$ from a rectangular region $R$, then we apply a sequence of guillotine 
cuts that step by step separates all items in $L$ from $R\setminus L$. Since inside $L$ the items are placed nicely, one can show that there exist such cuts 
that don't intersect any item in $R\setminus L$} \ama{(see Figure~\ref{fig:seqL}). See Appendix \ref{sec:omit} for full proof.}
\end{proof}

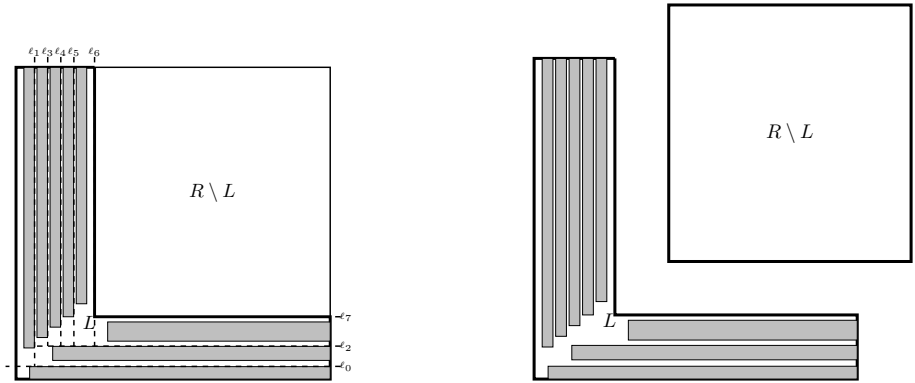
\begin{figure}[h]
	\captionsetup[subfigure]{justification=centering}
	\hspace{-10pt}
	\begin{subfigure}[b]{.5\textwidth}
		\centering
		\resizebox{!}{4.5cm}{
		\begin{tikzpicture}		
			\draw[ thick ] (0,0) rectangle (6,6);	
			\draw[ultra thick](1.5,6)--(1.5,1.2)--(6,1.2)--(6,0)--(0,0)--(0,6)--(1.5,6);	
			
			\draw[solid, fill=lightgray] (.26,0) rectangle (6,.24); 
			\draw[solid, fill=lightgray] (.7,.36) rectangle (6,.63);
			\draw[solid, fill=lightgray] (1.75,.73) rectangle (6,1.1);

			\draw[solid, fill=lightgray] (.15,.6) rectangle (.35,6);
			\draw[solid, fill=lightgray] (.4,.8) rectangle (.60,6);
			\draw[solid, fill=lightgray] (.65,1) rectangle (.85,6);
			\draw[solid, fill=lightgray] (.9,1.2) rectangle (1.1,6);
			\draw[solid, fill=lightgray] (1.15,1.45) rectangle (1.35,6);

			\draw[dashed, thick] (6.2, .245)--(-.2,0.245);
			\draw (6.3, .245) node {\tiny \textbf{$\ell_0$}};	
			
			\draw[dashed, thick] (0.355, 6.2)--(0.355,.245);
			\draw (.355, 6.3) node {\tiny \textbf{$\ell_1$}};	

			\draw[dashed, thick] (6.2, .635)--(.355, .635);
			\draw (6.3, .635) node {\tiny \textbf{$\ell_2$}};	

			\draw[dashed, thick] (0.605, 6.2)--(0.605,.635);
			\draw (.605, 6.3) node {\tiny \textbf{$\ell_3$}};	

			\draw[dashed, thick] (0.855, 6.2)--(0.855,.635);
			\draw (.855, 6.3) node {\tiny \textbf{$\ell_4$}};	

			\draw[dashed, thick] (1.105, 6.2)--(1.105,.635);
			\draw (1.105, 6.3) node {\tiny \textbf{$\ell_5$}};	

			\draw[dashed, thick] (1.5, 6.2)--(1.5,.635);
			\draw (1.5, 6.3) node {\tiny \textbf{$\ell_6$}};	

			\draw[dashed, thick] (6.2, 1.2)--(1.5,1.2);
			\draw (6.3, 1.2) node {\tiny \textbf{$\ell_7$}};	

			\draw (3.75,3.6) node {\large \textbf{$R \setminus L$}};
			\draw (1.4,1.1) node {\large \textbf{$L$}};
		\end{tikzpicture}}
		\label{fig:seqL1}
	\end{subfigure}
	\begin{subfigure}[b]{.5\textwidth}
		\centering
		\resizebox{!}{5cm}{
		\begin{tikzpicture}				
			\draw[ultra thick] (2.5, 2.2) rectangle (7,7);		
			\draw[ultra thick](1.5,6)--(1.5,1.2)--(6,1.2)--(6,0)--(0,0)--(0,6)--(1.5,6);	
			
			\draw[solid, fill=lightgray] (.26,0) rectangle (6,.24); 
			\draw[solid, fill=lightgray] (.7,.36) rectangle (6,.63);
			\draw[solid, fill=lightgray] (1.75,.73) rectangle (6,1.1);

			\draw[solid, fill=lightgray] (.15,.6) rectangle (.35,6);
			\draw[solid, fill=lightgray] (.4,.8) rectangle (.60,6);
			\draw[solid, fill=lightgray] (.65,1) rectangle (.85,6);
			\draw[solid, fill=lightgray] (.9,1.2) rectangle (1.1,6);
			\draw[solid, fill=lightgray] (1.15,1.45) rectangle (1.35,6);

			\draw (4.75,4.6) node {\large \textbf{$R \setminus L$}};
			\draw (1.4,1.1) node {\large \textbf{$L$}};
	
		\end{tikzpicture}}
		\label{fig:seqL2}
	\end{subfigure}
	\caption{\ama{Partition of rectangle $R$ into $L$ and $R\setminus L$ when items inside $L$ are packed nicely. $\ell_0, \dots, \ell_7$ (dashed lines) are a sequence of guillotine cuts that ultimately separate out the items in $L$ from $R$.}
	}
	\label{fig:seqL}
\end{figure}

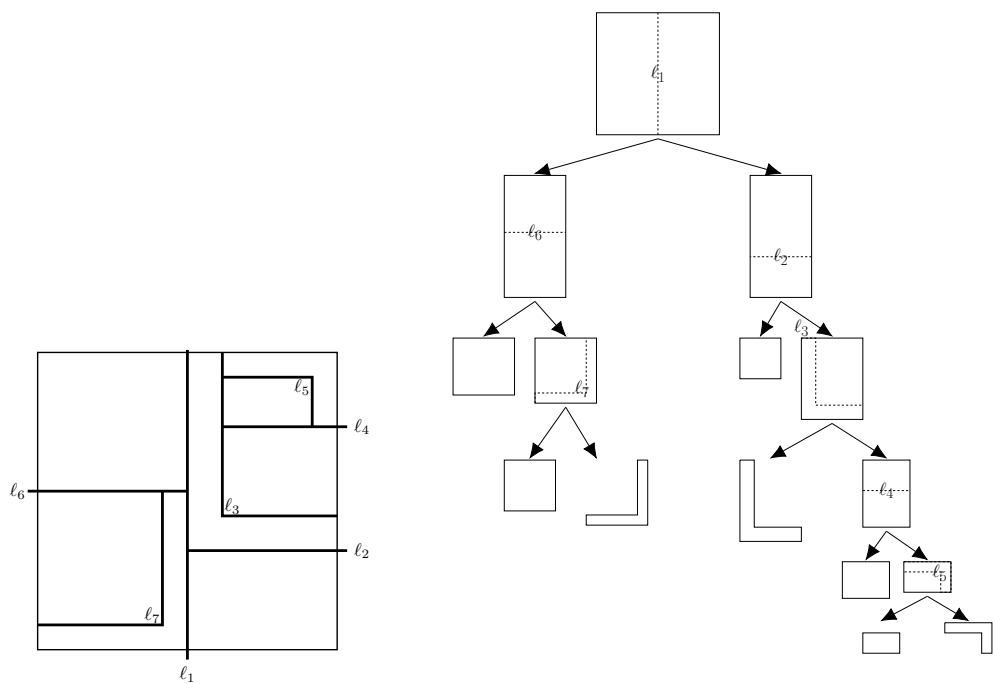
\begin{figure}[h]
	\captionsetup[subfigure]{justification=centering}
	\hspace{-10pt}
	\begin{subfigure}[b]{.47\textwidth}
		\centering
		\resizebox{!}{4.5cm}{
		\begin{tikzpicture}			
			\draw[thick] (0,0) rectangle (6,6);	

			\draw[ultra thick] (3,-0.2)--(3,6.05);
			\draw (3,-.5) node {\large \textbf{$\ell_1$}};

			\draw[ultra thick] (-.2,3.2)--(3,3.2);
			\draw (-.4, 3.2) node {\large \textbf{$\ell_6$}};

			\draw[ultra thick] (0, .5)--(2.5,.5)--(2.5,3.2);
			\draw (2.3,.7) node {\large \textbf{$\ell_7$}};

			\draw[ultra thick] (3,2)--(6.2,2);
			\draw (6.5,2) node {\large \textbf{$\ell_2$}};

			\draw[ultra thick] (3.7, 6)--(3.7,2.7)--(6,2.7);

			\draw (3.9,2.9) node {\large \textbf{$\ell_3$}};

			\draw[ultra thick] (3.7,4.5)--(6.2, 4.5);
			\draw (6.5,4.5) node {\large \textbf{$\ell_4$}};

			\draw[ultra thick] (5.5, 4.5)--(5.5, 5.5)--(3.7, 5.5);
			\draw (5.3,5.3) node {\large \textbf{$\ell_5$}};
	\end{tikzpicture}}
	\label{fig:pseudocut}
	\end{subfigure}
	\begin{subfigure}[b]{.47\textwidth}
		\centering
		\resizebox{!}{8.5cm}{
		\begin{tikzpicture}
			\draw[thick] (-3, -3) rectangle (3,3);	
			\draw[dashed] (0,-3)--(0,3);
			\draw (0,0) node {\Huge \textbf{$\ell_1$}};
			\draw[-{Latex[length=7mm, width=6mm]}](0,-3.2)--(-6, -4.9);
			\draw[-{Latex[length=7mm, width=6mm]}](0,-3.2)--(6, -4.9);

			\draw[thick] (-7.5,-5) rectangle (-4.5, -11);
			\draw[dashed] (-7.5,-7.8)--(-4.5,-7.8);
			\draw (-6,-7.8) node {\Huge \textbf{$\ell_6$}};
			\draw[-{Latex[length=7mm, width=6mm]}](-6,-11.2)--(-4.5, -12.9);
			\draw[-{Latex[length=7mm, width=6mm]}](-6,-11.2)--(-8.5, -12.9);

			\draw[thick] (4.5, -5) rectangle (7.5, -11);
			\draw[dashed] (7.5,-9)--(4.5,-9);
			\draw (6,-9) node {\Huge \textbf{$\ell_2$}};
			\draw[-{Latex[length=7mm, width=6mm]}](6,-11.2)--(8.5, -12.9);
			\draw[-{Latex[length=7mm, width=6mm]}](6,-11.2)--(5, -12.9);

			\draw[thick] (-6,-13) rectangle (-3, -16.2);
			\draw[-{Latex[length=7mm, width=6mm]}](-4.5,-16.4)--(-6.25, -18.9);
			\draw[-{Latex[length=7mm, width=6mm]}](-4.5,-16.4)--(-3, -18.9);
			\draw[dashed](-6,-16.2)--(-6, -15.7)--(-3.5, -15.7)--(-3.5, -13);
			\draw (-3.7,-15.5) node {\Huge \textbf{$\ell_7$}};

			\draw[thick] (-10,-13) rectangle (-7, -15.8);

			\draw[thick] (6,-13) rectangle (4, -15);

			\draw[thick] (10,-13) rectangle (7, -17);
			\draw[-{Latex[length=7mm, width=6mm]}](8.5,-17.2)--(5.5, -18.9);
			\draw[-{Latex[length=7mm, width=6mm]}](8.5,-17.2)--(11.15, -18.9);
			\draw[dashed](7,-13)--(7.7, -13)--(7.7, -16.3)--(10, -16.3);
			\draw (7,-12.5) node {\Huge \textbf{$\ell_3$}};

			\draw[thick] (-7.5, -19)rectangle(-5,-21.5);
			\draw[thick] (-.5, -19)--(-.5,-22.2)--(-3.5,-22.2)--(-3.5,-21.7)--(-1, -21.7)--(-1,-19)--(-.5,-19);

			\draw[thick] (4, -19)--(4, -23)--(7, -23)--(7, -22.3)--(4.7, -22.3)--(4.7,-19)--(4,-19);

			\draw[thick] (10, -19) rectangle (12.3, -22.3);
			\draw[-{Latex[length=7mm, width=6mm]}](11.15,-22.5)--(10.15, -23.9);
			\draw[-{Latex[length=7mm, width=6mm]}](11.15,-22.5)--(13.15, -23.9);
			\draw[dashed](10,-20.5)--(12.3, -20.5);
			\draw (11.15,-20.5) node {\Huge \textbf{$\ell_4$}};

			\draw[thick] (9, -24) rectangle (11.3, -25.8);

			\draw[thick] (12, -24) rectangle (14.3, -25.5);
			\draw[-{Latex[length=7mm, width=6mm]}](13.15,-25.7)--(10.9, -26.9);
			\draw[-{Latex[length=7mm, width=6mm]}](13.15,-25.7)--(15.15, -26.9);
			\draw[dashed](12,-24)--(14.3, -24)--(14.3, -25.5)--(13.8, -25.5)--(13.8,-24.5)--(12, -24.5);
			\draw (13.75,-24.55) node {\Huge \textbf{$\ell_5$}};

			\draw[thick] (10, -27.5) rectangle (11.8, -28.5);
			\draw[thick] (14, -27)--(16.3, -27)--(16.3, -28.5)--(15.8,-28.5)--(15.8, -27.5)--(14,-27.5)--(14,-27);

	\end{tikzpicture}}
	\label{fig:tree}
	\end{subfigure}
	\caption{(a) A pseudo-guillotine cutting sequence. The first cut is $l_1$, and then the resulting right piece is further subdivided by $\ell_2$, $\ell_3$, $\ell_4$ and $\ell_5$. Similarly, $\ell_6$, $\ell_7$ subdivide the left piece. Note that $\ell_3$, $\ell_5$ and $\ell_7$ are not guillotine cuts, but they cut out the corresponding $\Lc$-compartments. 
(b) step by step pseudo-guillotine cutting sequence corresponding to Figure (a). Dashed line at each level indicates a partition of a rectangle into two regions (two boxes, or one box and one $\Lc$-shaped).}
	\label{fig:pseudonice}
\end{figure}

%

\subsection{Near-optimal structured solutions}

Our main technical contribution is to show that there exists a $(1+\eps)$-approximate
solution whose items can be placed nicely inside a set of compartments
$\C$ that admit a pseudo-guillotine cutting sequence. By Lemma~\ref{lem:pseudo-enough}
there is a guillotine cutting sequence for them.
\begin{lem}
\label{lem:structured-OPT}There exists a set $OPT'\subseteq\R$ and
a partition of $K$ into a set of $O_{\eps}(1)$ 
\footnote{The notation $O_{\eps}(f(n))$ means that the implicit constant hidden by the big $O$ notation can depend on $\eps$.}
compartments $\C$
such that
\begin{itemize}
\item $|OPT'|\ge(1-\eps)|OPT|$,
\item the compartments $\C$ admit a pseudo-guillotine cutting sequence,
\item the items in $OPT'$ can be placed nicely inside the compartments $\C$.
\end{itemize}
\end{lem}

We will prove Lemma~\ref{lem:structured-OPT} in Section \ref{sec:bldecom-1}. Our
main algorithm works as follows. First, we guess the $O_{\eps}(1)$
compartments $\C$ due to Lemma~\ref{lem:structured-OPT} in time $(nN)^{O_{\eps}(1)}$
(note that we can assume w.l.o.g.~that they have integral coordinates).
Then we place items nicely inside $\C$ while maximizing the cardinality
of the placed items. For this we use a $(1+\eps)$-approximation
algorithm which is a slight adaptation of a recent algorithm in \cite{GalSocg21}
for the 2GK problem (i.e., without requiring that the computed solution
is guillotine separable). In fact, we simplify some steps of that
algorithm since our compartments are very simple.

\begin{lem}
\label{lem:algorithm}Given a set of compartments $\C$. In time $(nN)^{O_{\eps}(1)}$
we can compute a set of items $ALG\subseteq\R$ that \aw{are} placed
nicely inside $\C$ such that $|ALG|\ge(1-\eps)|OPT'|$ for any
set of items $OPT'$ that can be placed nicely inside the compartments
$\C$.
\aw{Inside each compartment $C\in \C$ the set $ALG$ admits an $O_\eps(\log (nN))$-stage packing}.
\end{lem}
We will prove Lemma~\ref{lem:algorithm} in Section~\ref{sec:algorithm} and Appendix~\ref{apx:algo-weighted}. Then, Lemmas~\ref{lem:pseudo-enough},
\ref{lem:structured-OPT}, and \ref{lem:algorithm} imply our main
theorem for the cardinality case. 
\ari{\aw{Due to Lemma~\ref{lem:structured-OPT},}
our pseudo-guillotine cutting sequence has $O_{\eps}(1)$  leaf nodes \aw{and each of them is} either a box- or an $\Lc$-compartment. 
The packing algorithm due to Lemma~\ref{lem:algorithm} gives a $O_{\eps}(\log (nN))$-stage packing inside each compartment.
This yields globally a $O_{\eps}(\log (nN))$-stage packing.}
As mentioned before, we will describe
our $(1+\eps)$-approximation algorithm for the weighted case
in Appendix \ref{apx:algo-weighted}.
\begin{thm}
There is a $(1+\eps)$-approximation algorithm for \tggk~with a running time of $(nN)^{O_{\eps}(1)}$
\aw{that computes an $O_\eps(\log (nN))$-stage packing.}
\end{thm}

\aw{We obtain a similar result also for the rotational case: our structural result from Lemma~\ref{lem:structured-OPT}
still holds and the algorithm due to Lemma~\ref{lem:algorithm} needs only some minor modifications.}
\begin{thm}
There is a $(1+\eps)$-approximation algorithm for \tggkr~with a running time of $(nN)^{O_{\eps}(1)}$
\aw{that computes an $O_\eps(\log (nN))$-stage packing.}
\end{thm}

\section{Existence of near-optimal structured solutions}

\label{sec:bldecom-1} In this section, we prove Lemma~\ref{lem:structured-OPT}
in the cardinality case, i.e., assuming that \ama{$\profit(i)=1$} for
each item $i\in\R$. We prove that  there exists a $(1+\eps)$-approximate
solution whose items can be placed nicely inside a set of compartments
$\C$ that admit a pseudo-guillotine cutting sequence. Note that Lemma~\ref{lem:structured-OPT}
trivially holds if $\left|\opt\right|\le O_{\eps}(1)$ and hence we
assume for the remainder of this section that $\left|\opt\right|$
is larger than any given constant (which in particular implies that
we can drop any set of $O_{\eps}(1)$ items from $\opt$ while losing
only a factor of $1+\eps$).

\arir{changed $\epsl N/2$ to $\epsl N/4$}

\begin{figure}[t]
	\captionsetup[subfigure]{justification=centering}
	\hspace{-10pt}
	\begin{subfigure}[b]{.5\textwidth}
		\centering
		\resizebox{!}{5cm}{
		\begin{tikzpicture}			
			\draw[thick] (0,0) rectangle (6,6);	
			\draw (1,1) rectangle (5,5.5);

			\draw (0.5, 0)--(0.5, 6.2);
			\draw (5.5, 0)--(5.5, 6.2);
			\draw (.5, .5)--(5.5, .5);
			\draw (.5, 1)--(5.5, 1);
			\draw (5, 1)--(5, 6);
			\draw (0.5, 5.5)--(5,5.5);

			\draw[dashed] (3, 1)--(3,5.5);

			\draw [thick,decorate,decoration={brace,amplitude=8pt}] (-0.2,0) -- (-.2,1);
			\draw (-.8,.5) node {\large \textbf{$h_B$}};

			\draw (3.5,.5) node {\large \textbf{$\ell_2$}}; 				
			\draw (3.75,1) node {\large \textbf{$\ell_3$}}; 		
			\draw (3.7,5.5) node {\large \textbf{$\ell_5$}}; 
			\draw (2,.25) node {\large \textbf{$B_2$}}; 	
			\draw (2.2,.75) node {\large \textbf{$B_3$}}; 	
			\draw (5,3) node {\large \textbf{$\ell_4$}};
			\draw (5.75,2.7) node {\large \textbf{$B_1$}};
			\draw (5.25,3.5) node {\large \textbf{$B_4$}};
			\draw (.25,3.5) node {\large \textbf{$B_0$}}; 		
			\draw (.75,2.2) node {\large \textbf{$B_6$}}; 		
			\draw (1,3) node {\large \textbf{$\ell_6$}}; 		 	
			\draw (0.5,5) node {\large \textbf{$\ell_0$}}; 		
			\draw (5.5,5.3) node {\large \textbf{$\ell_1$}}; 		
			\draw (2.3,5.75) node {\large \textbf{$B_5$}}; 	
			\draw (2.3,3) node {\large \textbf{$B_7$}}; 	
			\draw (3,3) node {\large \textbf{$\ell_7$}};
 	
			\draw [thick,decorate,decoration={brace,amplitude=8pt}] (-0.2,5.5) -- (-.2,6);
			\draw (-.8,5.75) node {\large \textbf{$h_T$}}; 		 		

			\draw [thick,decorate,decoration={brace,amplitude=8pt}] (0,6.2) -- (1.5,6.2);
			\draw (.75,6.8) node {\large \textbf{$\eps_{large}N/4$}}; 
			\draw [thick,decorate,decoration={brace,amplitude=8pt}] (1,4.7) -- (0,4.7);
			\draw (.75,4.2) node {\large \textbf{$w_L$}}; 
			\draw [thick,decorate,decoration={brace,amplitude=8pt}] (6,4.7) -- (5,4.7);
			\draw (5.75,4.2) node {\large \textbf{$w_R$}}; 

			\draw [thick,decorate,decoration={brace,amplitude=8pt}] (4.5,6.2) -- (6,6.2);
			\draw (5.25,6.8) node {\large \textbf{$\eps_{large}N/4$}}; 	 	

			\draw [thick,decorate,decoration={brace,amplitude=8pt}] (6.05, 1.5) -- (6.05, 0);
			\draw (7.2, .75) node {\large \textbf{$\eps_{large}N/4$}};

			\draw [thick,decorate,decoration={brace,amplitude=8pt}] (6.05, 6) -- (6.05, 4.5);
			\draw (7.2, 5.25) node {\large \textbf{$\eps_{large}N/4$}}; 
		\end{tikzpicture}}
		\caption{}
		\label{fig:Lrearrange1}
	\end{subfigure}
	\begin{subfigure}[b]{.5\textwidth}
		\centering
		\resizebox{!}{5cm}{
		\begin{tikzpicture}						
			\draw[thick] (0,0) rectangle (6,6);
			\draw (0.5,0) -- (.5, 6);
			\draw (1,0) -- (1, 6);
			\draw (1.5,1) -- (1.5, 6);
			\draw (2,1.5) -- (2, 6);
		
			\draw (1,.5) -- (6, .5);
			\draw (1,1) -- (6, 1);
			\draw (1.5,1.5) -- (6, 1.5);

			\draw[dashed] (4,1.5) -- (4, 6);

			\draw (.25,3) node {\large \textbf{$B_0$}};
			\draw (.75,3) node {\large \textbf{$B_1$}};
			\draw (1.25,3.5) node {\large \textbf{$B_4$}};
			\draw (1.75, 3.75) node {\large \textbf{$B_6$}};

			\draw (3.5,.25) node {\large \textbf{$B_2$}};
			\draw (3.5,.75) node {\large \textbf{$B_3$}};
			\draw (3.75,1.25) node {\large \textbf{$B_5$}};

			\draw (3.5,3.75) node {\large \textbf{$B_7$}};
			\draw (4,3.75) node {\large \textbf{$\ell_7$}};

			\draw [thick,decorate,decoration={brace,amplitude=8pt}] (0,6.2) -- (2,6.2);
			\draw (1,6.8) node {\large \textbf{$w_L+w_R$}}; 	 	

			\draw [thick,decorate,decoration={brace,amplitude=8pt}] (6.05, 1.5) -- (6.05, 0);
			\draw (7, .75) node {\large \textbf{$h_T+h_B$}};

		\end{tikzpicture}}
		\caption{}
		\label{fig:Lrearrange2}
	\end{subfigure}
	\caption{Transformation to obtain an $\Lc$-compartment.}
	\label{fig:Lrearrange}
\end{figure}
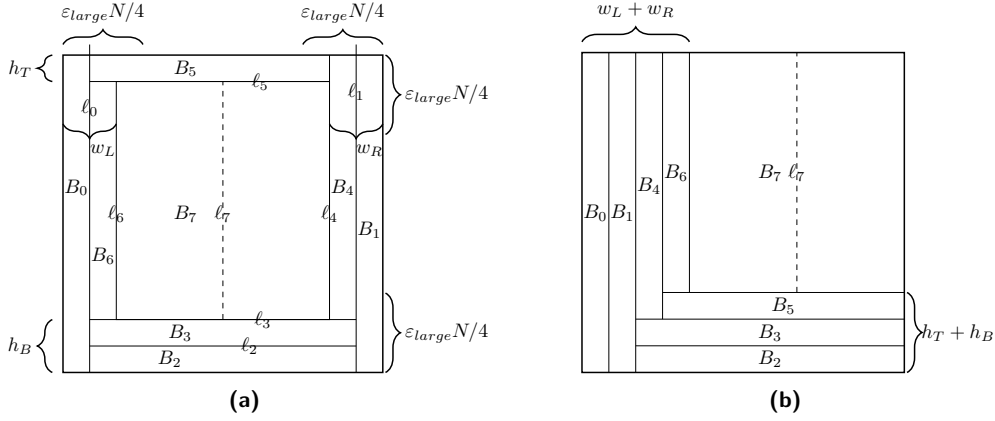

Consider an optimal solution $\opt$ and a corresponding guillotine
cutting sequence $S$. \aw{Temporarily, we remove from the packing the
items in $\optsm$; we will put back most of them later.} We identify a set of cuts of $S$ as follows.
Let $\ell_{0}$ denote the first cut of $S$. Assume w.l.o.g.~that
$\ell_{0}$ is vertical. If the distance of $\ell_{0}$ to the left
\emph{and }to the right edge of $K$ is at least $\ari{\epsl N/4}$ then we
stop. Otherwise $\ell_{0}$ cuts $K$ into two rectangles $R_{1},R_{2}$
and assume w.l.o.g.~that the width of $R_{1}$ is at most $\ari{\epsl N/4}$.
Now we consider how $S$ continues within $R_{2}$. We continue recursively.
Assume inductively that we identified a set of cuts $\ell_{0},...,\ell_{k-1}$
of $S$ and suppose that $\ell_{k-1}$ is vertical cut with distance
less than $\ari{\epsl N/4}$ to the left or the right edge of $K$, or that
$\ell_{k-1}$ is horizontal cut with distance less than $\ari{\epsl N/4}$
to the top or the bottom edge of $K$. Assume w.l.o.g.~that $\ell_{k-1}$
is vertical with distance less than $\ari{\epsl N/4}$ to the left edge of
$K$. Then the cut $\ell_{k-1}$ yields two rectangles $R_{1},R_{2}$,
and assume that $R_{1}$ lies on the left of $R_{2}$. Then we define
$\ell_{k}$ to be the next cut of $S$ within $R_{2}$. If the distance
of $\ell_{k}$ to the top \emph{and }the bottom edge of $K$ is at
least $\ari{\epsl N/4}$ then we stop. Otherwise we continue iteratively.
Eventually, this procedure must stop, let $\ell_{0},...,\ell_{k}$
denote the resulting sequence. Let $B_{0},...,B_{k-1}$ denote the
rectangles that are cut off by $\ell_{0},...,\ell_{k-1}$ and into
which we did \emph{not }recurse when we defined $\ell_{1},...,\ell_{k}$.
Let $B_{k}$ denote the rectangle that is cut by $\ell_{k}$. Then
each rectangle $B_{i}$ with $i\in\{1,...,k-1\}$ satisfies that $w(B_{i})\le \ari{\epsl N/4}$
or $h(B_{i})\le \ari{\epsl N/4}$ and in particular cannot contain both horizontal
and vertical items. Also, the items of $\opt$ inside $B_{i}$ are
guillotine separable. The important insight is that we can rearrange
the rectangles $B_{0},...,B_{k}$ (while moving their items accordingly)
such that $B_{0},...,B_{k-1}$ lies in an $\Lc$-compartment $L\subseteq K$
such that $K\setminus L$ is a rectangle, i.e., $L$ lies at the boundary
of $K$ \ama{as shown in the Figure~\ref{fig:Lrearrange}.}

\begin{lem}
\label{lem:move-items}There exists an $\Lc$-compartment $L\subseteq K$
such that $K\setminus L$ is a rectangle and we can \ama{rearrange} the
rectangles $B_{0},...,B_{k}$ such that
\begin{itemize}
\item $B_{0},...,B_{k-1}$ fit non-overlappingly into $L$,
\item there is a guillotine cutting sequence for $B_{0},...,B_{k-1}$,
\item $B_{k}$ fits into $K\setminus L$.
\end{itemize}
\end{lem}

\begin{proof}
\aw{Following the cutting sequence $S$ as described, let us assume that $B_k:= [w_L, N-w_R] \times [h_B, N-h_T]$, where $0 \le w_R, w_L, h_T, h_B \le \epsl N/4$.
Therefore, the cuts $\ell_{1},...,\ell_{k-1}$ separate of a ring-like region $Q:= ([0,w_L]\times[0,N]) \cup ([N-w_R, N]\times[0,N]) \cup ([0, N]\times[0, h_B]) \cup  ([0, N]\times[N-h_T, h_T])$ (see Figure \ref{fig:Lrearrange}).
Note that some of the values $w_R, w_L, h_T, h_B$ might be 0.}
The rectangles $B_{0},...,B_{k-1}$ fit in $Q$ and we \aw{want to show} that we can rearrange the rectangles in $B_{0},...,B_{k-1}$  into an $\Lc$-compartment $L\subseteq K$ such that $L:=([0, w_L+w_R]\times[0,N] )\cup ([0,N]\times[0, h_B+h_T])$ and there is a guillotine cutting sequence for $B_{0},...,B_{k-1}$.
Clearly, $B_{k}$ fits into $K\setminus L$. 
We prove the claim by induction on $k$. The base case is trivial. 
W.l.o.g. assume the vertical cut $\ell_0$ that divides $K$ into $B_0, R'$, where  $B_0$ lies on the left of $R'$. 
Hence, $B_0 := [0,b_0]\times [0, N]$ and $R':=K \setminus B_0$.
We use induction on $R'$ to find a packing of $B_{1},...,B_{k-1}$ in $L':=[b_0, w_L+w_R]\times[0,N] \cup [0,N]\times[0, h_B+h_T]$. Therefore, adding $B_0$ to $L'$ yields the desired $\Lc$-compartment $L$. For the guillotine cutting sequence for $B_{0},...,B_{k-1}$, we follow $\ell_0$ \aw{and afterwards} the guillotine cutting sequence for $B_{1},...,B_{k-1}$ \aw{obtained by} induction \aw{from} $R'$.
\aw{The other cases, i.e., when $B_0$ lies right or top or bottom of $R'$, follow analogously.}
\end{proof}
We adjust the packing of $\opt$ according to Lemma~\ref{lem:move-items},
i.e., for each rectangle $B_{i}$ with $i\in\{0,...,k\}$ we move
its items according to where $B_{i}$ was moved due to the lemma.
The resulting packing inside $L$ might not be nice. However, we can fix this by dropping at
most $O_{\eps}(1)$ items and subdividing $L$ into $O_{\eps}(1)$
box-compartments and a smaller $\Lc$-compartment $L'\subseteq L$ that 
\emph{lies at the outer boundary of }$L$, i.e., such that $L\setminus L'$
is again an $\Lc$-compartment and $h(L')=h(L)$ and $w(L')=w(L)$. 

\begin{lem}
\label{lem:make-L-nice}Given an $\Lc$-compartment $L$ containing a set
of items $\R(L)$. There exists a partition of $L$ into one $\Lc$-compartment
$L'\subseteq L$ and $O_{\eps}(1)$ box-compartments $\B(L)$
such that 
\begin{itemize}
\item $L'$ lies at the outer boundary of $L$, 
\item the box-compartments in $\B(L)$ are guillotine separable, and 
\item there is a nice placement of a set of items $\R'(L)\subseteq\R(L)$
with $|\R'(L)|\ge(1-\eps)|\R'(L)|-O_{\eps}(1)$ inside $\B(L)$
and $L'$.
\end{itemize}
\end{lem}
 \begin{proof}[Proof sketch.] \awr{Wrote this proof sketch. Since it is less technical than the real proof, I hope that it is easier to digest.}

\begin{figure}[h]
	\captionsetup[subfigure]{justification=centering}
	\hspace{-10pt}
	\begin{subfigure}[b]{.32\textwidth}
		\centering
		\resizebox{!}{3.2cm}{		
		\begin{tikzpicture}			
			\draw[ultra thick](2.5,6)--(2.5,2.5)--(6,2.5)--(6,0)--(0,0)--(0,6)--(2.5,6);

			\draw[solid, fill=lightgray] (.41,0) rectangle (4.45,.22);
			\draw[solid, fill=lightgray] (4.5,.09) rectangle (5.4,.26);
			\draw[solid, fill=lightgray] (5.5,.12) rectangle (6,.29); 

			\draw[solid, fill=lightgray] (.40,.3) rectangle (6,.45);

			\draw[solid, fill=lightgray] (.67,.52) rectangle (6,.7);

			\draw[solid, fill=lightgray] (1.01,.79) rectangle (2.3,1.1);
			\draw[solid, fill=lightgray] (2.45,.91) rectangle (3.3,1.18);
			\draw[solid, fill=lightgray] (3.7,.8) rectangle (6,1.11);
 
			\draw[solid, fill=lightgray] (1.75,1.25) rectangle (6,1.4);
			\draw[solid, fill=lightgray] (2.01,1.43) rectangle (6,1.6);  
			\draw[solid, fill=lightgray] (2.25,1.75) rectangle (6,1.9);
			\draw[solid, fill=lightgray] (2.52,1.95) rectangle (6,2.1);
			\draw[solid, fill=lightgray] (2.54,2.25) rectangle (6,2.45);
			\draw[ultra thick](0,0)--(.25,0)--(.25,.18)--(.4,.18)--(.4,.43)--(.67,.43)--(.67,.74)--(1.01,.74)--(1.01,1.19)--(1.75,1.19)--(1.75,1.48)--(2.01,1.48)--(2.01,1.75)--(2.25,1.75)--(2.25,2.02)--(2.5,2.02)--(2.5,2.5)--(6,2.5)--(6,0)--(0,0);
			
			\draw [thick, dashed] (.4,0.25)--(6.5, .25);
			\draw [thick, dashed] (.67,.5)--(6.5, .5);
			\draw [thick, dashed] (1.01,.75)--(6.5, .75);
			\draw [thick, dashed] (1.01,1)--(6.5, 1);
			\draw [thick, dashed] (1.75,1.25)--(6.5, 1.25);
			\draw [thick, dashed] (2.01,1.5)--(6.5, 1.5);
			\draw [thick, dashed] (2.25,1.75)--(6.5, 1.75);
			\draw [thick, dashed] (2.25,2)--(6.5, 2);
			\draw [thick, dashed] (2.5,2.25)--(6.5, 2.25);
			
 			\draw (3,-.2) node {\large \textbf{$e_1$}};
 			\draw (5,2.7) node {\large \textbf{$e_2$}};		
			\filldraw (0,0) circle (2pt);
			\filldraw (2.5,2.5) circle (2pt);
 			\draw (-.2,-.3) node {\large \textbf{$p_1$}};
 			\draw (2.75,2.75) node {\large \textbf{$p_2$}};
			\draw (6.4,.03) node {\scriptsize \textbf{$B_{\tiny 0}$}};
			\draw (6.7,2.38) node {\scriptsize \textbf{$B_{\tiny 1/\eps^2-1}$}};
			\draw (1.25,3.5) node {\large \textbf{$P_V$}};
			\draw (4.35,2.7) node {\large \textbf{$P_H$}};

		\end{tikzpicture}}
		\caption{}
		\label{fig:Lnice1}
	\end{subfigure}
	\begin{subfigure}[b]{.32\textwidth}
		\centering
		\resizebox{!}{3.2cm}{
		\begin{tikzpicture}						
			\draw[ultra thick](2.5,6)--(2.5,2.5)--(6,2.5)--(6,0)--(0,0)--(0,6)--(2.5,6);

			\draw[solid, fill=lightgray] (.41,0) rectangle (4.45,.22);
			\draw[solid, fill=lightgray] (4.5,.09) rectangle (5.4,.26);
			\draw[solid, fill=lightgray] (5.5,.12) rectangle (6,.29); 

			\draw[solid, fill=lightgray] (.40,.3) rectangle (6,.45);

			\draw[solid, fill=lightgray] (.67,.52) rectangle (6,.7);

			\draw[solid, fill=gray] (1.01,.79) rectangle (2.3,1.1);
			\draw[solid, fill=gray] (2.45,.91) rectangle (3.3,1.18);
			\draw[solid, fill=gray] (3.7,.8) rectangle (6,1.11);
 
			\draw[solid, fill=lightgray] (1.75,1.25) rectangle (6,1.4);
			\draw[solid, fill=lightgray] (2.01,1.43) rectangle (6,1.6);  
			\draw[solid, fill=lightgray] (2.25,1.75) rectangle (6,1.9);
			\draw[solid, fill=lightgray] (2.52,1.95) rectangle (6,2.1);
			\draw[solid, fill=lightgray] (2.54,2.25) rectangle (6,2.45);
			\draw[ultra thick](0,0)--(.25,0)--(.25,.18)--(.4,.18)--(.4,.43)--(.67,.43)--(.67,.74)--(1.01,.74)--(1.01,1.19)--(1.75,1.19)--(1.75,1.48)--(2.01,1.48)--(2.01,1.75)--(2.25,1.75)--(2.25,2.02)--(2.5,2.02)--(2.5,2.5)--(6,2.5)--(6,0)--(0,0);
			
			\draw [thick, dashed] (.4,0.25)--(6, .25);
			\draw [thick, dashed] (.67,.5)--(6, .5);
			\draw [thick, dashed] (1.01,.75)--(6, .75);
			\draw [pattern = north west lines] (1.01,.75)--(1.01,1.19)--(1.75,1.19)--(1.75,1.25)--(6,1.25)--(6,.75)--(1.01,.75);
			\draw [thick, dashed] (1.01,1)--(6, 1);
			\draw [thick, dashed] (1.75,1.25)--(6, 1.25);
			\draw [thick, dashed] (2.01,1.5)--(6, 1.5);
			\draw [thick, dashed] (2.25,1.75)--(6, 1.75);
			\draw [thick, dashed] (2.25,2)--(6, 2);
			\draw [thick, dashed] (2.5,2.25)--(6, 2.25);

			\draw (6.4,.03) node {\scriptsize \textbf{$B_{\tiny 0}$}};
			\draw (6.7,2.38) node {\scriptsize \textbf{$B_{\tiny 1/\eps^2-1}$}};
			\draw (1.25,3.5) node {\large \textbf{$P_V$}};
			\draw (4.35,2.7) node {\large \textbf{$P_H$}};
			\draw [thick,decorate,decoration={brace,amplitude=8pt}] (6.05, 1.25) -- (6.05, .75);
			\draw (7.1,1) node {\scriptsize \textbf{$2.\eps^2.h(P_H)$}};
 			\draw (3,-.2) node {\large \textbf{$e_1$}};
 			\draw (5,2.7) node {\large \textbf{$e_2$}};		
			\filldraw (0,0) circle (2pt);
			\filldraw (2.5,2.5) circle (2pt);
 			\draw (-.2,-.3) node {\large \textbf{$p_1$}};
 			\draw (2.75,2.75) node {\large \textbf{$p_2$}};

		\end{tikzpicture}}
		\caption{}
		\label{fig:Lnice2}
	\end{subfigure}
	\begin{subfigure}[b]{.32\textwidth}
		\centering
		\resizebox{!}{3.2cm}{
		\begin{tikzpicture}						
			\draw[ultra thick](2.5,6)--(2.5,2.5)--(6,2.5)--(6,0)--(0,0)--(0,6)--(2.5,6);

			\draw[solid, fill=lightgray] (.41,0) rectangle (4.45,.22);
			\draw[solid, fill=lightgray] (4.5,.09) rectangle (5.4,.26);
			\draw[solid, fill=lightgray] (5.5,.12) rectangle (6,.29); 

			\draw[solid, fill=lightgray] (.40,.3) rectangle (6,.45);

			\draw[solid, fill=lightgray] (.67,.52) rectangle (6,.7);

			\draw[solid, fill=lightgray] (1.75,.8) rectangle (6,.95);
			\draw[solid, fill=lightgray] (2.01,1.03) rectangle (6,1.2);  
			\draw[solid, fill=lightgray] (2.25,1.3) rectangle (6,1.45);
			\draw[solid, fill=lightgray] (2.52,1.55) rectangle (6,1.7);
			\draw[solid, fill=lightgray] (2.54,1.8) rectangle (6,2);
			\draw[ultra thick](0,0)--(.25,0)--(.25,.18)--(.4,.18)--(.4,.43)--(.67,.43)--(.67,.74)--(1.01,.74)--(1.01,1.19)--(1.75,1.19)--(1.75,1.48)--(2.01,1.48)--(2.01,1.75)--(2.25,1.75)--(2.25,2.02)--(2.5,2.02)--(2.5,2.5)--(6,2.5)--(6,0)--(0,0);
			
			\draw [pattern = north east lines] (0.25,0)--(.25, .18)--(.4,.18)--(.4, .43)--(.67,.43)--(.67,.74)--(6,.74)--(6,0)--(.4,0);
			\draw [thick] (1.01,1)--(6, 1);
			\draw [thick] (1.75,1.25)--(6, 1.25);
			\draw [thick] (2.01,1.5)--(6, 1.5);
			\draw [thick] (2.25,1.75)--(6, 1.75);
			\draw [thick] (2.25,2)--(6, 2);
			\draw [thick] (2.5,2.25)--(6, 2.25);

			\draw [thick] (2.5,1.75)--(2.5, 2);
			\draw [thick] (2.25,1.5)--(2.25, 1.75);
			\draw [thick] (2.01,1.25)--(2.01, 1.5);
			\draw [thick] (1.75,1.)--(1.75, 1.25);

			\draw (6.7,2.38) node {\scriptsize \textbf{$B_{\tiny 1/\eps^2-1}$}};
			\draw (1.25,3.5) node {\large \textbf{$P_V$}};
			\draw (4.35,2.7) node {\large \textbf{$P_H$}};
			\draw [thick,decorate,decoration={brace,amplitude=8pt}] (6.05, .75) -- (6.05, 0);
			\draw (6.7,.375) node {\scriptsize \textbf{$P_H'$}};
 			\draw (3,-.2) node {\large \textbf{$e_1$}};
 			\draw (5,2.7) node {\large \textbf{$e_2$}};		
			\filldraw (0,0) circle (2pt);
			\filldraw (2.5,2.5) circle (2pt);
 			\draw (-.2,-.3) node {\large \textbf{$p_1$}};
 			\draw (2.75,2.75) node {\large \textbf{$p_2$}};
		\end{tikzpicture}}
		\caption{}
		\label{fig:Lnice3}
	\end{subfigure}

	\begin{subfigure}[b]{.32\textwidth}
		\centering
		\resizebox{!}{3.2cm}{
		\begin{tikzpicture}						
			\draw[ultra thick](2.5,6)--(2.5,2.5)--(6,2.5)--(6,0)--(0,0)--(0,6)--(2.5,6);

			\draw[solid, fill=lightgray] (.41,0) rectangle (4.45,.22);
			\draw[solid, fill=lightgray] (4.5,.09) rectangle (5.4,.26);
			\draw[solid, fill=lightgray] (5.5,.12) rectangle (6,.29); 

			\draw[solid, fill=lightgray] (.40,.3) rectangle (6,.45);

			\draw[solid, fill=lightgray] (.67,.52) rectangle (6,.7);

			\draw[solid, fill=lightgray] (1.75,.8) rectangle (6,.95);
			\draw[solid, fill=lightgray] (2.01,1.03) rectangle (6,1.2);  
			\draw[solid, fill=lightgray] (2.25,1.3) rectangle (6,1.45);
			\draw[solid, fill=gray] (2.52,1.55) rectangle (6,1.7);
			\draw[solid, fill=gray] (2.54,1.8) rectangle (6,2);
			\draw[ultra thick](0,0)--(.25,0)--(.25,.18)--(.4,.18)--(.4,.43)--(.67,.43)--(.67,.74)--(1.01,.74)--(1.01,1.19)--(1.75,1.19)--(1.75,1.48)--(2.01,1.48)--(2.01,1.75)--(2.25,1.75)--(2.25,2.02)--(2.5,2.02)--(2.5,2.5)--(6,2.5)--(6,0)--(0,0);
			
			\draw [thick, dashed] (1.01,.75)--(6.2, .75);
			\draw [thick] (0.25,0)--(.25, .18)--(.4,.18)--(.4, .43)--(.67,.43)--(.67,.74)--(6,.74)--(6,0)--(.4,0);
			\draw [thick, dashed] (1.75,1.5)--(6.2, 1.5);
			\draw [thick, dashed] (2.5,2.25)--(6.2, 2.25);

			\draw[pattern = north west lines] (2.01, 1.5)--(2.01, 1.75)--(2.25, 1.75)--(2.25, 2.02)--(2.5,2.02)--(2.5,2.25)--(6,2.25)--(6, 1.5)--(2.01,1.5);

			\draw (1.25,3.5) node {\large \textbf{$P_V$}};
			\draw (4.35,2.7) node {\large \textbf{$P_H$}};
			\draw [thick,decorate,decoration={brace,amplitude=8pt}] (6.25, 2.25) -- (6.25, 1.5);
			\draw (7.1,1.875) node {\scriptsize \textbf{$\frac{6}{\eps}~boxes$}};
			\draw [thick,decorate,decoration={brace,amplitude=8pt}] (6.05, .75) -- (6.05, 0);
			\draw (6.7,.375) node {\scriptsize \textbf{$P_H'$}};
 			\draw (3,-.2) node {\large \textbf{$e_1$}};
 			\draw (5,2.7) node {\large \textbf{$e_2$}};		
			\filldraw (0,0) circle (2pt);
			\filldraw (2.5,2.5) circle (2pt);
 			\draw (-.2,-.3) node {\large \textbf{$p_1$}};
 			\draw (2.75,2.75) node {\large \textbf{$p_2$}};

		\end{tikzpicture}}
		\caption{}
		\label{fig:Lnice4}
	\end{subfigure}
	\begin{subfigure}[b]{.32\textwidth}
		\centering
		\resizebox{!}{3.2cm}{
		\begin{tikzpicture}						
			\draw[ultra thick](2.5,6)--(2.5,2.5)--(6,2.5)--(6,0)--(0,0)--(0,6)--(2.5,6);

			\draw[solid, fill=lightgray] (1.41,0) rectangle (5.45,.22);
			\draw[solid, fill=lightgray] (4.45,1.6) rectangle (5.35,1.79);
			\draw[solid, fill=lightgray] (5.4,1.6) rectangle (5.9, 1.79);

			\draw[solid, fill=lightgray] (.40,.3) rectangle (6,.45);

			\draw[solid, fill=lightgray] (.67,.52) rectangle (6,.7);

			\draw[solid, fill=lightgray] (1.75,.8) rectangle (6,.95);
			\draw[solid, fill=lightgray] (2.01,1.03) rectangle (6,1.2);  
			\draw[solid, fill=lightgray] (2.25,1.3) rectangle (6,1.45);
			\draw[ultra thick](0,0)--(.25,0)--(.25,.18)--(.4,.18)--(.4,.43)--(.67,.43)--(.67,.74)--(1.01,.74)--(1.01,1.19)--(1.75,1.19)--(1.75,1.48)--(2.01,1.48)--(2.01,1.75)--(2.25,1.75)--(2.25,2.02)--(2.5,2.02)--(2.5,2.5)--(6,2.5)--(6,0)--(0,0);
			
			\draw [thick] (0.25,0)--(.25, .18)--(.4,.18)--(.4, .43)--(.67,.43)--(.67,.74)--(6,.74)--(6,0)--(.4,0);
			\draw (2.01, 1.5)--(2.01, 1.75)--(2.25, 1.75)--(2.25, 2.02)--(2.5,2.02)--(2.5,2.25)--(6,2.25)--(6, 1.5)--(2.01,1.5);

			\draw (1.25,3.5) node {\large \textbf{$P_V$}};
			\draw (4.35,2.7) node {\large \textbf{$P_H$}};
			\draw [thick,decorate,decoration={brace,amplitude=8pt}] (6.25, 2.25) -- (6.25, 1.5);
			\draw (7.1,1.875) node {\scriptsize \textbf{$\frac{6}{\eps}~boxes$}};
			\draw [thick,decorate,decoration={brace,amplitude=8pt}] (6.05, .75) -- (6.05, 0);
			\draw (6.7,.375) node {\scriptsize \textbf{$P_H'$}};
 			\draw (3,-.2) node {\large \textbf{$e_1$}};
 			\draw (5,2.7) node {\large \textbf{$e_2$}};		
			\filldraw (0,0) circle (2pt);
			\filldraw (2.5,2.5) circle (2pt);
 			\draw (-.2,-.3) node {\large \textbf{$p_1$}};
 			\draw (2.75,2.75) node {\large \textbf{$p_2$}};

			\draw [thick] (1.01,1)--(6, 1);
			\draw [thick] (1.75,1.25)--(6, 1.25);
			\draw [thick] (2.01,1.25)--(2.01, 1.5);
			\draw [thick] (1.75,1.)--(1.75, 1.25);
			\draw[->] (5.5,1.7) edge[out=90,in=270,->] (4,4);	\draw (4,4.2) node {\scriptsize \textbf{$Steinberg~Packing$}};/

		\end{tikzpicture}}
		\caption{}
		\label{fig:Lnice5}
	\end{subfigure}
	\begin{subfigure}[b]{.32\textwidth}
		\centering
		\resizebox{!}{3.2cm}{
		\begin{tikzpicture}						
			\draw[ultra thick](2.5,6)--(2.5,2.5)--(6,2.5)--(6,0)--(0,0)--(0,6)--(2.5,6);

			\draw[solid, fill=gray] (1.41,0) rectangle (5.45,.22);
			\draw[solid, fill=lightgray] (4.45,1.6) rectangle (5.35,1.79);
			\draw[solid, fill=lightgray] (5.4,1.6) rectangle (5.9, 1.79);

			\draw[solid, fill=lightgray] (.40,.3) rectangle (6,.45);

			\draw[solid, fill=lightgray] (.67,.52) rectangle (6,.7);

			\draw[solid, fill=lightgray] (1.75,.8) rectangle (6,.95);
			\draw[solid, fill=lightgray] (2.01,1.03) rectangle (6,1.2);  
			\draw[solid, fill=lightgray] (2.25,1.3) rectangle (6,1.45);
			\draw[ultra thick](0,0)--(.25,0)--(.25,.18)--(.4,.18)--(.4,.43)--(.67,.43)--(.67,.74)--(1.01,.74)--(1.01,1.19)--(1.75,1.19)--(1.75,1.48)--(2.01,1.48)--(2.01,1.75)--(2.25,1.75)--(2.25,2.02)--(2.5,2.02)--(2.5,2.5)--(6,2.5)--(6,0)--(0,0);
			
			\draw [thick] (0.25,0)--(.25, .18)--(.4,.18)--(.4, .43)--(.67,.43)--(.67,.74)--(6,.74)--(6,0)--(.4,0);

			\draw (2.01, 1.5)--(2.01, 1.75)--(2.25, 1.75)--(2.25, 2.02)--(2.5,2.02)--(2.5,2.25)--(6,2.25)--(6, 1.5)--(2.01,1.5);
			\draw [thick] (1.01,1)--(6, 1);
			\draw [thick] (1.75,1.25)--(6, 1.25);
			\draw [thick] (2.01,1.25)--(2.01, 1.5);
			\draw [thick] (1.75,1.)--(1.75, 1.25);

			\draw[thick] (.35, 0.18)--(.35,6);
			\draw[thick] (.7, 0.74)--(.7,6);
			\draw[thick] (1.05, 1.19)--(1.05,6);
			\draw[thick] (1.4, 1.19)--(1.4,6);
			\draw[thick] (1.75, 1.48)--(1.75,6);
			\draw[thick] (2.1, 1.75)--(2.1,6);

			\draw[thick] (0, .18)--(.4,.18);
			\draw[thick] (.35, .74)--(2.1,.74);
			\draw[thick] (.7, 1.19)--(1.01,1.19);
			\draw[thick] (1.4, 1.48)--(1.75,1.48);
			\draw[thick] (1.75, 1.75)--(2.1,1.75);

			\draw[ultra thick, dashed] (.35,.18)--(6.1,.18);

			\draw (1.25,3.5) node {\large \textbf{$P_V$}};
			\draw (4.35,2.7) node {\large \textbf{$P_H$}};
			\draw [thick,decorate,decoration={brace,amplitude=8pt}] (6.05, .75) -- (6.05, 0.18);
			\draw (6.9,.465) node {\scriptsize \textbf{$P_{H,top}'$}};
			\draw [thick,decorate,decoration={brace,amplitude=8pt}] (6.05, .18) -- (6.05, 0);
			\draw (7.2,.09) node {\scriptsize \textbf{$P_{H, bottom}'$}};
			\draw (.185,5.5) node {\scriptsize \textbf{$P_V'$}};
 			\draw (3,-.2) node {\large \textbf{$e_1$}};
 			\draw (5,2.7) node {\large \textbf{$e_2$}};		
			\filldraw (0,0) circle (2pt);
			\filldraw (2.5,2.5) circle (2pt);
 			\draw (-.2,-.3) node {\large \textbf{$p_1$}};
 			\draw (2.75,2.75) node {\large \textbf{$p_2$}};

		\end{tikzpicture}}
		\caption{}
		\label{fig:Lnice6}
	\end{subfigure}
	\caption{\ama{Processing done in Lemma \ref{lem:make-L-nice} to obtain a nice packing in L-compartment }
}
	\label{fig:Lnice}
\end{figure}
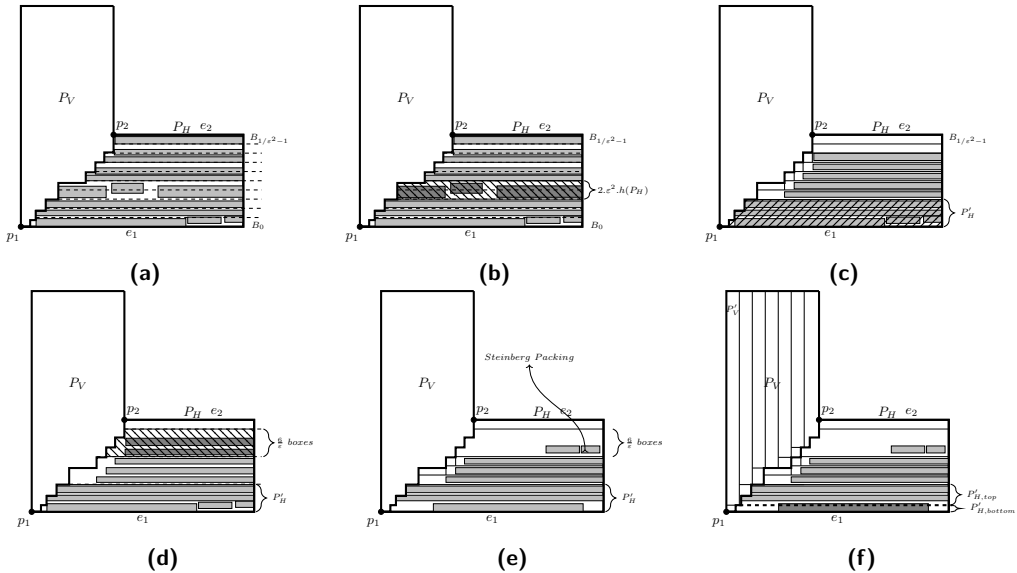

\aw{Since it is sufficient to place $(1-\eps)|\R'(L)|-O_{\eps}(1)$
items, we can} drop $O_{\eps}(1)$ items. So w.l.o.g. assume that
$\R(L)$ contains only skewed items \aw{(i.e., we remove all large
items)}. Intuitively, we partition $L$ into two polygons $P_{H}$
and $P_{V}$ that are separated via a monotone axis-parallel curve
connecting the two vertices of $L$ at the bend of $L$, such that
$P_{H}$ contains all horizontal items placed inside $L$ and $P_{V}$
contains all vertical items inside $L$, see Figure~\ref{fig:Lnice1}. We rearrange
the items in $P_{H}$ and $P_{V}$ separately, starting with $P_{H}$.
Denote by $\R(P_{H})\subseteq\R(L)$ the items of $\R(L)$ placed
inside $P_{H}$. 

We place $1/\eps^{2}$ boxes inside $P_{H}$ of height $\eps^{2}h(P_{H})$
each, stacked one on top of the other. We define their width maximally
large such that they are still contained inside $P_{H}$ (note that
some area of $P_{H}$ is then not covered by these boxes), see Figure~\ref{fig:Lnice2}.
Denote by $\left\{ B_{0},...,B_{1/\eps^{2}-1}\right\} $ these boxes
in this order, such that $B_{0}$ touches the longer horizontal edge
of $P_{H}$. With a shifting argument, we can show that there are
two consecutive boxes $B_{j^{*}},B_{j^{*}+1}$ with $j^{*}\le1/\eps$
that intersect with at most an $O_{\eps}(1)+\ari{O(\eps|\R(P_{H})|)}$
items in $\R(P_{H})$. We remove these items. Let $P'_{H}\subseteq P_{H}$
denote the part of $P_{H}$ underneath $B_{j^{*}}$ (see Figure~\ref{fig:Lnice3}).
We move down by $\eps^{2}h(P)$ units each item in $\R(P_{H})$ that
intersect one of the boxes $B_{j^{*}+2},...,B_{1/\eps^{2}-1}$ and
we remove all $O_{\eps}(1)$ items from $\R(L)$ that intersect
more than one box. Note that then the moved items fit into the boxes
$\B':=\left\{ B_{j^{*}+1},...,B_{1/\eps^{2}-2}\right\} $. 

Using another shifting step, we delete all items in $6/\eps$
consecutive boxes of $\B'$; since there are $\Omega(1/\eps^{2})$
boxes in $\B'$ this costs only a factor $1+O(\eps)$ in the profit.
We use the empty space to place in it all items in $P'_{H}$ that
are shorter than the shorter horizontal edge of $P_{H}$, see Figure~\ref{fig:Lnice5}.
One can show that they can be placed into this empty space using Steinberg's
algorithm~\cite{steinberg1997strip} (maintaining guillotine separability) since the available space is
much larger than the area of the items to be placed. For the remaining
items in $P'_{H}$ one can show that the width of each of them is
more than half of the width of $L$. Hence, we can assume w.l.o.g.
that they are placed nicely within $P'_{H}$. Again, we remove all
items that intersect more than one box after this movement, which
are at most $O_{\eps}(1)$ items. Denote by $\B_{hor}$ the resulting
set of boxes.

We do a symmetric procedure for $P_{V}$, yielding a set of boxes
$\B_{ver}$ and a nicely packed region $P'_{V}$. Intuitively, we want to define $L'$ as $P'_{H}\cup P'_{V}$.
However, $P'_{H}\cup P'_{V}$ might not have exactly the shape of
an $\Lc$-compartment. Nevertheless, one can show that we can subdivide
one of these polygons, say $P'_{H},$ along a horizontal line into
two subpolygons $P'_{H,\mathrm{top}},P'_{H,\mathrm{bottom}}$ (with
$P'_{H,\mathrm{top}}$ lying on the top of $P'_{H,\mathrm{bottom}}$)
such that 
\begin{itemize}
\item we can place the items in $P'_{H,\mathrm{top}}$ into another set
of $O_{\eps}(1)$ boxes $\B'_{hor}$ that are non-overlapping
with $\B_{hor}\cup\B_{ver}$, and 
\item $L':=P'_{H,\mathrm{bottom}}\cup P'_{V}$ forms an $\Lc$-compartment,
see Figure~\ref{fig:Lnice6}.
\end{itemize}
Then the items are nicely placed inside $L'$. To each of
the $O_{\eps}(1)$ boxes $B\in\B_{hor}\cup\B'_{hor}\cup\B_{ver}$
we apply a standard routine that removes some of the items inside
$B$ and partitions $B$ into smaller boxes, such that the remaining items 
inside these smaller boxes are nicely placed.
\end{proof}
\awr{another figure or a different figure for Lemma 8 would be good}


Therefore, we define that the first cuts of our pseudo-guillotine
cutting sequence $S'$ looks as follow: we first separate $K$ into
$L'$ and $K\setminus L'$ and then separate the boxes in $\B(L)$.
 Then we apply a guillotine cut to the rectangular area $K\setminus L$ that corresponds
to $\ell_{k}$ (since we moved the items in $B_{k}$ we need to adjust
$\ell_{k}$ accordingly), which yields two rectangular areas $R_{1},R_{2}$.
With each of them we continue recursively, i.e., we apply the same
routine that we had applied to $K$ above. 

\aw{
We do
not recurse further if for a considered rectangular area $R$ it
holds that $h(R)<\epsl N$ or $w(R)<\epsl N$. In this case $R$ contains
only horizontal or only vertical items, respectively. However, these
items might not be packed nicely. Thus, we apply to $R$ a similar routine
as in Lemma~\ref{lem:make-L-nice}. In a sense, $R$ behaves like a degenerate $\Lc$-compartment with only
four edges. Also note that $R$ is a box-compartment.}

\begin{lem} \label{lem:make-boxes-nice}Given a box-compartment $B$
containing a set of items $\R(B)$ with $h(B)<\epsl N$ or $w(B)<\epsl N$, 
 there exists a partition of $B$ into $O_{\eps}(1)$ box-compartments
$\B(B)$ such that 
\begin{itemize}
\item the box-compartments in $\B(B)$ are guillotine separable, and 
\item there is a nice placement of a set of items $\R'(B)\subseteq\R(B)$
with $|\R'(B)|\ge(1-\eps)|\R'(B)|-O_{\eps}(1)$ inside $\B(B)$. 
\end{itemize}
\end{lem} 

\aw{
It remains to put back the (small) items in $\optsm$. Intuitively,
we assign them to the empty space in our $O_{\eps}(1)$ constructed
compartments. More formally, we subdivide our compartments further 
into smaller compartments by guillotine cuts, some of the resulting
compartments are empty, and into those we assign the small items
with the Next-Fit-Decreasing-Height algorithm} \ama{ [Lemma~\ref{lem:NFDH}]}. \aw{ 
 For each of these compartments we 
ensure that their height and width is $\epss N/\eps$. 
There might be empty space that is not used in this way, however,
we can ensure that its total
area is very small, e.g., at most $O(\eps^2 N^2)$. This allows us to pack essentially
all items in $\optsm$ (handling a few special cases differently, e.g., if 
the total area of the items in 
$\optsm$ is very small)}. \ama{See Appendix~\ref{smallrec} for details.
}

Let $S'$ denote the resulting pseudo-guillotine cutting sequence.
We need to argue that this yields in total $O_{\eps}(1)$ compartments.
This follows easily since every time we identify a sequence of cuts
$\ell_{0},...,\ell_{k}$ of $S$, we construct exactly one $\Lc$-compartment
and $O_{\eps}(1)$ box-compartments. Also, after each such operation,
we recurse on rectangular areas $R_{1},R_{2}$ that are at least by
$\ari{\epsl N/4}$ units thinner or shorter (i.e., by at least $\ari{\epsl N}/4$
units smaller in one of the two dimensions) than the rectangular area
that we had started with when we constructed $\ell_{0},...,\ell_{k}$
(which is the whole knapsack $K$ in the first iteration). 
Also, when we do not recurse further we subdivide the remaining region
into $O_{\eps}(1)$ box-compartments. Each resulting compartment is subdivided into 
$O_{\eps}(1)$ smaller compartments when we place the small items.
Hence, the depth
of the binary tree $T$ defining the pseudo-guillotine cutting sequence
$S'$ is $O_{\eps}(1)$ and thus we define at most $O_{\eps}(1)$
compartments in total. In particular, we applied Lemmas~\ref{lem:make-L-nice} and \ref{lem:make-boxes-nice}
at most $O_{\eps}(1)$ times and, therefore, the constructed solution
contains at least $(1-\eps)|\opt|-O_{\eps}(1)$ items. 



\section{Assigning items into compartments}\label{sec:algorithm}

For proving Lemma~\ref{lem:algorithm}, we need to provide an algorithm
that, given a set of compartments $\C$, computes a solution $ALG\subseteq\R$
with $\profit(ALG)\ge(1-\eps)\profit(OPT')$ that can also be
placed nicely in $\C$ (where $OPT'\subseteq\R$ is the subset of
$\R$ of maximum profit that can be placed nicely in the compartments
in $\C$). 

First, we guess for each box-compartment $B\in\C$ which case of Definition~\ref{def:structured-boxes}
applies, i.e., whether $B$ contains only a single large item, or only
horizontal items, or only vertical items, or only small items. For
each box-compartment $B\in\C$ for which we guessed that it contains
only one large item, we simply guess this item. We can do this deterministically in time $O(n^{|\C|})=n^{O_{\eps}(1)}$
for all such box-compartments $B\in\C$.

Then, for assigning the small items, we  use a standard reduction
to the Generalized Assignment Problem (GAP) \ama{[Lemma~\ref{lem:GAP}]} for selecting a near-optimal
set of small items and an assignment of these items into the corresponding
box-compartments. Inside of each box-compartment $B$ we place the
items with the Next-Fit-Decreasing-Height algorithm~\cite{coffman1980performance} which
results in a \ama{$2$-stage} guillotine separable packing for the items inside $B$.

\begin{lem}
\label{lem:small-items}Given a set of box compartments $\B$
such that a set of items $I_{small}^{*}\subseteq I_{small}$ can be
placed non-overlappingly inside $\B$,  in $n^{O_{\eps}(1)}$ time 
we can we can compute a set of items $I'_{small}\subseteq I_{small}$
with $\profit(I'_{small})\ge(1-\eps)\profit(I{}_{small}^{*})$
and a nice placement of the items in $I'_{small}$ inside $\B$
which is guillotine separable with $O_\eps(1)$ stages.
\end{lem}
Let $\C_{skew}\subseteq\C$ denote the compartments in $\C$ into
which skewed items are placed in $\opt'$ (which in particular contains
all $\Lc$-compartments in $\C$). It remains to select a profitable
set of items from $\Rsk$ that can be placed nicely in the compartments
in $\C_{skew}$. For this task, we use a recent algorithm in~\cite{GalSocg21}
which is a routine for \tgk~which takes as input (in our terminology)
a set of box- and $\Lc$-compartments, and also compartments of more
general shapes (e.g., with the shapes of a U or a Z). In time $(nN)^{O_{\eps}(1)}$, 
it computes a subset of the input items of maximum total profit, up
to a factor of $1+\eps$, that can be placed non-overlappingly
inside the given compartments. In fact, it first partitions the given
compartments such that there exists a profitable solution for the
smaller compartments inside of which the items are placed nicely (according
to our definition). Then it computes a $(1+\eps)$-approximation
of the most profitable subset of items that can be placed nicely.

In our setting, we can skip the first step since also in $\opt'$
the items are placed nicely inside the compartments $\C_{skew}$.
Hence, we execute directly the second part the algorithm in~\cite{GalSocg21}.
In fact, a simpler version of that routine is sufficient for our purposes
since we have only box- and $\Lc$-compartments. 
\ari{The algorithm in~\cite{GalSocg21} \aw{can handle also the case where rotations by 90 degree are allowed, and the same holds
for the routine in Lemma~\ref{lem:small-items}}. Thus our result works for the case with rotations as well.}
We refer to Appendix~\ref{apx:algo-weighted}
for a complete and self-contained description of this routine, adapted to the guillotine setting.
In particular, inside each compartment its solution is guillotine separable 
with $O_\eps(\log nN)$ stages.

\section{Power of stages in guillotine packing}
\label{sec:stages}
\begin{figure}[h]
\centering
\resizebox{!}{4cm}{
\begin{tikzpicture}
\draw[ultra thick] (0,0) rectangle (64,64);
\draw[fill=lightgray!40] (0,0) rectangle (64,1);
\draw[fill=lightgray!40] (1,1) rectangle (64,3);
\draw[fill=lightgray!40] (3,3) rectangle (64,7);
\draw[fill=lightgray!40] (7,7) rectangle (64,15);
\draw[fill=lightgray!40] (15,15) rectangle (64,31);				
\draw[fill=lightgray!40] (0,1) rectangle (1,64);
\draw[fill=lightgray!40] (1,3) rectangle (3,64);
\draw[fill=lightgray!40] (3,7) rectangle (7,64);
\draw[fill=lightgray!40] (7,15) rectangle (15,64);
\draw[fill=lightgray!40] (15,31) rectangle (31,64);
\end{tikzpicture}}
\caption{Hard example for Theorem \ref{thm:hardstage}.}
\label{hardStages}
\end{figure}
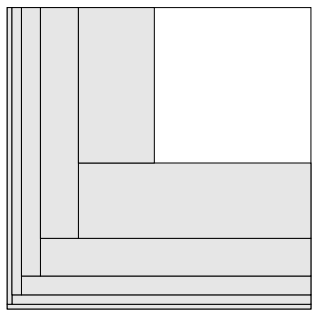

\aw{Recall that our two algorithms compute packings with $O_{\eps}(\log (nN))$-stages. This raises the question whether one can obtain $(1+\eps)$-approximate solutions with fewer stages. In particular, for the related guillotine 2BP  and guillotine 2SP problems there are APTASs 
whose solutions use $O(1)$-stage packings \cite{BansalLS05,seiden2005two}.}
\aw{However, we show that in contrast for \tggk~sometimes $\Omega(\log N)$ stages are necessary already for a better approximation ratio than $2$, even if there are only skewed items.}
\awr{shortened discussion such that it does not look too much like related work}

%

\begin{thm}
\label{thm:hardstage}
For any constant $0 <\eps<\frac{1}{2}$, \aw{there is a family of instances of \tggk~with only skewed items for which any $(2-\eps)$-approximate solution 
requires $k=\Omega (\eps \log N)$ stages.
}
%
\end{thm}

Figure \ref{hardStages} shows the hard example, see Appendix \ref{sec:smallstage} for more details. 

\newpage
\bibliography{bibliography}


\appendix

\section{Weighted case}\label{apx:algo-weighted}
Before we start this section, we would like to introduce some terminologies. Vertical part of an $\Lc$-compartment is the maximal vertical rectagular region which can be included in the $\Lc$-compartment. Similarly the horizontal part of an $\Lc$-compartment is the maximal horizontal rectagular region which can be included in the $\Lc$-compartment. 

We prove Lemma~\ref{lem:algorithm} in the weighted case. Our algorithm is an adaptation
of an algorithm in \cite{GalSocg21} for 2GK (in fact, it is a simpler version
of that algorithm, using that all our compartments are box- or $\Lc$-compartments
and hence they are relatively simple).

First, we guess for each box-compartment $B\in\C$ which case of Definition~\ref{def:structured-boxes}
applies, i.e., whether $B$ contains only one single item, or only
horizontal items, or only vertical items, or only small items. For
each box-compartment $B\in\C$ for which we guessed that it contains
only one item, we simply guess this item. We can do this in time $n^{|\C|}=n^{O_{\eps}(1)}$
for all box-compartments $B\in\C$. 

Then, for assigning the small items, we  use a standard reduction
to the generalized assignment problem for selecting a near-optimal
set of small items and an assignment of these items into the corresponding
box-compartments. Inside of each box-compartment $B$ we place the
items with the Next-Fit-Decreasing-Height algorithm \cite{coffman1980performance} which
results in a guillotine separable packing for the items inside $B$.

\begin{lem}
\label{lem:small-items-1}Given a set of box-compartments $\B$ such
that a set of items $I_{small}^{*}\subseteq I_{small}$ can be placed
non-overlappingly inside $\B$. Then in time $n^{O_{\eps}(1)}$
we can we can compute a set of items $I'_{small}\subseteq I_{small}$
with $\profit(I'_{small})\ge(1-6\eps)\profit(I{}_{small}^{*})$ and a nice placement
of the items in $I'_{small}$ inside $\B$.
\end{lem}
\begin{proof}
Let us first formulate an instance of generalized assignment problem (GAP).  For each compartment $C_j$ we create a bin $j$ in the GAP instance with capacity equal to its area. For each item $i\in I_{small}$, we create an item in the GAP instance with $p_{ij}=\profit(i)$ and $s_{ij}$ equal to area of item $i$ if $w(i)\leq \eps w(C_j)$ and $h(i)\leq \eps h(C_j)$ and equal to some arbitrary large value greater than area of $C_j$ otherwise.  Now by applying Lemma \ref{lem:GAP} on the GAP instance, we get a subset $I_{small}''\subseteq I_{small}$ which are assigned to the boxes in $\B$ such that $\profit(I_{small}'')\geq (1-3\eps)\profit(I_{small}^{*})$,  and sum of areas of items assigned to a box $B$ doesn't cross the area of $B$. Also for $\forall i \in I_{small}''$, $w(i)\leq \eps w(C)$ and $h(i)\leq \eps h(C)$ where $C$ is the box to which $i$ is assigned to. Let the items assigned to a box $B\in \B$ be $I_B$. In the non-increasing order of profit by area ratio we choose items from $I_B$ until the total area of chosen items crosses $(1-2\eps-\eps^2)\cdot w(B)\cdot h(B)$. Let the set of items which were chosen be $I_B'$. Clearly $\profit(I_B')\geq (1-2\eps-\eps^2)  \profit(I_B)$. Now by applying Lemma \ref{lem:NFDH}, we can pack all the items of $I_B'$ in $B$ and the packing is a 2 stage packing which is guillotine separable. Applying this procedure to every box, we get a packing of set of items $I_{small}'$ into the boxes such that $\profit(I_{small}')\geq (1-2\eps-\eps^2)\profit(I_{small}'') \geq (1-2\eps-\eps^2)(1-3\eps)\profit(I_{small}^{*})\geq (1-6\eps)\profit(I_{small}^{*})$    
\end{proof}

The hard part is the assignment of the skewed items. Let $\C_{skew}\subseteq\C$
denote the compartments in $\C$ into which skewed items are placed
in $\opt'$. We partition $\Rsk$ into two subsets $\Rh$ and $\Rl$
where intuitively $\Rh$ contains input items with relatively high
profit, and $\Rl=\Rsk\setminus\Rh$. Let $c=O_{\eps}(1)$ be a
constant to be defined later.

\ari{W.l.o.g. assume $|\opt'| \ge c \cdot \log(nN)$. Otherwise, we can add dummy items $i$ with  
$\profit(i)=0$.}

\begin{lem}
\label{lem:shifting-weighted}There is a solution $\opt''\subseteq\opt'$
and an item $i^{*}\in I$ such that $\Rh=\{i\in\Rsk|p(i)>p(i^{*})\}$
and $\Rl=\{i\in\Rsk|p(i)\le p(i^{*})\}$ satisfy that 
\begin{itemize}
\item $p(\opt'')\ge(1-O(\eps))p(\opt')$, 
\item $p(\opt'')\ge c\cdot\log(nN)\cdot \profit(i^{*})$, 
\item $\left|\opt''\cap\Rh\right|\le c\cdot\log(nN)$, and 
\item for each item $i\in\opt''\cap\Rsk$ it holds that $p(i)\ge\frac{\eps}{2n}\max_{i'\in\R}p(i')$. 
\end{itemize}
\end{lem}
\begin{proof}
Consider the items in $\opt' $ in the order of non-increasing profits.
Let $i^{*}$ to be the first item in $\opt'$ in this order such that the total profit from items $i\in\opt'$ with $p(i)>p(i^{*})$ is $\ge c\cdot\log(nN)\cdot \profit(i^{*})$.
Then trivially, the second and third property is true. 
Also, we know that $\profit(\opt')\ge\frac{1}{2}\max_{i'\in\R}p(i')$
since otherwise $\opt'$ could not be a $(1+\eps)$-approximate
solution. We drop each item $i\in\opt'$ for which $p(i)<\frac{\eps}{2n}\max_{i'\in\R}p(i')$
holds. The total profit of these dropped items is thus at most $n\cdot\frac{\eps}{2n}\max_{i'\in\R}p(i')\le\eps\cdot\profit(\opt')$.
We define $\opt''$ to be the resulting solution. 
\end{proof}
We guess $i^{*}$ and hence know the sets $\Rh$ and $\Rl$.
\begin{lem}
\label{lem:color-coding-1}In time $(nN)^{O_{\eps}(c)}$ we can
guess a partition $\left\{ \Rh(C)\right\} _{C\in\C}$ of $\Rh$ such
that for each $C\in\C$ the set $\Rh(C)$ contains all items in $\opt''$
that are placed inside $C$.
\end{lem}
\begin{proof}
We employ the color coding technique \cite{alon1995color} where we seek to color the items in
$\Rh$ such that no two items
in $\opt''\cap\Rh$ have the same color. Let us denote $k'=|\opt''\cap\Rh|$ 
If we color the items randomly with $k=c\cdot\log N$ colors, then the probability that no items in $\opt''\cap\Rh$ have the same color is ${k \choose k'}\frac{k'!k^{n-k'}}{k^n} \geq \big(\frac{k}{k'}\big)^{k'}\cdot \frac{k'!}{k^{k'}}\geq \frac{k'!}{k'^{k'}}\geq e^{-k'}\ge\left(\frac{1}{N}\right)^{O(c)}$.
The last inequality follows from Lemma \ref{lem:shifting-weighted}. 
Using splitters \cite{cygan2015parameterized} we can derandomize this to guess such a coloring
deterministically in time $N^{O(c)}$. Once we have guessed the coloring in which each item of $\opt''\cap\Rh$ is colored differently, then for each color $d\in\{1,...,c\cdot\log N\}$
we guess which of the $O_{\eps}(1)$ compartments in $\C'$ contains
an item of color $d$ so that the condition in the lemma is guaranteed to be achieved, which can be done deterministically in time $2^{O_{\eps}(c\cdot\log N)}=N^{O_{\eps}(c)}$ by exploring all possibilities. 
\end{proof}

We group the items in $\Rl$ into $O_{\eps}(\log(nN))$ groups
with one set of groups for the items in $\Rl\cap\Rho$ and another
set of groups for the items in $\Rl\cap\Rve$. We group the items
in $\Rl\cap\Rho$ (resp. in $\Rl\cap\Rve$) according to their densities
which are defined as the ratio between their profit and heights (resp. widths).
Formally, for each $\ell\in\mathbb{Z}$ we define $\Rho^{(\ell)}:=\{i\in\Rl\cap\Rho|\frac{p(i)}{\height(i)}\in[(1+\eps)^{\ell},(1+\eps)^{\ell+1})\}$
and $\Rve^{(\ell)}:=\{i\in\Rl\cap\Rve|\frac{p(i)}{\width(i)}\in[(1+\eps)^{\ell},(1+\eps)^{\ell+1})\}$
and observe that for only $O(\log(nN)/\eps)$ values $\ell$ the
respective sets $\Rho^{(\ell)},\Rve^{(\ell)}$ are non-empty
as $p(i)\ge\frac{\eps}{2n}\max_{i'\in\R}p(i')$ and $\width(i) \in [N]$. Intuitively,
for each $\ell$ the items in $\Rho^{(\ell)}$ (in $\Rve^{(\ell)}$)
essentially all have the same density. Next, for each group $\Rho^{(\ell)}$,
$\Rve^{(\ell)}$ we guess an estimate for $p\left(\Rho^{(\ell)}\cap\opt''\right)$
and $p\left(\Rve^{(\ell)}\cap\opt''\right)$, respectively. 
We call these estimates $\mathrm{opt}_{hor}^{(\ell)}$ and $\mathrm{opt}_{ver}^{(\ell)}$, respectively.
We do
this in time $(nN)^{O_{\eps}(1)}$ adapting a technique from \cite{chekuri2005polynomial}. 
\begin{lem}
In time $(nN)^{O_{\eps}(1)}$, we can guess values $\mathrm{opt}_{hor}^{(\ell)},\mathrm{opt}_{ver}^{(\ell)}$
for each $\ell$ with $\Rho^{(\ell)}\cup\Rve^{(\ell)}\ne\emptyset$
such that 
\begin{itemize}
\item $\sum_{\ell}\mathrm{opt}_{hor}^{(\ell)}+\mathrm{opt}_{ver}^{(\ell)}\ge(1-O(\eps))p(\opt'')$
and 
\item $\mathrm{opt}_{hor}^{(\ell)}\le p\left(\opt''\cap\Rho^{(\ell)}\right)$
and $\mathrm{opt}_{ver}^{(\ell)}\le p\left(\opt''\cap\Rho^{(\ell)}\right)$
for each $\ell$. 
\end{itemize}
\end{lem}
\begin{proof}
First, as $p(\opt'')$ is a value between $p_{\max}:=\max_{i'\in\R}p(i')$
and $n\cdot p_{\max}$ we can guess a $(1+\eps)$-approximation
$P^{*}$ for $p(\opt'')$ in time $O(\log_{1+\eps}n)$, i.e.,
$(1-\eps)p(\opt'')\le P^{*}\le p(\opt'')$. Assume w.l.o.g.~that
for some $\hat{L}=O(\log(nN)/\eps)$ it holds that for each $\ell\notin[\hat{L}]$
the sets $\Rho^{(\ell)}$ and $\Rve^{(\ell)}$ are empty.

Now, for each $\ell\in[\hat{L}]$ let $\hat{k}_{hor}^{(\ell)}$ be 
be the biggest integer such that $\hat{k}_{hor}^{(\ell)}\frac{\eps}{\hat{L}}\cdot P^{*}\leq p(\opt''\cap I_{hor}^{\ell})$.
We define $\hat{k}_{ver}^{(\ell)}$ accordingly. We guess all $O(\hat{L})$
values $\hat{k}_{hor}^{(\ell)},\hat{k}_{ver}^{(\ell)}$ for each $\ell$
as follows. Let $w_{hor}$ be a binary string with at most $O(\hat{L}/\eps)$
digits, out of which exactly $\hat{L}+1$ entries are 1's and for
each $\ell$ the substring of $w_{hor}$ between the $\ell$-th and
$(\ell+1)$-th 1 has exactly $k_{hor}^{(\ell)}$ digits (all 0's).
We can guess $w_{hor}$ in time $2^{O(\hat{L}/\eps)}=(nN)^{O(1/\eps)}$,
infer all values $\hat{k}_{hor}^{(\ell)}$ from it, and define $\mathrm{opt}_{hor}^{(\ell)}=\hat{k}_{hor}^{(\ell)}\cdot\eps P^{*}/\hat{L}$
for each $\ell$. We have that $p(\opt'')-\left(\sum_{\ell}opt_{hor}^{(\ell)}+opt_{ver}^{(\ell)}\right)\le O(\eps)p(\opt'')$, 
since we make a mistake of at most $\frac{\eps}{\hat{L}}\cdot P^{*}\le\frac{\eps}{\hat{L}}\cdot p(\opt'')$
for each of the $\hat{L}$ groups.

We guess the values $\mathrm{opt}_{ver}^{(\ell)}$ in a similar fashion. 
\end{proof}
In other words, one can show that in order to obtain a $(1+O(\eps))$-approximation,
it suffices to pack, for each $\ell$, items from $\Rho^{(\ell)}$ with
a total profit of $\mathrm{opt}_{hor}^{(\ell)}$ (for which items
of a total height of $\frac{\mathrm{opt}_{hor}^{(\ell)}}{(1+\eps)^{\ell+1}}$
suffice) and items from $\Rve^{(\ell)}$ with a total profit of $\mathrm{opt}_{ver}^{(\ell)}$
(resp. a total width of $\frac{\mathrm{opt}_{ver}^{(\ell)}}{(1+\eps)^{\ell+1}}$ suffice).
For each $\ell$, consider the items in $\Rho^{(\ell)}$ ordered non-decreasingly
by width. We slice these items into slices of height 1 each, i.e.,
for each item $i\in\Rho^{(\ell)}$ we introduce $h_{i}$ items of
height 1 and profit $(1+\eps)^{\ell}$ each. Note that the sum
of the profits of the slices introduced for $i$ is in $[\frac{1}{1+\eps}p(i),p(i))$
(and hence this quantity is essentially $p(i)$). In the resulting
order of the slices (i.e., the order inherited from our ordering of
the items in $\Rho^{(\ell)}$) we select the first $\frac{\mathrm{opt}_{hor}^{(\ell)}}{(1+\eps)^{\ell+1}}$
slices, i.e., the $\frac{\mathrm{opt}_{hor}^{(\ell)}}{(1+\eps)^{\ell+1}}$
slices of smallest width. Let $\hat{\R}_{hor}^{(\ell)}$ denote the
resulting set of slices. We do a similar procedure with the items
in $\Rve^{(\ell)}$ for each $\ell$, resulting in a set of slices
$\hat{\R}_{ver}^{(\ell)}$. 
\begin{lem}
\label{lem:slices-fit}There exists a nice packing for the slices
in $\bigcup_{\ell}\hat{\R}_{hor}^{(\ell)}\cup\hat{\R}_{ver}^{(\ell)}$
inside $\C_{skew}$.
\end{lem}
\begin{proof}
Consider a value $\ell$. For each item $i\in\R_{hor}^{(\ell)}$ we
have that $\frac{p(i)}{\height(i)}\in[(1+\eps)^{\ell},(1+\eps)^{\ell+1})$
and thus in particular $p(i)\le(1+\eps)^{\ell+1}\height(i)$.
Therefore, $\mathrm{opt}_{hor}^{(\ell)}\le\profit(\R_{hor}^{(\ell)}\cap\opt'')\le(1+\eps)^{\ell+1}h(\R_{hor}^{(\ell)}\cap\opt'')$
and thus $h(\R_{hor}^{(\ell)}\cap\opt'')\ge\frac{\mathrm{opt}_{hor}^{(\ell)}}{(1+\eps)^{\ell+1}}$.
A similar statement holds for $\R_{ver}^{(\ell)}$.
\end{proof}
Next, we do standard linear grouping \cite{vl81} with the slices in each set $\hat{\R}_{hor}^{(\ell)}/\hat{\R}_{ver}^{(\ell)}$
so that the resulting slices have only $1/\eps$ different widths/heights, 
respectively. Formally, consider a set $\hat{\R}_{hor}^{(\ell)}$.
We sort the slices in $\hat{\R}_{hor}^{(\ell)}$ non-increasingly
by width and then partition them into $1/\eps+1$ groups such
that each of the first $1/\eps$ groups contains $\left\lceil \frac{1}{1/\eps+1}|\hat{\R}_{hor}^{(\ell)}|\right\rceil $
slices and the last group contains the remaining slices (which might
be less than $\left\lceil \frac{1}{1/\eps+1}|\hat{\R}_{hor}^{(\ell)}|\right\rceil $
but not more). Let $\hat{\R}_{hor}^{(\ell)}=\hat{\R}_{hor,1}^{(\ell)}\dot{\cup}...\dot{\cup}\hat{\R}_{hor,1/\eps+1}^{(\ell)}$
denote the resulting partition. We drop the slices in $\hat{\R}_{hor,1}^{(\ell)}$
(which are at most $\eps\cdot\left|\hat{\R}_{hor}^{(\ell)}\right|$
many). Then, for each $j\in\{2,...,1/\eps+1\}$ we increase the
width of the slices in $\hat{\R}_{hor,j}^{(\ell)}$ to the width of
the widest slice in $\hat{\R}_{hor,j}^{(\ell)}$. By construction,
the resulting slices have $1/\eps$ different widths. Let $\tilde{\R}_{hor}^{(\ell)}$
denote the resulting set and let $\tilde{\R}_{hor}^{(\ell)}=\tilde{\R}_{hor,1}^{(\ell)}\dot{\cup}...\dot{\cup}\tilde{\R}_{hor,1/\eps}^{(\ell)}$
denote a partition of $\tilde{\R}_{hor}^{(\ell)}$ according to the
widths of the slices, i.e., for each $j\in\{1,...,1/\eps\}$ the
set $\tilde{\R}_{hor,j}^{(\ell)}$ contains the rounded slices from
$\hat{\R}_{hor,j+1}^{(\ell)}$. We do this procedure for each $\ell$
and a symmetric procedure for the group $\hat{\R}_{ver}^{(\ell)}$
for each $\ell$.
\begin{lem}
\label{lem:rounded-slices-fit}There exists a nice packing for the
slices in $\bigcup_{\ell}\tilde{\R}_{hor}^{(\ell)}\cup\tilde{\R}_{ver}^{(\ell)}$
inside $\C_{skew}$.
\end{lem}
\begin{proof}
Observe that for each $\ell$ and $j$, each slice in $\tilde{\R}_{hor,j}^{(\ell)}$
is at most as long as any slice in $\hat{\R}_{hor,j}^{(\ell)}$ and
$|\hat{\R}_{hor,j}^{(\ell)}|\ge|\tilde{\R}_{hor,j}^{(\ell)}|$. Hence,
we can assign the slices in $\tilde{\R}_{hor,j}^{(\ell)}$ to the
space that was occupied by the slices in $\hat{\R}_{hor,j}^{(\ell)}$
in the packing due to Lemma~\ref{lem:slices-fit}.
\end{proof}
Finally, for each set $\tilde{\R}_{hor,j}^{(\ell)}$ and each $C\in\C{}_{skew}$
we denote by $\tilde{\R}_{hor,j}^{(\ell)}(C)$ the number of slices
from $\tilde{\R}_{hor,j}^{(\ell)}$ that are assigned to $C$ in the
solution due to Lemma~\ref{lem:rounded-slices-fit}. We define $\tilde{\R}_{ver,j}^{(\ell)}(C)$
similarly for each set $\tilde{\R}_{ver,j}^{(\ell)}$ and each $C\in\C{}_{skew}$.
We would like to construct this solution and then use it as a template
for assigning the items in $\R_{hor}^{(\ell)}$ and $\R_{ver}^{(\ell)}$.

If we could guess all the values $\left|\tilde{\R}_{hor,j}^{(\ell)}(C)\right|$
and $\left|\tilde{\R}_{ver,j}^{(\ell)}(C)\right|$ we would know exactly
which slices need to be assigned into each compartment in $\C{}_{skew}$,
since the slices in each set $\tilde{\R}_{hor,j}^{(\ell)}(C)$ and
$\tilde{\R}_{ver,j}^{(\ell)}(C)$ are identical. It is not clear how
to guess these $O(\log(nN))$ values exactly in polynomial time, but
we can guess them approximately in time $(nN)^{O_{\eps}(1)}$
while losing only a factor of $1+\eps$.
\begin{lem}
\label{lem:guess-slices-1}In time $(nN)^{O_{\eps}(1)}$ we can
guess values $\mathrm{opt}_{hor,j}^{(\ell)}(C),\mathrm{opt}_{ver,j}^{(\ell)}(C)$
for each $\ell\in\{0,...,\left\lfloor \log_{1+\eps}N\right\rfloor \}$,
$j\in\{1,...,1/\eps\}$, and $C\in\C{}_{skew}$ such that
\begin{itemize}
\item $\mathrm{opt}_{hor,j}^{(\ell)}(C)\le\left|\tilde{\R}_{hor,j}^{(\ell)}(C)\right|$
and $\mathrm{opt}_{ver,j}^{(\ell)}(C)\le\left|\tilde{\R}_{ver,j}^{(\ell)}(C)\right|$
for each $\ell,j,C$, and
\item $\sum_{C\in\C{}_{skew}}\mathrm{opt}_{hor,j}^{(\ell)}(C)\ge(1-\eps)\sum_{C\in\C{}_{skew}}\left|\tilde{\R}_{hor,j}^{(\ell)}(C)\right|$
for each $\ell,j$, and 
\item $\sum_{C\in\C{}_{skew}}\mathrm{opt}_{ver,j}^{(\ell)}(C)\ge(1-\eps)\sum_{C\in\C{}_{skew}}\left|\tilde{\R}_{ver,j}^{(\ell)}(C)\right|$
for each $\ell,j$.
\end{itemize}
\end{lem}
\begin{proof}
For each $\ell,j$ we define $\mathrm{opt}_{hor,j}^{(\ell)}(C)$ and
$\mathrm{opt}_{ver,j}^{(\ell)}(C)$ to be the largest integral multiple
of $\frac{\eps}{|\C|}\tilde{\R}_{hor,j}^{(\ell)}$ and $\frac{\eps}{|\C|}\tilde{\R}_{ver,j}^{(\ell)}$,
respectively. Then for each $\ell,j$ we can guess $\mathrm{opt}_{hor,j}^{(\ell)}(C)$
and $\mathrm{opt}_{ver,j}^{(\ell)}(C)$ in time $O_{\eps}(1)$,
and hence we can guess these values for all $\ell,j$ in parallel
in time $2^{O_{\eps}(\log(nN))}$.
\end{proof}

Consider a compartment $C\in\C{}_{skew}$. From each set $\tilde{\R}_{hor,j}^{(\ell)}$
we assign exactly $\mathrm{opt}_{hor,j}^{(\ell)}(C)$ slices to $C$.
\ari{We group these items to $\mathrm{opt}_{hor,j}^{(\ell)}(C)/\epss N$
 \emph{pseudo-items}, each of height at most $\epss N$. This pseudo-items are
 skewed and will have total  profit $\mathrm{opt}_{hor,j}^{(\ell)}(C)\cdot(1+\eps)^{\ell}$.}
 We do a similar procedure for the vertical items for $C$. Denote 
by $\Rlp(C)$ the constructed set of pseudo-items. Our goal is now
to pack the most profitable subset of $\Rh(C)\cup\Rlp(C)$ into $C$
such that they yield a nice packing.

\arir{made pseudo-items skewed.}

If $C$ is a box-compartment then this is very simple: assume w.l.o.g.~that
in $C$ horizontal items are placed. Then, we discard all items in
$\Rh(C)$ that are wider than $C$ (if we guessed correctly, no item
in $\Rlp(C)$ is wider than $C$). The remaining problem is equivalent
to an instance of knapsack where the capacity of the knapsack is the
height of $C$ and for each item $i\in\Rh(C)\cup\Rlp(C)$ there is
an item with size $h_{i}$ and profit $p_{i}$. We solve this problem
with an FPTAS for knapsack.

If $C$ is an $\Lc$-compartment we  essentially invoke an algorithm from
\cite{GalvezGHI0W17} as a black-box (after some straight forward adjustments of
the input items).
\begin{lem}[\cite{GalvezGHI0W17}]
Given an $\Lc$ compartment $C$ and a set of skewed items $I'_{skew}$.
In time $(nN)^{O(1)}$ we can compute the subset $I_{skew}^{*}\subseteq I'{}_{skew}$
of maximum profit that can be placed nicely inside $C$ and the corresponding
packing.
\end{lem}
\begin{proof}
Let us consider an $\Lc$-compartment $C$ and a set of items $I_{skew}'$. Let the height and width of the vertical part of $C$ be $h_v$ and $w_v$ respectively. Similarly let the height and width of the vertical part of $C$ be $h_h$ and $w_h$ respectively. Now we redefine the height of a vertical item $i$ as max$( h(i), h_v-h_h)$.  If $h(i)< h_v-h_h$, then 
$h_v>h(i)+h_h > 2h_h$ (as $h(i)> h_h$).
So, $h_v/2>h_h$. 
Thus, $h_v-h_h>h_v/2$.
Similarly if $h(i)\geq h_v-h_h$ then $h(i)> h_v-h(i)$ as $h(i)>h_h$. This implies that $h(i)>h_v/2$. Similarly we can redefine the width of a horizontal item  $j$ as max$( w(j), w_h-w_v)$ and we would get similar inequalities as we got for item $i$. In order to correctly apply the pseudo-polynomial time algorithm \cite{GalvezGHI0W17} of finding a nice-packing in an $\Lc$-compartment $C$ with maximum profit, the items have to be long (that is if the item is horizontal then the width is greater than $w_h/2$, similar analogy for the vertical items) with respect to the compartment which we are ensuring by redefining the dimensions and also we have ensured that the items whose dimension was increased after redefining it, don't interact with the items in the other part. Now applying the pseudo-polynomial time algorithm of \cite{GalvezGHI0W17} on the items with redefined dimension concludes the proof of this lemma.
\end{proof}
This yields a packing for some items in $\Rh(C)\cup\Rlp(C)$ inside
$C$, for each $C\in\C{}_{skew}$. For each $\ell$ denote by $\Rlph^{(\ell)}$
and $\Rlpv^{(\ell)}$ the horizontal and vertical pseudo-items that
resulted from slices in $\tilde{\R}_{hor}^{(\ell)}$ and $\tilde{\R}_{ver}^{(\ell)}$,
respectively, and that are included in the computed packing. Since
for each compartment $C\in\C_{skew}$ we computed the (essentially)
most profitable solution for the input items $\Rh(C)\cup\Rlp(C)$,
we have that the total profit of all packed (pseudo-)items is at least
$(1-O(\eps))\profit(\opt'')$. 

Finally, we assign items from $\Rl$ into the space occupied by the
items in $\bigcup_{\ell}\Rlph^{(\ell)}\cup\Rlpv^{(\ell)}$. 

\begin{lem}
\label{lem:round-slices}For each $\ell$, we  can assign items from $\R_{hor}^{(\ell)}$
with a total profit of at least {$\frac{1}{1+O(\eps)}\profit(\Rlph^{(\ell)})$$-O_{\epss, \eps}(1) |\C| \profit(i^{*})$}
nicely into the space occupied by the pseudo-items in $\Rlph^{(\ell)}$.
A symmetric statement holds for $\R_{ver}^{(\ell)}$ for each $\ell$. 
\end{lem}
\begin{proof}\arir{check the proof.}
Consider a group $\R_{hor}^{(\ell)}$. 
For a $C\in\C{}_{skew}$, let $\Rlph^{(\ell)} \cap C$ also denote the set of pseudo-items inside $C$. 
$\Rlph^{(\ell)}$ be the union of all such pseudo-items. 
We interpret each pseudo-item in $\Rlph^{(\ell)}$
as a container into which we can assign items from $\R_{hor}^{(\ell)}$,
stacked on top of each other. One can show that there is a fractional
solution that assigns items from $\R_{hor}^{(\ell)}$ with a total
profit of at least $\frac{1}{1+O(\eps)}\profit(\Rlph^{(\ell)})$.
This follows since we have such a solution
for the slices, which in fact represents a fractional solution. 
Let $X_{ij}$ denote the fractional extent to which item $i \in \R_{hor}^{(\ell)}$ is assigned 
to the container $j$ (corresponding to $j$th  pseudo-item $s_j$).
then the following LP has a feasible solution:
\begin{align}
\sum_{i \in \R_{hor}^{(\ell)}} \height(i) \cdot  X_{ij} &\le \height(s_j)
&& \forall s_j \in \Rlph^{(\ell)} \nonumber \\
\sum_{i \in \R_{hor}^{(\ell)}} \profit(i) \cdot  X_{ij} &\ge \frac{1}{1+O(\eps)}\profit(\Rlph^{(\ell)}) && \nonumber \\
\sum_{s_j \in \Rlph^{(\ell)}} X_{ij}  &\le 1 && \forall i \in \R_{hor}^{(\ell)}, \nonumber \\
0\le X_{ij}  &\le 1 && \forall i \in \R_{hor}^{(\ell)}, s_j \in {\Rlph^{(\ell)} } \nonumber 
\end{align}

The first  type of constraints ensures a feasible packing, second type ensures a high profit solution, third type ensures 
each item is packed at most once. 
Now using rank lemma, number of variables in the support of the LP solution is at most (the number of items in $\R_{hor}^{(\ell)}$ + number of psedo-items in $\Rlph^{(\ell)}$ + 1).
Now we drop all items that are assigned fractionally to two or more pseudo-items (as well as those who are not appearing in the support) and remove them from the LP. Then  for each pseudo-item we have only one item fractionally assigned.  Otherwise, we can take the convex combination of the fractionally assigned items to obtain another feasible solution and the solution is not an extreme point solution. Thus each pseudo-item has only one fractionally assigned item in it. 
Hence, in total only $O_{\eps, \epss}(1)|\C|$
number of items are fractionally assigned. 
we simply drop all items that are assigned fractionally to some container.
In total, we drop $O_{\eps, \epss}(1)|\C|$ items for the group $\R_{hor}^{(\ell)}$.
\end{proof}

We define a similar procedure for groups of vertical items $\R_{ver}^{(\ell)}$.
We apply Lemma~\ref{lem:round-slices} to each group $\R_{hor}^{(\ell)}$
and $\R_{ver}^{(\ell)}$ for each $\ell$. Thus, the total profit
of the obtained solution is at least
\[
\sum_{\ell}\left(\frac{1}{1+O(\eps)}\left(\profit(\Rlph^{(\ell)})+\profit(\Rlpv^{(\ell)})\right)- O_{\eps, \epss}(1)|\C|\profit(i^{*})\right).
\]
By choosing $c$ sufficiently large, we can ensure that $\sum_{\ell} O_{\eps, \epss}(1)|\C|\profit(i^{*})\le\eps\cdot\profit(\opt'')$.
Therefore, our packing has a total profit at least $(1-O(\eps))\profit(\opt'')$.

\section{Limitation of small stage guillotine cutting}
\label{sec:smallstage}

\begin{figure}[b]
		\captionsetup[subfigure]{justification=centering}
		\hspace{-10pt}
	\begin{subfigure}[b]{.25\textwidth}
			\centering
			\resizebox{!}{3cm}{
			\begin{tikzpicture}
				\draw[ultra thick] (0,0) rectangle (64,64);

				\draw[fill=lightgray!40] (0,0) rectangle (64,1);
				\draw[fill=lightgray!40] (1,1) rectangle (64,3);
				\draw[fill=lightgray!40] (3,3) rectangle (64,7);
				\draw[fill=lightgray!40] (7,7) rectangle (64,15);
				\draw[fill=lightgray!40] (15,15) rectangle (64,31);
				
				\draw[fill=lightgray!40] (0,1) rectangle (1,64);
				\draw[fill=lightgray!40] (1,3) rectangle (3,64);
				\draw[fill=lightgray!40] (3,7) rectangle (7,64);
				\draw[fill=lightgray!40] (7,15) rectangle (15,64);
				\draw[fill=lightgray!40] (15,31) rectangle (31,64);
			\end{tikzpicture}}
			\caption{}
			\label{fig:bad_container}
		\end{subfigure}	
		\begin{subfigure}[b]{.25\textwidth}
			\centering
			\resizebox{!}{3cm}{
			\begin{tikzpicture}
				\draw[thick] (0,0) rectangle (10,10);
				\draw[dashed] (1,0)--(1,10);
				\draw[dashed] (2,0)--(2,10);
				\draw [dashed] (8,0)--(8,10);
				\draw [dashed] (9,0)--(9,10);
				\draw [solid,fill=lightgray] (0.2,0.2) rectangle (0.8,9.8);
				\draw [solid,fill=lightgray] (1.2,0.2) rectangle (1.8,9.8);
				\draw [solid,fill=lightgray] (8.2,0.2) rectangle (8.8,9.8);
				\draw [solid,fill=lightgray] (9.2,0.2) rectangle (9.8,9.8);
				\draw [solid,fill=lightgray] (2.1,0.1) rectangle (7.9,1);
				\draw [solid,fill=lightgray] (2.1,9.9) rectangle (7.9,9);
				\draw [solid,fill=lightgray] (2.2,1.2) rectangle (3.2,8.8);
				\draw [solid,fill=lightgray] (3.5,1.2) rectangle (4.5,8.8);
				\draw [solid,fill=lightgray] (4.8,1.2) rectangle (5.8,8.8);
				\draw [solid,fill=lightgray] (6,1.2) rectangle (7,8.8);
				\draw [solid,fill=lightgray] (7.2,1.2) rectangle (7.8,8.8);
			\end{tikzpicture}}
			\caption{}
			\label{k_a}
		\end{subfigure}
		\begin{subfigure}[b]{.25\textwidth}
			\centering
			\resizebox{!}{3cm}{
			\begin{tikzpicture}
				\draw[thick] (0,0) rectangle (10,10);
				\draw[dashed] (1,0)--(1,10);
				\draw[dashed] (2,0)--(2,10);
				\draw [dashed] (3,0)--(3,10);
				\draw [dashed] (4,0)--(4,10);
				\draw [solid,fill=lightgray] (0.2,0.2) rectangle (0.8,9.8);
				\draw [solid,fill=lightgray] (1.2,0.2) rectangle (1.8,9.8);
				\draw [solid,fill=lightgray] (2.2,0.2) rectangle (2.8,9.8);
				\draw [solid,fill=lightgray] (3.2,0.2) rectangle (3.8,9.8);
				\draw [solid,fill=lightgray] (4.1,0.1) rectangle (9.9,1);
				\draw [solid,fill=lightgray] (4.1,9.9) rectangle (9.9,9);
				\draw [solid,fill=lightgray] (4.2,1.2) rectangle (5.2,8.8);
				\draw [solid,fill=lightgray] (5.5,1.2) rectangle (6.5,8.8);
				\draw [solid,fill=lightgray] (6.8,1.2) rectangle (7.8,8.8);
				\draw [solid,fill=lightgray] (8,1.2) rectangle (9,8.8);
				\draw [solid,fill=lightgray] (9.2,1.2) rectangle (9.8,8.8);
			\end{tikzpicture}}
			\caption{}
			\label{k_b}
		\end{subfigure}
		\begin{subfigure}[b]{.25\textwidth}
			\centering
			\resizebox{!}{3cm}{
			\begin{tikzpicture}
				\draw[thick] (0,0) rectangle (10,10);
				\draw [solid,red] (0,0) rectangle (4,10);
				\draw [solid,fill=lightgray] (0.2,0.2) rectangle (0.8,9.8);
				\draw [solid,fill=lightgray] (1.2,0.2) rectangle (1.8,9.8);
				\draw [solid,fill=lightgray] (2.2,0.2) rectangle (2.8,9.8);
				\draw [solid,fill=lightgray] (3.2,0.2) rectangle (3.8,9.8);
				\draw [solid,fill=lightgray] (4.1,0.1) rectangle (9.9,1);
				\draw [solid,fill=lightgray] (4.1,9.9) rectangle (9.9,9);
				\draw [solid,fill=lightgray] (4.2,1.2) rectangle (5.2,8.8);
				\draw [solid,fill=lightgray] (5.5,1.2) rectangle (6.5,8.8);
				\draw [solid,fill=lightgray] (6.8,1.2) rectangle (7.8,8.8);
				\draw [solid,fill=lightgray] (8,1.2) rectangle (9,8.8);
				\draw [solid,fill=lightgray] (9.2,1.2) rectangle (9.8,8.8);
			\end{tikzpicture}}
			\caption{}
			\label{k_c}
		\end{subfigure}
		\caption{(a): Arrangement of items in set $\R$, (b): Dashed lines represent 1st stage of guillotine cutting, (c): Rearrangement of vertical strips, (d): Red box denotes the created container. }
	\end{figure}
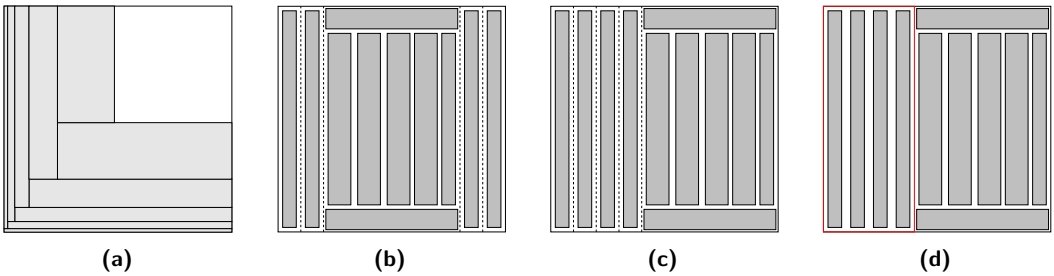

Let $N=2^{k+1}, k \in \mathbb{N}$.  Let us define a set $\R$ of long rectangles consisting of $2k$ rectangles. \\
For each $j \in [k]$, we have a vertical rectangle $V_{j}$ of height $N - (2^{j}-1)$ and width $2^{j-1}$, and a horizontal rectangle $H_{j}$ of height $2^{j-1}$ and width $N-(2^{j-1}-1)$. In an $N \times N$ knapsack place $V_{j}$ in an axis-parallel fashion such that the topmost horizontal edge of $V_{j}$ touches the topmost horizontal edge of knapsack and the left vertical  edge of $V_{j}$ is at a distance of $2^{j-1}-1$ from the left vertical edge of the knapsack. Similarly, place $H_{j}$ in an axis-parallel fashion such that the right vertical edge of $H_{j}$ touches the right vertical edge of knapsack and the bottom-most horizontal edge of $H_{j}$ is at a distance of $2^{j-1}-1$ from the bottom-most horizontal edge of knapsack. This packing consists of all the rectangles in $\R$ and is guillotine separable as shown in Figure \ref{fig:bad_container}.


\begin{lem}\label{lem_k1} \cite{GalvezGHI0W17}
For any constant $0 <\eps<\frac{1}{2}$ and a \tggkc~instance $\R$ to packed in an $N \times N$ knapsack,  any container packing of $\R' \subseteq \R$ such that $|\R|\leq(2 - \eps)|\R'|$ must use $\Omega (\eps \log N)$ containers. 
\end{lem}

\begin{lem}\label{lem_k2}
Given a $k$-stage packing of a set of {\em long} rectangles $\R' \subseteq \R$, there exists a container packing of $\R'$ which uses $k$ containers and the rectangles are guillotine separable.	
\end{lem}
\begin{proof}
Consider a $k$-stage packing of a set of rectangles $\R' \subseteq \R$. Let the first stage cuts be vertical. Let the vertical cuts in the first stage cuts divide the knapsack into vertical strips as shown in Figure \ref{k_a}. At most one vertical strip has width greater than $\frac{N}{2}$. Now rearrange the strips in such a way that the vertical strip $S_{1}$ with width greater than $\frac{N}{2}$ touches the right edge of the knapsack as shown in Figure \ref{k_b}. Then merge the vertical strips to the left of strip $S_{1}$ into a single container $C_{1}$ as shown in Figure \ref{k_c}. Note that the container $C_{1}$ touches both the left edge of the knapsack and the left edge of $S_{1}$. Now freeze the rectangles in $C_{1}$. Now consider the horizontal cuts through the vertical strip $S_{1}$ during the second stage of guillotine cutting sequence. These cuts divide $S_{1}$ into horizontal strips with same width as that of $S_{1}$. At most one horizontal strip has height greater than $\frac{N}{2}$. Now rearrange the horizontal strips in such a way that the horizontal strip $S_{2}$ with height greater than $\frac{N}{2}$ touches the topmost edge of the $S_{1}$. Then merge the vertical strips below the strip $S_{2}$ into a single container $C_{2}$. Note that the container $C_{2}$ touches both the bottom-most edge of the knapsack and the bottom-most edge of $S_{2}$. Now freeze the rectangles in $C_{2}$. Now consider $S_{2}$ as a smaller version of the original knapsack and iteratively create containers in similar way as mentioned above during each stage in the guillotine cutting sequence. So, during each stage we create exactly one new container and therefore in total we create a container packing $\R'$ which uses $k$ containers and the rectangles are guillotine separable.
\end{proof}

\begin{thm}
For any constant $0 <\eps<\frac{1}{2}$, any $k$-stage packing of $\R' \subseteq \R$ such that $|\R|\leq(2 - \eps)|\R'|$ is possible only if  $k=\Omega (\eps \log N)$ , where $N$ is the width of the knapsack.
\end{thm}
\begin{proof}
Lemma \ref{lem_k1} and \ref{lem_k2} together imply that for any constant $0 <\eps<\frac{1}{2}$, any $k$-stage packing of $\R' \subseteq \R$ such that $|\R|\leq(2 - \eps)|\R'|$ is possible only if  $k=\Omega (\eps \log N)$ , where $N$ is the width of the knapsack.
\end{proof}

\section{Omitted Proofs }\label{sec:omit}
\begin{lem}
(Restatement of Lemma \ref{Lguillotine})
Consider a set of items $I_{L}\subseteq I$ that is placed nicely
inside an $\Lc$-compartment $L$. Then $I_{L}$ is guillotine separable.
\end{lem}
\begin{proof}
W.l.o.g. assume the vertical arm of $L$ is on top and left of the horizontal arm of $L$ as shown in figure \ref{fig:lguil}.
Consider $w(L)$ and $h(L)$ to be the width of vertical arm and height of horizontal arm of $L$, respectively. We use strong induction to the prove the claim. Base case, considers all degenerate $\Lc$-compartments with just one arm i.e. $w(L) = 0 $ or $h(L) = 0$. These are just stacks of items which is trivially guillotine separable. For induction hypothesis assume that all nicely packed $\Lc$-compartments $L'$ are guillotine separable for width $w(L')$ and height $h(L')$ such that $w(L') < w(L)$ and $h(L') \leq h(L)$ or $h(L') < h(L)$ and $w(L') \leq w(L)$. For this, we first prove that there exists an item touching the boundary of $L$ which can be separated by a guillotine cut. Consider the left-most vertical item $i_v$ in the vertical arm of $L$. We extend the right boundary of $i_v$ to obtain an end-to-end cut $l_v$ (may be cutting one or more items in $L$). If $l_v$ doesn't cut through any items of $L$, then the we stop as the claim follows trivially. Otherwise, we observe that since the items are placed \emph{nicely} inside $L$, no items in $I_L \cap I_{ver}$ can be cut by $l_v$. This implies $l_v$ intersects only with a non-empty subset of items in $I_{hor}'\subseteq I_L \cap I_{hor}$. Let the top-most intersected horizontal item in $I_{hor}'$ be $i_h$. We then extend the top boundary of $i_h$ to obtain an end-to-end cut $l_h$. Clearly because of nice packing $l_h$ doesn't intersect any items in $I_L \cap I_{hor}$. Also, from the construction it follows that $l_h$ cannot intersect any item of $I_L \cap I_{ver}$ as all items are above it. This gives us a guillotine cut separating a rectangle containing items in $I_{hor}'$ and $L \backslash I_{hor}'$ (can be a degenerate $\Lc$-compartment with just one leg). We consequently get a smaller, nicely packed $\Lc$-compartment $L' =L \backslash I_{hor}'$ with $w(L') =w(L) - w(I_{hor}') < w(L)$ and $h(L') = h(L)$. From induction assumption we deduce that $L'$ is guillotine separable. Appending $l_h$ to the guillotine cut-sequence of $L'$ gives us a guillotine cutting sequence for $L$. Clearly nice packing of skew items ensures that all the cuts in this guillotine cutting sequence lie completely inside the area of $L$ (maybe coincident with boundary of $L$). This concludes the proof.
\end{proof}

\begin{lem}
(Restatement of Lemma \ref{lem:pseudo-enough})
Let $\C$ be a set of compartments inside
$K$ that admit a pseudo-guillotine cutting sequence. Let $\R'\subseteq\R$
be a set of items that are placed nicely inside the compartments in
$\C$. Then there is a guillotine cutting sequence for $\R'$.
\end{lem}
\begin{proof}
We now justify the two operations of pseudo-guillotine cutting sequence. First operation applies a horizontal or vertical guillotine cut, say $l$, that separates rectangle $R$ into two disjoint rectangles $R_{1},R_{2}$ and then continues recursively with $R_{1}$ and $R_{2}$. This operation trivially maintains the guillotine cutting sequence with first cut being $l$ and each of $R_{1}$ and $R_2$ can be considered independently for a guillotine cutting. Harder part is to define a guillotine sequence for second operation which partitions $R$ into a boundary $\Lc$-compartment $L$ and $R\setminus L$. For that we first observe that any sequence of guillotine cuts inside $R$ can be considered independently of rest of the plane. Now, we use Lemma \ref{Lguillotine} to obtain a guillotine cutting sequence $S_L$ separating nicely packed items in $L$  with guillotine cuts completely lying inside the area of $L$-region. This separates out $L$ from $R\setminus L$ without intersecting any items of $R \setminus L$. Now the guillotine cutting sequence of $R\setminus L$ can be considered independently. This proves our claim. Furthermore to obtain the complete guillotine cutting sequence, say $S_K$, for the given set of items $\R'$, we do a Depth First Search (DFS) of the tree of pseudo-guillotine cutting sequence. Every internal node of tree represents an operation. For node labelled as first operation, we can add the considered cut $l$ to the $l_K$ and continue with its child nodes in a DFS fashion. For a node labelled as second operation, we can add the corresponding guillotine sequence $S_L$ of $L$ to $S_K$ and continue to the other child representing $R \setminus L$ in a DFS manner. By the end of DFS run, we get the guillotine cutting sequence $S_K$ of items $\R'$ in $K$. 
\end{proof}

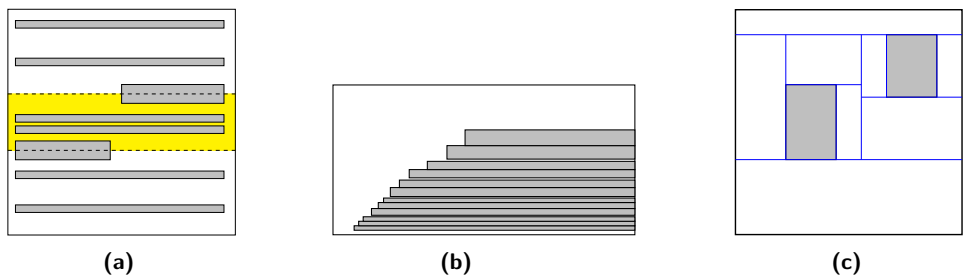
\begin{figure}[b]
		\captionsetup[subfigure]{justification=centering}
		\hspace{-10pt}
	\begin{subfigure}[b]{.25\textwidth}
			\centering
			\resizebox{!}{3cm}{
			\begin{tikzpicture}
			\draw [fill=yellow,yellow] (0,2.25)rectangle (6,3.75);
				\draw (0,0) rectangle (6,6);
			
			\draw [solid,fill=lightgray] (3,3.5) rectangle (5.7,4); 
			\draw [solid,fill=lightgray] (0.2,2.5) rectangle (2.7,2); 
			\draw [solid,fill=lightgray] (0.2,2.7) rectangle (5.7,2.9);
			\draw [solid,fill=lightgray] (0.2,3) rectangle (5.7,3.2);		
			\draw [solid,fill=lightgray] (0.2,5.5) rectangle (5.7,5.7);
			\draw [solid,fill=lightgray] (0.2,0.6) rectangle (5.7,0.8);	
			\draw [solid,fill=lightgray] (0.2,1.5) rectangle (5.7,1.7);
			\draw [solid,fill=lightgray] (0.2,4.5) rectangle (5.7,4.7);	
			
			\draw[dashed] (0,2.25) -- (6,2.25);	
			\draw[dashed] (0,3.75)--(6,3.75);
			\end{tikzpicture}}
			\caption{}
		\end{subfigure}	
\hspace{20pt}
		\begin{subfigure}[b]{.25\textwidth}
			\centering
			\resizebox{!}{2cm}{
			\begin{tikzpicture}
				\draw (0,0) rectangle (4,2);

			\draw[solid, fill=lightgray] (.28,.06) rectangle (4,.12); 
			\draw[solid, fill=lightgray] (.34,.12) rectangle (4,.18); 
			\draw[solid, fill=lightgray] (.40,.18) rectangle (4,.24);
			\draw[solid, fill=lightgray] (.51,.26) rectangle (4,.35);
			\draw[solid, fill=lightgray] (.6,.35) rectangle (4,.43); 
			\draw[solid, fill=lightgray] (.67,.43) rectangle (4,.49);
			\draw[solid, fill=lightgray] (.76,.51) rectangle (4,.63); 
			\draw[solid, fill=lightgray] (.88,.63) rectangle (4,.73);
			\draw[solid, fill=lightgray] (1.01,.76) rectangle (4,.87); 
			\draw[solid, fill=lightgray] (1.25,.87) rectangle (4,.98);
			\draw[solid, fill=lightgray] (1.51,1.01) rectangle (4,1.19); 
			\draw[solid, fill=lightgray] (1.75,1.19) rectangle (4,1.4);
			\end{tikzpicture}}
			\caption{}
		\end{subfigure}
		\hspace{40pt}
		\begin{subfigure}[b]{.25\textwidth}
			\centering
			\resizebox{!}{3cm}{
			\begin{tikzpicture}
				\draw[ultra thick] (0,0) rectangle (9,9);
				\draw[solid,fill=lightgray] (6,5.5) rectangle (8,8);
				\draw[solid,fill=lightgray] (2,3) rectangle (4,6);
				\draw[thick,color=blue] (0,8)--(9,8);
				\draw[thick,color=blue] (0,3)--(9,3);
				\draw[thick,color=blue] (2,8)--(2,3);
				\draw[thick,color=blue] (5,8)--(5,3);
				\draw[thick,color=blue] (2,6)--(5,6);
				\draw[thick,color=blue] (4,6)--(4,3);
				\draw[thick,color=blue] (5,5.5)--(9,5.5);
				\draw[thick,color=blue] (6,5.5)--(6,8);
				\draw[thick,color=blue] (8,5.5)--(8,8);	
			\end{tikzpicture}}
			\caption{}
		\end{subfigure}
		
		\caption{(a): Yellow strip is the least profitable strip and the items which overlap partially or completely will be removed.(b): Items are nicely packed in a box compartment (c): Gray items can not be dropped. So using the blue guillotine cuts, we divide the knapsack into $O_{\eps,\epsl}(1)$ box compartments.}
\label{box-to-comp}
	\end{figure}

\begin{lem} 
(Restatement of Lemma \ref{lem:make-boxes-nice})
Given a box-compartment $B$
containing a set of items $\R(B)$ with $h(B)<\epsl N$ or $w(B)<\epsl N$.
There exists a partition of $B$ into $O_{\eps}(1)$ box-compartments
$\B(B)$ such that 
\begin{itemize}
\item the box-compartments in $\B(B)$ are guillotine separable, and 
\item there is a nice placement of a set of items $\R'(B)\subseteq\R(B)$
with $|\R'(B)|\ge(1-\eps)|\R'(B)|-O_{\eps}(1)$ inside $\B(B)$. 
\end{itemize}
\end{lem} 
\begin{proof}
If the box-compartment contains only one rectangle we don't do anything. Otherwise let us consider a box-compartment $B$ which contains more than one horizontal rectangle (similar procedure can be applied to compartments with vertical rectangles). Now we remove the cheapest strip $S_1$ of height of $3\eps\cdot h(B)$. Note that the profit of a strip is calculated by adding the profits of the items lying completely inside the strip. So the profit of the strip $S_1$ is at most $O(\eps)\cdot \profit(B)$. Now remove the items in the strip $S_1$ and also the items which partially overlap with $S_1$. Note that the number of items which partially overlap with $S_1$ is $O_{\epsl}$. Now by using the techinque of resource augmentation (See Lemma \ref{lemma1_old}), we can repack $(1-O(\eps))\profit(B)$ in a region of size $w(B)\times (1-2\eps)h(B)$ such that this region contains $O_{\eps,\epsl}(1)$ nicely packed box compartments that are guillotine separable. See figure \ref{box-to-comp} for more clarity of the process. Now this works well in the unweighted case. But if we are dealing with a weighted case then we can't drop the $O_{\epsl}(1)$ items which we were dropping as it might contain a lot of profit. To handle the weighted case we use the shifting argumentation and for this we refer the reader to Section \ref{shift-arg}. 
\end{proof}

In this section, we give a detailed proof of Lemma \ref{lem:make-L-nice}. 
First, we prove the following lemma, which will be useful in the proof.

\begin{lem}\label{lem:A}
	There exists a set $OPT'\subseteq \opt$, $p(\opt') > (1-O(\eps))p(\opt)$ and a partition of $K$ 
	into a set of $O_{\eps}(1)$ compartments such that for every $\Lc$-compartment $L$ it holds that
	for each item $i$ contained in the horizontal  $P_H$ (resp. vertical arm $P_V$) of $L$, $\height(i)\le \eps^{4} \cdot\height(P_H)$ 
	(resp. $\width(i)\le \eps^{4} \cdot\width(P_V)$).
\end{lem}
\begin{proof}
	For an $\Lc$ compartment $L$, we use $P_H$ (resp. $P_V$) to denote the horizontal (resp. vertical) arm of $L$.
	In the cardinality case, we simply drop all the items of $\height(i) > \eps^{4} \cdot\height(P_H)$. 
	Number of such items can be at most $\frac{1}{\eps^4 \eps_{large}}$ which is a constant.
	Thus we get the required packing of rest of the items $\opt'$ s.t. $|\opt'| \leq (1-O_{\eps \eps_{large}}(1))|\opt|$.
	
	For weighted case, we cannot drop these $\frac{1}{\eps^4 \eps_{large}}$ items as they might actually carry most of the profit of knapsack. Thus, we now aim to pack back this constant number of items dropped. We do this by employing an averaging argument, that recursively partitions the knapsack for $\frac{1}{\eps}$ times, obtainig a set of items which can be dropped and have a profit of at most $O(\eps).p(\opt)$. Formally, we employ the Shifting Argumentation as defined in Section \ref{shift-arg} which recursively procesess the partitioning such that all the profitable items lie completely inside box compartments. We refer to Section \ref{shift-arg} for the Shifting Argmuntation for further details of the procedure. This way we ensure that for every $\Lc$-compartment, items contained in the horizontal arm $P_H$ (resp. vertical arm $P_V$) have $\height(i)\le \eps^{4} \cdot\height(P_H)$ 	(resp. $\width(i)\le \eps^{4} \cdot\width(P_V)$) without losing more that $O(\eps)$ fraction of profit.

\end{proof}

Hence, we assume from now on that for every $\Lc$-compartment $L$, it holds that
	for each item $i$ contained in the horizontal arm $P_H$ (resp. vertical arm $P_V$) of $L$, $\height(i)\le \eps^{4} \cdot\height(P)$ 
	(resp. $\width(i)\le \eps^{4} \cdot\width(P)$).

\begin{lem}
(Restatement of Lemma \ref{lem:make-L-nice})
Given an $\Lc$-compartment $L$ containing a set
of items $\R(L)$. There exists a partition of $L$ into one $\Lc$-compartment
$L'\subseteq L$ and $O_{\eps}(1)$ box-compartments $\B(L)$
such that 
\begin{itemize}
\item $L'$ lies at the outer boundary of $L$, 
\item the box-compartments in $\B(L)$ are guillotine separable, and 
\item \arn{there is a nice placement of a set of items $\R'(L)\subseteq\R(L)$
with $\profit(\R'(L))\ge(1-O(\eps))\profit(\R'(L))$ inside $\B(L)$}
and $L'$.
\end{itemize}
\end{lem}
\begin{proof}
Let $P_H$ and $P_V$ be the horizontal and vertical arms $L$, respectively.
So $P_H$ is defined via two horizontal edges $e_{1}=\overline{p_{1}p'_{1}}$
and $e_{2}=\overline{p_{2}p'_{2}}$, and additionally one monotone
axis-parallel curves connecting $p_{1}=(x_{1},y_{1})$ with $p_{2}=(x_{2},y_{2})$. 
Assume w.l.o.g. that $x_{1}\le x_{2}$ and that $y_{1}<y_{2}$. 
Let $h(P_H)$ denote the the height of $P_H$ which we define as the distance
between $e_{1}$ and $e_{2}$. Intuitively, we place $1/\eps^{2}$
boxes inside $P_H$ of height $\eps^{2}h(P_H)$ each, stacked one
on top of the other, and of maximum width such that they are contained
inside $P_H$. Formally, we define $1/\eps^{2}$ boxes $B_{0},...,B_{1/\eps^{2}-1}$
such that for each  $j\in\{0,...,1/\eps^{2}-1\}$ the bottom
edge of box $B_{j}$ has the $y$-coordinate $y_{1}+j\cdot\eps^{2}h(P_H)$
and the top edge of $B_{j}$ has the $y$-coordinate $y_{1}+(j+1)\cdot\eps^{2}h(P_H)$. 
For each such $j$ we define
the $x$-coordinate of the left edge of $B_{j}$ maximally small 
such that $B_{j}\subseteq P_H$.


In the weighted
case, from Lemma \ref{lem:A} each item $i$ contained in $P_H$ satisfies that $\height(i)\le \eps^{4} \cdot\height(P)$
and therefore each item $i$ contained in $P_H$ intersects at most
2 stripes in $\left\{ S_{0},...,S_{1/\eps^{2}-1}\right\} $.
Therefore, by pigeon hole principle, there exist two boxes $B_{j^{*}}, B_{j^{*}+1} \in\left\{ B_{0},...,B_{1/\eps-1}\right\} $
with the property that the stripe $S'_{j^{*}}:=[0,N]\times[y_{1}+j^{*}\cdot\eps^{2}h(P_H),y_{1}+(j^{*}+2)\cdot\eps^{2}h(P_H)]$
(containing $B_{j^{*}}, B_{j^{*}+1}$) intersects at most $2\eps|\opt'(P_H)|$
of the remaining items in $\opt'(P_H)$. We drop all items in $\opt'(P_H)$
that are intersected by $S'_{j^{*}}$. Next, we move down all items
in $\opt'(P_H)$ that intersect the boxes $B_{j^{*}+2},...,B_{1/\eps^{2}-1}$
by $\eps^{2}h(P_H)$ units. Note that then they fit into the area
defined by the union of the boxes $B_{j^{*}+1},...,B_{1/\eps^{2}-2}$.
We assign to $B_{j^{*}}$ all items that intersect
a horizontal edge of a box in $\left\{ B_{j^{*}+1},...,B_{1/\eps^{2}-1}\right\} $.
This can be done since each such item has a height of at most $\eps^{4} \cdot\height(P_H)$
which implies that $\eps^{4} \cdot\height(P)/\eps^{2}\le\eps^{2}\cdot\height(P_H)$.

Let  $w(P_H)$ and  $w'(P_H)$  be  the length of $e_{1}$ and $e_{2}$, resp. 
Due to the definition of $\Lc$-compartment, $w(P_H)\ge \epsl N$ and $w'(P_H)\ge w(P_H)-\epsl N/2 \ge w(P_H)/2$.
Next, we would
like to ensure that below the box $B_{j^{*}}$ there is no item $i$
with $\width(i)<w'(P_H)$ (we want to achieve this since then we can
stack the items underneath $B_{j^{*}}$ on top of each other) and
no small item intersects the boundary of a box.

Therefore, consider the topmost $1/\eps^{2}-1/\eps$ boxes.
We group them into $1/6\eps-1$ groups with $6/\eps$ boxes
each, i.e., for each $k\in\{0,...,1/6\eps-2\}$ we define a group
$\B_{k}:=\{B_{j}|j\in\{1/\eps+6k/\eps,...,1/\eps+6(k+1)/\eps-1\}\}$.
Note that each group $\B_{k}$ contains exactly $6/\eps$ boxes
and below $B_{j^{*}}$ there are at most $1/\eps$ boxes. By the
pigeon hole principle, there is a value $k^{*}\in\{0,...,1/\eps-2\}$
such that the the boxes in the group $\B_{k^{*}}$ intersect with
items of total weight of $O(\eps)w(\opt'(P_H))$ where $w(\opt'(P_H))$
denotes the weight of the remaining items in $\opt'(P_H)$. Therefore,
we drop all items that intersect a box in $\B_{k^{*}}$.

Consider all items $i\in\opt'(P_H)\cap\Rsk$ that intersect at least
one of the stripes in $\left\{ S_{0},...,S_{j^{*}-1}\right\} $ and
that satisfy that $\width(i)\le w'(P_H)$. Due to Steinberg's algorithm~\cite{steinberg1997strip}
they fit into a box of height $3\eps\cdot h(P_H)$ and width $w'(P_H)$.
Therefore, all but $O(1/\epsl\cdot1/\eps)$ of them fit into $3/\eps$
boxes in $\B_{k^{*}}$. We assign them to these boxes $3/\eps$
boxes in $\B_{k^{*}}$.



In the weighted case, the total area of items $i$ that intersect
the boundary of a box in $\left\{ B_{0},...,B_{1/\eps^{2}-1}\right\} $
or that are contained in one of the stripes in $\left\{ S_{0},...,S_{j^{*}-1}\right\} $
with $\height(i)\le \eps^{4} \cdot\height(P_H)$ and $\width(i)\le\epss\cdot\width(P_H)$
is bounded by $\eps\cdot h(P_H)\cdot w(P_H)+O(w(P_H)h(P_H)\epss/\eps^{2})\le O(\eps)\cdot h(P_H)\cdot w'(P_H)$.
Hence, we can pack them into the remaining boxes in $\B_{k^{*}}$ like before using Steinberg's algorithm~\cite{steinberg1997strip}.

Now we define $P'_H$ as the acute piece induced by $e_{1}$, the bottom
edge of $B_{j^{*}}$, and the respective part of the two monotone
axis-parallel curves connecting $p_{1}=(x_{1},y_{1})$ with $p_{2}=(x_{2},y_{2})$
and the right boundary edge of $P_H$,
respectively. Each remaining item $i\in\opt'(P_H)$ intersecting $P'_H$
satisfies that $\width(i)\ge w'(P_H)$. Therefore, we can stack these
items on top of each other (using that $w'(P_H)>w(P_H)/2$).

We obtain that each remaining item from $\opt'(P_H)$ is assigned to
a box in $\left\{ B_{j^{*}},...,B_{1/\eps^{2}-1}\right\} $ or
lies in $P'_H$. We define $\opt'_{1}(P_H)$ to be the former set of items
and $\opt'_{2}(P_H)$ to be the latter set. Finally, we apply Lemma \ref{lem:make-boxes-nice} 
to each box $B\in\left\{ B_{j^{*}},...,B_{1/\eps^{2}-1}\right\} $
in order to partition $B$ further and such that the items assigned
to $B$ are nicely packed inside $B$.

We do a symmetric procedure for $P_{V}$, yielding a set of boxes
$\B_{ver}$. Intuitively, we want to define $L'$ as $P'_{H}\cup P'_{V}$.
However, $P'_{H}\cup P'_{V}$ might not have exactly the shape of
an $\Lc$-compartment. Nevertheless, one can show that we can subdivide
one of these polygons, say $P'_{H},$ along a horizontal line into
two subpolygons $P'_{H,\mathrm{top}},P'_{H,\mathrm{bottom}}$ (with
$P'_{H,\mathrm{top}}$ lieing on the top of $P'_{H,\mathrm{bottom}}$)
such that 
\begin{itemize}
\item we can place the items in $P'_{H,\mathrm{top}}$ into another set
of $O_{\eps}(1)$ boxes $\B'_{hor}$ that are non-overlapping
with $\B_{hor}\cup\B_{ver}$ and 
\item $L':=P'_{H,\mathrm{bottom}}\cup P'_{V}$ forms an $\Lc$-compartment,
see Figure~\ref{fig:Lnice6}.
\end{itemize}
Then the items in $L'$ are nicely placed inside $L'$. To each of
the $O_{\eps}(1)$ boxes $B\in\B_{hor}\cup\B'_{hor}\cup\B_{ver}$
we apply a standard routine that removes some of the items inside
$B$ and partitions $B$ into smaller boxes, such that inside these
smaller boxes the remaining items are nicely placed.
It is easy to check that the resulting packing satisfy guillotine separability.
\end{proof}

\subsection{Adding small rectangles}\label{smallrec}
\begin{figure}[h]
		\captionsetup[subfigure]{justification=centering}
		\hspace{-10pt}
	\begin{subfigure}[]{.5\textwidth}
			\centering
			\resizebox{!}{3cm}{
			\begin{tikzpicture}
				\draw[ultra thick] (0,0) rectangle (9,9);
				\draw[dashed] (0,9) rectangle (10,-1);
				\draw[dashed](0,7)--(10,7);
				\draw[dashed](0,5)--(10,5);
				\draw[dashed](0,3)--(10,3);
				\draw[dashed](0,1)--(10,1);	
				\draw[dashed](2,9)--(2,-1);
				\draw[dashed](4,9)--(4,-1);
				\draw[dashed](6,9)--(6,-1);
				\draw[dashed](8,9)--(8,-1);
			\end{tikzpicture}}
			\caption{}
			\label{fig:small-grid-1}
		\end{subfigure}	
		\begin{subfigure}[]{.5\textwidth}
			\centering
			\resizebox{!}{3cm}{
			\begin{tikzpicture}
				\draw (0,9)--(3,9)--(3,4)--(9,4)--(9,0)--(0,0)--(0,9);
				\draw[ultra thick] (0,0) rectangle (9,9);
				\draw[dashed] (0,9) rectangle (10,-1);
				\draw[dashed](0,7)--(10,7);
				\draw[dashed](0,5)--(10,5);
				\draw[dashed](0,3)--(10,3);
				\draw[dashed](0,1)--(10,1);	
				\draw[dashed](2,9)--(2,-1);
				\draw[dashed](4,9)--(4,-1);
				\draw[dashed](6,9)--(6,-1);
				\draw[dashed](8,9)--(8,-1);
			\end{tikzpicture}}
			\caption{}
			\label{fig:small-grid-2}
		\end{subfigure}
		
		\caption{(a): Grid made for a rectangular region (b): Grid for a $\Lc$-region}
	\end{figure}
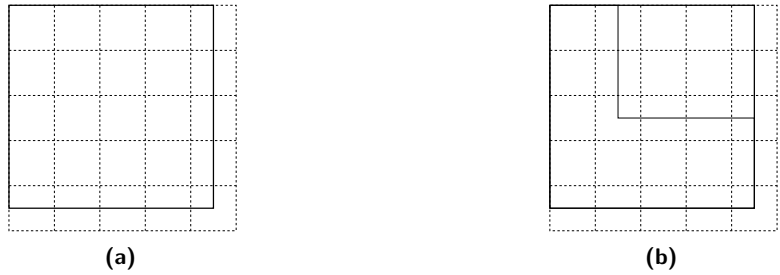

\begin{lem}\cite{GalvezGHI0W17}\label{dead}Consider a rectangular region $R$ of size $a\times b$ where $a,b\leq N$. Now assume that $R$ consists of only box compartments in which the skewed items are nicely packed. Now if we build a grid of width $\eps'N=\frac{\epss}{\eps}\cdot N$ (as shown in Figure \ref{fig:small-grid-1}) and delete any cell of the grid that overlaps with the compartment or intersects with one of the boundaries of $R$,  then total area of the deleted (also called {\em dead}) grid cells is min$\{ (1-\eps)a(R),a(\opt_{com}')+(\epsau+2\eps')a(R)\}$ where $a(\opt_{com}')$ denotes the  total area of  items in the compartments in $R$.
\end{lem}
Note that Lemma \ref{dead} is applicable when there are no large items. If now if we have a $\Lc$-region $L'$ consisting of $O_{\eps}(1)$ box-compartments $\B(L)$ (for example in the packing mentioned in Lemma \ref{lem:make-L-nice}) then we can consider a dummy minimal rectangular region (see Figure \ref{fig:small-grid-2}, the smaller rectangular region to the right of the $\Lc$-compartment can be considered as a pseudo item in order to apply the above lemma)  that can contain the $\Lc$-region $ L'$ and apply Lemma \ref{dead} and we can get similar results.

 We denote the cells which are not dead as free cells. Now take the knapsack consider those regions which don't contain $\Lc$-compartment. Now find the regions which are guillotine separable and do not overlap with any compartment. Let us denote these regions as guillotine free region. By appropriately choosing $\eps,\epsl,\epss,\epsau$ (as shown in subsection \ref{cnt}) we can ensure that the total area of guillotine free regions which don't have both the dimensions greater than or equal to $\eps' N$ is at most $O(\eps^{2} N^2)$. Also the total area of guillotine free region is greater than or equal to the total area of free cells.
\begin{lem}\label{bigL}
Consider an $\Lc$- compartment $C$ such that the items in it are nice packed and the length of any of its 6 edges is atmost $N$. Now we build a grid of width $(\eps'+\epss )N$ in a dummy minimal rectangular region that can contain (see Figure \ref{fig:small-grid-2}). We delete a cell if it overlaps with a rectangle or intersects with one of the 6 edges of $C$. Then the total area of deleted grid cells is at most $a(\opt_{C}')+O(\eps'\cdot N^{2})$ where $a(\opt_{C}')$ is the total area of items in $C$.
\end{lem}
\begin{proof}Let us assume that the vertical part $S_{1}$ of $C$ is top-left  of horizontal part $S_{2}$ of $C$. Also assume the items in $S_1$ are placed in non-increasing order of height from left to right (similar analogy for vertical rectangles). Now consider the first column of grid cells from left. The area of dead cells in first column is at most $(\eps'+\epss)N^{2}$. In the second column of grid cells from left if there are $x$ grids cells which are dead then the length of all the long  rectangles in the first column is at least $(x-1)\cdot (\eps'+\epss)\cdot N$. So the area of dead cells in second column is at most area of rectangles in the first column plus $(\eps'+\epss)^{2}N^{2}$. Adding up the area of all the dead cells in this manner and repeating the same for horizontal rectangles we get total area of deleted grid cells to be at most $a(\opt_{C}')+O(\eps'\cdot N^{2})$. 
 \end{proof}

A box is said to be $\eps'$-large if both dimensions are greater than or equal to $\eps' N$. Now we describe a procedure to obtain $\eps'$ large boxes (to pack the small rectangles) from the free cells in an $\Lc$- compartment $C$. Note that these $\eps'$ large boxes when considered as pseudorectangles are guillotine separable. Consider the first column of grid cells from the left and assume that we have free cells in this column. Separate the horizontal rectangle, if any (using a horizontal guillotine cut), which overlaps with this first column. Now consider the rightmost vertical cut which passes through this column. The distance of this cut from the right-hand side of this column is at most $\epss N$. So we obtain $\eps'$ large boxes in the first column which are guillotine separable.  Continue the same process for other columns which overlap with the vertical part of the boundary-$\Lc$. Free cells which are in the horizontal part at the end of this process are guillotine separable. Now the guillotine cuts which we used to separate out the $\eps'$ large boxes can be used to divide $C$ into $O_{\eps'}(1)$ guillotine separable box compartments in which the items are nice packed and possibly a $\Lc$- compartment in which the items are nicely packed.

Now $\eps'$-large boxes and guillotine free regions which were created have an area of at least $(\max\{ N^2-a(\opt_{comp})-a(\opt_L)+\eps a(\opt_{comp}),a(\opt_{small})\}-O(\eps^{2}N^{2}))$. Here $a(\opt_{comp})$ denotes the total area of box compartments and $a(\opt_L)$ denotes the total area of items in $\Lc$-compartments. So if $a(\opt_{small})\geq \frac{\eps N^{2}}{4}$ we can pack almost all the profitable small rectangles using NFDH in the guillotine separable  $\eps'$-large boxes created. If $a(\opt_{small})$<$\frac{\eps N^{2}}{4}$ and $N^2-a(\opt_{comp})-a(\opt_L)+\eps a(\opt_{comp})>\frac{\eps N^{2}}{4}$, we can pack all the small rectangles as area of guillotine separable regions is more than $\frac{\eps N^{2}}{4}$. Now if $a(\opt_{small})$<$\frac{\eps N^{2}}{4}$ and $N^2-a(\opt_{comp})-a(\opt_L)+\eps a(\opt_{comp})\leq\frac{\eps N^{2}}{4}$, then we can remove some least profitable items from $\opt_L$ (profit of such items is at most $O(\eps)\profit(\opt)$ as $a(\opt_L)\geq N^2/2$ and number of $\Lc$-compartments is at most $O(1/\epsl)$) and create enough space to pack all the items in $\opt_{small}$.
This ensures a packing of profit $(1-O(\eps))\profit(\opt_{small})$ from small items. 
Lemma \ref{dead} is applicable when there are no large items and the constant number of items which we remove have a profit of at most $O(\eps)\profit(\opt)$. So if these items are profitable then we can use the shifting argumentation of \cite{GalvezGHI0W17} and create a box compartment for each such item and repeat the entire process of adding small items in the remaining regions of knapsack which don't contain the box compartments which we created for the profitable items which can't be dropped. The analysis now is same as in the case of the cardinality case. The only difference is that $N^2$ will get replaced by $N^2/c'$ where $c'$ is a constant which depends only on $\epsl$ and $\eps$ only.  See section \ref{shift-arg} for a brief overview of shifting argumentation.

\subsection{Relationship between different constants}\label{cnt}
 We set $\epsau$ to be $\eps^{2}$ and $\epsau'=\epsau/2$. By choosing a big enough constant $\lambda$ we set function $f(x)$ in the lemma \ref{class} to be $\frac{x^{120+1/\eps}}{2^{1/\eps}(2K'(\epsau'^{\lambda^{1/\eps}})+\frac{1}{\epsau'})^{2}}$. Note than $\lambda$ doesn't depend on $\epsau'$ or $\eps$. Here $K'(x)=(\frac{2}{x^{10}}+(\frac{3}{x^{2}}(M(x))^{2}2^{2M(x)})^{\epsau'})$ $\lceil \log_{1+\epsau'}(1/x)\rceil/\eps^{2}$ where $M(x)=\frac{1+2x}{x^{2}}$. Now we show that all our constraints are satisfied. By lemma \ref{class} we get a pair $(\epss,\epsl)$ such that $\epss \leq \frac{\epsl^{120+1/\eps}}{2^{1/\eps}(2K'(\epsau'^{\lambda^{1/\epsau'}})+1/\epsau')^{2}}$.  Let the total area of guillotine free regions with at least one of the dimensions less than $\eps' N$ be $A_{< \eps'}$. Then $A_{< \eps'} < c\cdot (2/\epsl)^{1/\eps}\cdot(2K'(\delta)+1/\epsau')^{2}\cdot\frac{1}{\eps^{c_1}}\cdot\frac{1}{\epsl^{c_2}}\cdot\frac{\epss}{\eps}N^{2}<\eps^{2}N^{2}$. Here $\delta > \epsau'^{\lambda^{1/\eps}}$, $c$ is a constant independent of $\eps$ and $c_1,c_2\leq 10$. Loose upper bound for $A_{< \eps'}$ is found by multiplying a multiple of total number of compartments by maximum area of guillotine free region which isn't $\eps'$ large.

\section{Shifting Argumentation }\label{shift-arg}
	In the cardinality case, we can simply drop a  $O_{\eps,\epsl}(1)$ number of items and thereby obtain a packing of rest of the items $\opt'$ s.t. $|\opt'| \leq (1-O_{\eps}(1))|\opt|$.


But if we are dealing with a weighted case then we can't drop the $O_{\epsl,\eps}(1)$ items which we were dropping as it might contain a lot of profit. So instead we employ a shifting argumentation very similar to the one used in \cite{GalSocg21}. We define a pairwise disjoint set of items $K(t)$ where $0\leq t < 1/\eps$. Also for $0\leq t < 1/\eps$, let $\mathcal{K}(t):=\bigcup \limits  _{i=0}^t K(t)$. Note that we begin with iteration $t=0$ and $K(0)$ is the set of items of $O_{\epsl,\eps}(1)$ items which we dropped in the cardinality.  If $\profit(K(0)) \leq \eps\cdot\profit(\opt)$, then we stop and drop all items of $K(0)$ as the rest of items will have profit at least  $(1-\eps)p(\opt)$. Otherwise, we perform the following procedure recursively. Assume that we are at the $t^{th}$ iteration and we have computed $\mathcal{K}(t)$. Denote $G(t)$ as the non-uniform grid formed by extending the $x$, $y$ coordinates of the items in $\mathcal{K}(t)$(which we obtained from previous $t-1$ iterations). Consider $C_{t}$ to be the set of grid cells of $G(t)$. We classify items in the knapsack as small, large, horizontal, vertical as before depending on its intersection with the cells it overlaps. Let $\opt(C)$ denote the set of items from $OPT$ intersecting with grid cell C. This gives us the following partition of items inside cell $C$:
\begin{enumerate}
\item $\opt_{ver}(C) = \{ i~|~ \forall i\in C$ s.t. $ \height(i \cap C)$ $>$ $\eps_{large}$ $\height(C)$ and $\width(i \cap C)$ $\leq$ $\eps_{small}$ $\width(C)$ $\}$, 

\item $\opt_{hor}(C)= \{ i~ | ~\forall i\in C$ s.t. $\height(i \cap C)$ $\leq$ $\eps_{small}$ $\height(C)$ and $\width(i \cap C)$ $>$ $\eps_{large}$ $\width(C)$ $\}$, 

\item $\opt_{large}(C)$ $= \{ i~ |~ \forall i\in C$ s.t. $\height(i \cap C)$ $>$ $\eps_{large}$ $\height(C)$ and $\width(i \cap C)$ $>$ $\eps_{large}$ $\width(C)$ $\}$, 

\item $\opt_{small}(C)$ $= \{ i ~| ~\forall i\in C$ s.t.  $\height(i \cap C)$ $\leq$ $\eps_{small}$ $\height(C)$ and $\width(i \cap C)$ $\leq$ $\eps_{small}$ $\width(C)$ $\}$, 

\item $\opt_{intermediate}(C)$ $= \{ i ~| ~\forall i\in C$ s.t. $\height(i \cap C)$ $\in$ $(\eps_{small}$ $\height(C)$,  $(\eps_{large}$ $\height(C)]$ or  $\width(i \cap C)$ $\in$ $(\eps_{small}$ $\width(C)$,  $(\eps_{large}$ $\width(C)]$ $\}$, 

\item $\opt_{skew}(C)$ $=$  $\opt_{hor}(C) \cup \opt_{ver}(C)$.
\end{enumerate}

Using Lemma~\ref{class}, we can fix $\eps_{small}$ and $\eps_{large}$ according to our convenience ensuring that the 
profit of items classified as intermediate for any cell is at most 
$\eps. \profit(\opt)$ such that $\eps$ $\geq$ $\eps_{large}$ $\geq$ $f(\eps_{small})$ $\geq$ $\omega_{\eps}(1)$, 
for some positive increasing function $f(.)$ . We add items from $\opt_{large}(C_t)$ (not contained in $\mathcal{K}(t)$) to $K(t+1)$ for all $C$ 
not fully covered by any item in $\mathcal{K}(t)$. This adds a constant number of items to $K(t+1)$. For simplification, we can stretch this non-uniform
 grid to a uniform $N\times N$ grid where each cell is of same dimension. Clearly, we can map any decomposition done in this uniform grid back to original knapsack, mapping back any guillotine cuts in the uniform grid as well, thereby conserving guillotine separability of items. After this process, each item in $\opt_{skew}(C)\setminus \mathcal{K}(t)$ 
has its longer dimension (width for $\opt_{hor}$ and height for $\opt_{ver}$) at least $\frac{\eps_{large}N}{1+2\dot|\mathcal{K}(t)|}$.

Now we start the process of partitioning the knapsack into $O_{\epsl,\eps}(1)$ $\Lc$- and $\Bc$-compartments as mentioned in Section \ref{sec:bldecom-1} by considering only skew items in the cells i.e. $\opt_{skew}(C_t) = \cup_{\forall C \in C_t}\opt_{skew}(C)$ and the items in $\mathcal{K}(t)$. Next we show how to the handle the items in $\mathcal{K}(t)$. If there is an item $i'$ from $\mathcal{K}(t)$ which is in a $\Lc$-compartment, then let's first separate the $\Lc$-compartment using the pseudo guillotine cuts and then consider a guillotine cut such that you are left with a rectangular strip and a smaller $\Lc$-compartment. If the guillotine cut was vertical (resp. horizontal) then the height (resp. width) of the rectangular strip must be equal to the height (resp. width) of the $\Lc$-compartment. Recursively repeat this procedure until you find a rectangular strip $S$ which contains the item $i'$.  Merge the strips which were removed before the strip $S$ into a single $\Lc$-compartment and make $S$ a box-compartment. Repeat this for every item of  $\mathcal{K}(t)$ which is in a $\Lc$-compartment and it will lead to additional $O_{\epsl,\eps}(1)$ $\Lc$- and $\Bc$-compartments.

Next we show how to handle the items from $\mathcal{K}(t)$ packed in $\Bc$-compartments created till now. W.l.o.g let $i'$ be such an item in a horizontal $\Bc$-compartment, say $B_1$. Let the first stage cuts be horizontal. Consider the topmost cut below $i'$ and the bottom most cut above $i'$ (these cuts are allowed to coincide with the edges of the $\Bc$-compartment $B_1$). Using these two cuts we divide the $\Bc$-compartment into at most 3 $\Bc$-compartments. Now consider the smaller $\Bc$-compartment; say $B_2$; which contains $i'$. The first stage cuts in $B_2$ will be vertical and by using 2 vertical cuts. Consider the rightmost cut to the left of $i'$ and the leftmost cut to the right of $i'$ (these cuts are allowed to coincide with the edges of the $\Bc$-compartment $B_2$). Using these two cuts we divide the $\Bc$-compartment into at most 3 $\Bc$-compartments. Now recursively perform the procedure described on the smaller $\Bc$-compartment $B_3$ which contains $i'$ until item $i'$ becomes a container. See Figure \ref{box-to-comp}(c) for more clarity. Note that there can be at most  $O_{\epsl,\eps}(1)$ number of stage cuts in $B_1$ and each stage cut we create 2 extra box compartments. Repeat this for every item of  $\mathcal{K}(t)$ which is in a $\Bc$-compartment and it will lead to additional $\Bc$-compartments. 

Now we can apply the same procedure which we applied during the cardinality case on the current set of $\Lc$-compartments and $\Bc$-compartments. If it is ensured that the $O_{\epsl,\eps}(1)$ items which we will drop now is disjoint from $\mathcal{K}(t)$, then we consider these $O_{\epsl,\eps}(1)$ items as $K(t+1)$ and proceed to the next iteration.

This way we repeatedly perform the same process for at most ${1}/{\eps}$ iterations ensuring $\mathcal{K}(t)$ being untouchable set of items in $t^{th}$ iteration as before. Now, we exploit the fact that sets $K(0), K(1), ..., K(t)$ are pairwise disjoint and have constant number of items. Due to averaging argument, we can claim that there is a $t<1/\eps$ such that $p(K(t)) \leq \eps\cdot p(\opt)$. Then we drop $K(t)$ and we have the required packing in the weighted case.
\section{One Sided Resource Augmentation}\label{resource_aug}
	In this section we show that the proof techniques used in \cite{GalvezGHI0W17} for packing rectangles with resource augmentation maintain the guillotine separability of the rectangles. This is one of the key techniques which was used to prove Lemma \ref{lem:make-boxes-nice}. We will be using compartments and containers interchangeably for the rest of this paper.
	\begin{lem}
	\label{lemma1_old}
	(Resource Augmentation Packing Lemma\cite{GalvezGHI0W17}) Let $\R'$ be a collection of rectangles that can be packed into a box of size $a \times b$, and $\epsau > 0$ be a given constant. Here $a$ denotes the height of the box and $b$ denotes the width. Then there exists a nice packing of $\R''\subseteq \R'$ inside a box of size $a\times(1+\epsau)b$ (resp.~$(1+\epsau)a\times b$) such that:
	\begin{enumerate}
	\item $\profit(\R'')\geq (1-O(\epsau))\profit(\R')$;
	\item the number of containers is $O_{\epsau}(1)$ and their sizes belong to a set of cardinality $n^{O_{\epsau}(1)}$ that can be computed in polynomial time;
	\item the total area of the the containers is at most $\area(\R')+\epsau ab$;  
	\end{enumerate}
	\end{lem}
	In the following subsections, we show how to extend the above lemma using the same proof techniques and show at the end of each proof technique that the set of rectangles is still guillotine separable. For simplicity, we assume $a=1$ and $b=1$, and all widths and heights of rectangles are in $(0,1]$.
	\begin{lem}\cite{GalvezGHI0W17}
	Let $\eps > 0$ and let $f(\cdot)$ be any positive increasing function such that $f(x)<x$ for all  $x$. Then, $\exists \delta, \mu \in \Omega_{\eps}(1)$,with $f(\eps)\geq\mu$ such that the total profit of all rectangles whose width or height lies in $(\mu,\delta]$ is at most $\eps\cdot \profit(\R')$.
	\end{lem}
	For now we use the value of $\eps$ required in this lemma as $\eps'_{ra}$ and later on define the function $f$ that is being used. By choosing appropriate $\mu,\delta$ we classify the rectangles for this section as follows
	\begin{itemize}
	\item short if $\width \leq \mu$;
	\item narrow if $\height \leq \mu$;
	\item wide if $\width > \delta$;
	\item high if $\height >\delta$;
	\item Vertical is the item which is short as well as high, i.e., $\height >\delta$ and $\width \leq \mu$;
	\item Horizontal is the item which is wide as well as narrow, i.e., $\height \leq \mu$ and $\width > \delta$.
	\end{itemize}

	\subsection{Shifting Argument\cite{GalvezGHI0W17}}
	\label{sec:reshift}
	The procedure defined in this subsection is used as a tool at various steps of our proof of Resource Augmentation Lemma. Thus we define the processing and establish that it maintains guillotine property with losing no more than a small profit fraction as follows:
	\label{shifting1}
	\subsubsection{Process}

	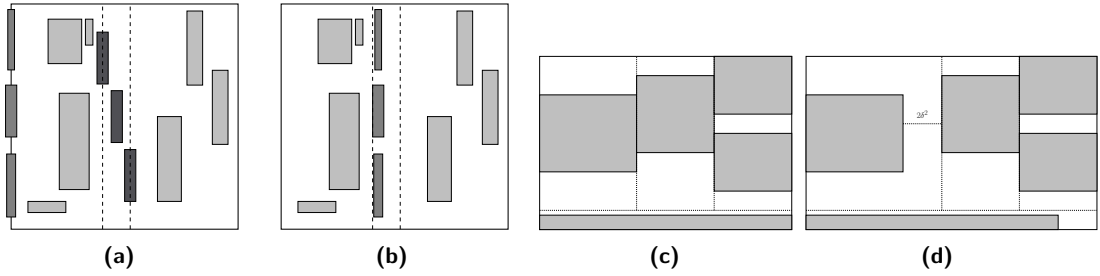
\begin{figure}[!tbh]
		\captionsetup[subfigure]{justification=centering}
		\hspace{-10pt}
		\begin{subfigure}[b]{.25\textwidth}
			\centering
			\resizebox{!}{3cm}{
			\begin{tikzpicture}
				\draw[thick] (0,0) rectangle (6.08,6.08);
				\draw[solid, fill = gray] (-.12,.33) rectangle (.12,2.03);
				\draw[solid, fill = gray] (-.15,2.49) rectangle (.15,3.89);
				\draw[solid, fill = gray] (-.09, 4.3) rectangle (.09,5.93);
				\draw[solid, fill = lightgray] (.45,.45) rectangle (1.465,.75);
				\draw[solid, fill = lightgray] (.99, 4.47) rectangle (1.89,5.67);
				\draw[solid, fill = lightgray] (1.29, 1.07) rectangle (2.09,3.67);
				\draw[solid, fill = lightgray] (1.99, 4.97) rectangle (2.19,5.67);

				\draw[solid, fill = darkgray] (3.04,.75) rectangle (3.34,2.15);
				\draw[solid, fill = darkgray] (2.68,2.34) rectangle (2.98,3.74);
				\draw[solid, fill = darkgray] (2.3,3.92) rectangle (2.6,5.32);
	
				\draw[dashed] (2.45,0) -- ( 2.45, 6.08);
				\draw[dashed] (3.19, 0) -- ( 3.19, 6.08);
				
				\draw[solid, fill = lightgray] (3.92,.75) rectangle (4.56,3.04);
				\draw[solid, fill = lightgray] (5.39,2.29) rectangle (5.81,4.29);
				\draw[solid, fill = lightgray] (4.71,3.89) rectangle (5.13,5.89);
	
			\end{tikzpicture}}
			\caption{}
			\label{figure2}
		\end{subfigure}
		\begin{subfigure}[b]{.25\textwidth}
			\centering
			\resizebox{!}{3cm}{
			\begin{tikzpicture}
				\draw[thick] (0,0) rectangle (6.08,6.08);
				
				\draw[solid, fill = lightgray] (.45,.45) rectangle (1.465,.75);
				\draw[solid, fill = lightgray] (.99, 4.47) rectangle (1.89,5.67);
				\draw[solid, fill = lightgray] (1.29, 1.07) rectangle (2.09,3.67);
				\draw[solid, fill = lightgray] (1.99, 4.97) rectangle (2.19,5.67);

				\draw[solid, fill = gray] (2.48,.33) rectangle (2.72,2.03);
				\draw[solid, fill = gray] (2.45,2.49) rectangle (2.75,3.89);
				\draw[solid, fill = gray] (2.51, 4.3) rectangle (2.69,5.93);
	
				\draw[dashed] (2.45,0) -- ( 2.45, 6.08);
				\draw[dashed] (3.19, 0) -- ( 3.19, 6.08);
				
				\draw[solid, fill = lightgray] (3.92,.75) rectangle (4.56,3.04);
				\draw[solid, fill = lightgray] (5.39,2.29) rectangle (5.81,4.29);
				\draw[solid, fill = lightgray] (4.71,3.89) rectangle (5.13,5.89);
				
			\end{tikzpicture}}
			\caption{}
			\label{figure1}
		\end{subfigure}
		\begin{subfigure}[b]{.25\textwidth}
			\centering
			\resizebox{!}{2.3cm}{
			\begin{tikzpicture}
				\draw[thick] (0,0) rectangle (26,18);
				
				\draw[solid, fill = lightgray] (0,6) rectangle (10,14);
				\draw[solid, fill = lightgray] (10,8) rectangle (18,16);
				\draw[solid, fill = lightgray] (18,4) rectangle (26,10);
				\draw[solid, fill = lightgray] (18,12) rectangle (26,18);
				\draw[dashed] (0,2) -- ( 26,2);
				\draw[dashed] ( 10,2) -- (10,18);
				\draw[dashed] ( 18,2) -- ( 18,18);
				\draw [solid,fill=lightgray] (0,0) rectangle (26,1.5);	
			\end{tikzpicture}}
			\caption{}
			\label{rounding}
		\end{subfigure}
		\begin{subfigure}[b]{.25\textwidth}
			\centering
			\resizebox{!}{2.3cm}{
			\begin{tikzpicture}
				\draw[thick] (0,0) rectangle (30,18);
				\draw[solid, fill = lightgray] (0,6) rectangle (10,14);
				\draw[solid, fill = lightgray] (14,8) rectangle (22,16);
				\draw[solid, fill = lightgray] (22,4) rectangle (30,10);
				
				\draw[solid, fill = lightgray] (22,12) rectangle (30,18);
				\draw[dashed] (10,11) -- ( 14,11);
				\draw[dashed] (0,2) -- ( 30,2);
				\draw[dashed] ( 14,2) -- (14,18);
				\draw[dashed] ( 22,2) -- ( 22,18);
				\draw(12,12) node {\Huge \textbf{$2\delta^{2}$}};
				\draw [solid,fill=lightgray] (0,0) rectangle (26,1.5);
			\end{tikzpicture}}
			\caption{}
			\label{fig h}
		\end{subfigure}
		\caption{Figure (a),(b) : Shifting - From original packing we remove a low profit subset of rectangles (dark gray). Then we make space for short-high rectangles that cross the left edge of the box (medium gray). Figure (c),(d) : Rounding - Strips which share boundary with vertical cuts are shifted by $2\delta^{2}$.}
	\end{figure}
	
	Consider a box $\mathcal{B}$ in a knapsack with width $\delta^{2}$ consisting of vertical items only and leftmost edge intersecting some vertical items as shown in Figure \ref{figure2}. Let us assume that the box as a pseudoitem is guillotine separable if we remove the vertical items intersecting the leftmost edge of $\mathcal{B}$ . Also assume that the items in the box are guillotine separable. Divide the box in vertical strips of width $\mu$. Using averaging argument, there must exist a strip $S$ where the weight of items which overlaps with it is at most 2$\mu \profit(\mathcal{B})$ $\leq \epsau'\profit(\mathcal{B})$, where $\profit(\mathcal{B})$ denotes the total profit inside the box $\mathcal{B}$. Now remove this low profit strip and place the items intersected by the leftmost edge of the box in the space created by the strip on top of one another. 
		\subsubsection{Guillotine Separability}
		
Since we delete all the rectangles lying either partially or completely on the strip $S$ in the box $\mathcal{B}$, there is no rectangle which intercepts  the  cuts along the boundary of strip $S$. Thus the strip $S$ can be guillotine separated from the rest of the box. Given that the box $\mathcal{B}$ is already guillotine separable, it follows that the process of shifting maintains the guillotine structure.
	\subsection{Rounding wide rectangles}\label{roundwiderects}
We describe a procedure for rounding width and coordinates of wide items to the nearest integral multiples of $\delta^{2}$ in this subsection. In the subsequent subsection \ref{box} we describe the procedure for packing vertical items as done in \cite{GalvezGHI0W17} thereby maintaining the guillotine property in the processing. We thus will be proving the following lemma:
\begin{lem}\label{verticalandbox}
Let $\R'$ be the set of guillotine separable items that can be packed into a $1\times1$ box. Then there exists a guillotine separable packing of a profitable subset $\R^{+}$ $\subset$ $\R'$ with profit of $\R^{+}$ is at least $(1-O(\epsau')$$.\profit(\R')$ in a $1\times(1+2\delta)$ box such that:
\begin{itemize}
	\item Every wide item in $\R^{+}$ has width rounded up to the nearest multiple of $\delta^{2}$ and is shifted such that its leftmost coordinate is also an integral multiple of $\delta^{2}$.
	\item Every box in $B$ has width $\delta^{2}$ and stores at least one vertical item packed in at most $1/\delta^{7}$ containers. 
\end{itemize}
\end{lem}
We prove the second point of Lemma \ref{verticalandbox} in the next subsection \ref{box}. For now we will be defining the processing done and prove the first point of the lemma.

	\subsubsection{Process}
	
We start with optimal packing of rectangles belonging to $\R'$. For now we remove the vertical rectangles in $\R'$ which will be repacked later.

Partition the wide rectangles into the groups $\{R_{0}, R_{1}, \dots, R_{k-1}\}$, where $k=\frac{1}{\delta}$. Let $g(R)$ denote the distance between the leftmost edge of the knapsack and the leftmost edge of $R$. Rectangles $R\in R_{i}$ have $\lfloor \frac{g(R)}{\delta}\rfloor=i$. Now we describe a procedure on how to round the wide rectangles to the nearest integer multiple of $\delta^{2}$ and shift them horizontally so that the starting and ending coordinates are integral multiple of $\delta^{2}$. Now consider rectangles in group $R_{0}$ from top to bottom and let these be $\{ R_{0_{1}}, R_{0_{2}}, \dots ,R_{0_{m}}\}$ where $R_{0_{1}}$ is the topmost and $R_{0_{m}}$ is the bottom-most rectangle.\\
Consider the sequence of guillotine cuts to separate all the rectangles in the original knapsack.  Let us denote the set of cuts used in the sequence of guillotine cuts by $G_{c}$. Now using the cuts in $G_{c}$ we create some pseudoitems as follows. During some stage of our cutting sequence we are doing horizontal (resp.~vertical) cuts in the subpiece $P$ containing $R_{0_{1}}$ then consider the topmost (resp.~leftmost) cut below (resp. right of) $R_{0_{1}}$ and divide the subpiece into 2 parts $P_{1}$ and $P_{2}$ such that $P_{1}$ contains $R_{0_{1}}$. Then make $P_{2}$ a pseudoitem.  Create these pseudoitems and stop this process if we get a subpiece $P'$ which has only $R_{0_{1}}$ in it. Now consider the psuedoitem $P_{1}$ to the right of subpiece $P'$ which shares a boundary vertically with subpiece $P'$. Similarly we define a set of pseudoitems  $\{ P_{2}, \dots ,P_{k''}\}$ where $\forall i \in \{2, \dots ,k\}$ and $P_{i}$ is right of $P_{i-1}$ and shares a boundary vertically with $P_{i-1}$. Note that $k'' \leq \frac{1}{\delta}$. Now shift these pseudoitems in the set $\{ P_{1},...,P_{k''}\}$ together to the right by $2\delta^{2}$ as done in Figure \ref{rounding} and Figure \ref{fig h}.\\
Now we show that for the rest of the rectangles in $R_{0}$ we can create a similar gap of $2\delta^{2}$ to the right of it such that there won't be any increase in the width of the $1\times(1+2\delta^{2})$ knapsack. Suppose we have created a space of $2\delta^{2}$ to the right of the rectangles$\{ R_{0_{1}}, \dots ,R_{0_{i-1}}\}$ such that  the width of $1\times(1+2\delta^{2})$ knapsack  has not increased. Consider the largest guillotine separable subpiece (using only the cuts in $G_{c}$ to separate it)  $P''$ in $1 \times 1$ knapsack which contains $R_{0_{i}}$ and does not contain any  rectangle from the set $\{ R_{0_{1}}, \dots ,R_{0_{i-1}}\}$. Now consider the $P''$ as a knapsack and do the same procedure of creating pseudoitems in $P''$ using the cuts of $G_{c}$ and shifting pseudoitems to the right by $2\delta^{2}$ as we did for the original knapsack which contained $R_{0_{1}}$ (Basically $R_{0_{i}}$ takes the role of $R_{0_{1}}$ in $P''$). If right edge of $P''$ touches the right edge of the original $1 \times 1$ knapsack, then we already have an extra space of  $2\delta^{2}$ to the right in the $1\times(1+2\delta^{2})$ knapsack. Otherwise, let $P_{1}''$ be the largest guillotine separable subpiece (using only the cuts in $G_{c}$ to separate it)  in $1 \times 1$ knapsack which is of same width as that of $P''$, is top of $P''$ and shares boundary with $P''$. Let $P_{2}''$ be the largest guillotine separable subpiece such it is of same height as height of $P''$ and $P_{1}''$ combined, is right of $P_{1}''$ and shares boundary with both $P''$ and $P_{1}''$. Then there is empty space of $2\delta^{2}$ to the right of $P''$ due to the similar shifting procedure (by creating pseudoitems using the cuts in $G_{c}$ which includes $P_{2}''$ and shifting them to the right by  $2\delta^{2}$) we did to create the space of $2\delta^{2}$ to the right of the rectangles in  the set $\{ R_{0_{1}}, \dots, R_{0_{i-1}}\} \cap P_{1}''$ for which we have to shift $P_{2}''$ to the right by  $2\delta^{2}$ compulsorily. So we can create a space of $2\delta^{2}$ to the right of $R_{0_{i}}$ without increasing the width of the original knapsack.\\
Now repeat the same procedure for the rest of the group of wide rectangles in the increasing order of the group numbers. Assuming that we have shifted items in $\{ R_{0}, \dots,R_{i-1}\}$ we demonstrate now how to shift the items in $R_{i}$. Find all the maximal guillotine separable subpieces (using only the cuts in $G_{c}$ to separate it) which contains rectangles from $R_{i}$ and doesn't contain rectangles from the set $\{ R_{0}, \dots, R_{i-1}\}$. Now considering each maximal subpiece as knapsack we can shift the items appropriately to it's right by $2\delta^{2}$ in the same way as we did for the items in $R_{0}$ for the original knapsack. 
	\begin{remark}\label{rem1}
	When we shift the items in $P_{i}'$ the groups $\{R_{0},R_{1}, \dots, R_{i-1}\}$ remain unaffected.
	\end{remark}
	\begin{remark}
	Rectangles in group $R_{i}$ is shifted at most $i$ times. So maximum number of times a rectangle is shifted is at most $1/\delta$ and hence the extra width required is $1/\delta\cdot2\delta^{2} = 2\delta$. So we can round the wide rectangles to the nearest integer multiple of $\delta^{2}$ and shift them horizontally so that the starting and ending coordinates are integral multiples of $\delta^{2}$.  
	\end{remark}
	\subsubsection{ Guillotine Separability}
		Basic idea considers the fact that we create pseudoitems using the cuts in $G_{c}$. The shifting of these pseudoitems is done horizontally right. Shifting horizontally doesn't disturb the guillotine separability as the guillotine cuts are always along the boundary of pseudoitems in the larger $1\times(1+2\delta)$ knapsack that is being created. Also when we create a space of $2\delta^{2}$ to the right for the rectangle, say $R'$, then the width of largest guillotine separable subpiece  (using only the cuts in $G_{c}$ to separate it)  $P_{R'}$  containing $R'$ in the original $ 1 \times 1$ knapsack increases by $2\delta^{2}$ in the larger $1\times(1+2\delta)$ knapsack. So we can round and shift the rectangle $R'$ so that the starting and ending coordinates are integral multiples of $\delta^{2}$ without disturbing the guillotine property. 
	\subsection{Containers for vertical rectangles\cite{GalvezGHI0W17}}
	\label{box}
	Now we will describe the procedure for packing vertical items as done in \cite{GalvezGHI0W17} and thus establishing the safety of guillotine property in the processing. We will thus be establishing the proof of second point of lemma \ref{verticalandbox} thereby completing the proof.
	
		\subsubsection{ Process }
	 Draw vertical lines spaced equally by $\delta^{2}$. In each vertical strip consider the maximal region which doesn't contain or overlap with any wide rectangle. We define a box for each such region which consists of at least one vertical rectangle and denote such a set of boxes by $B$. Also consider $M:=(1+2\delta)/\delta^{2}$. A vertical rectangle can overlap with at most $2$ vertical boxes and by considering these rectangles to be part of right of the two boxes and by the shifting argument as explained in subsection \ref{shifting1}, we can have vertical boxes with no rectangle intersecting the boundary. By using the strip packing mentioned in \cite{kenyon2000near}, we repack vertical items in a vertical box into at most $1/\delta^{7}$ containers. 
	\subsubsection{ Guillotine Separability }
	The argument follows similarly as in subsection \ref{shifting1}. Consider each of these box as a pseudoitem. It can be separated by a sequence of guillotine cuts as there are no items intersecting the boundary after processing by shifting argument. So on a whole, the knapsack is still guillotine separable. The repacking of vertical items maintains guillotine property as mentioned in \cite{kenyon2000near}. A pseudoitem can either be a single wide rectangle which is already guillotine separable or it can be strip packing of vertical rectangle which also follows guillotine property or it can be a set of items which follows guillotine property from original knapsack. Thus the new arrangement maintains guillotine separability property. 
	\subsection{Packing wide rectangles into O(1) containers }
	\begin{lem}\cite{GalvezGHI0W17}
	Given rectangles at the end of subsection \ref{box} and considering the vertical boxes as pseudoitem there exists a packing of wide rectangles with profit $(1-O(\epsau')\profit(OPT)$ into $(K+3/\delta^{3})M2^{M}$ containers where $K\leq O_{\epsau',\delta}(1)$
	\end{lem}
	Intuitively the proof involves obtaining a fractional packing with $O(1)$ containers and then showing an existence of an integral packing with a profitable set of items.
	\subsubsection{Fractional Packing}\label{fracpack} 
	We follow the same procedure of fractionally packing the items as done in \cite{GalvezGHI0W17}. Following the processing in section \ref{box}, we obtain a set of guillotine separable items packed in boxes and a set of wide rectangles. Since the items inside the boxes are guillotine separable, we can consider a box as a single pseudoitem and process it along with wide rectangles. Let us denote the set of pseudoitems as $B$ and the set of wide items as $W$. The algorithm $Alg_{frac}$ takes in input a subset $K$ of $B$ and returns their fractional packing as follows. We have $R_{slice}$ as the set $K \cup W$. Draw end-to-end horizontal lines along horizontal edges of all the items $\R$ $\in$ $R_{slice}$ slicing the rest of the rectangles of knapsack. Each such end-to-end horizontal slice so formed is termed as a slice. We then define $configuration$ as the set of intercepts of vertical edges of items with the horizontal line at the given $y$-coordinate. We slice the rectangles suitably to arrange stripe with the same configuration together. We then pack sliced rectangles into $M:=(1+2\delta)/\delta^{2}$ number of containers. Each stripe can have at most $2^{M-1}$ configurations, which implies each stripe can contain at most $M2^{M-1}$ containers. Thus the total number of containers is at most $2\times(|R_{slice}|+1)$$\times$$M2^{M-1}$. Since $|W|$$\leq$$\frac{1}{\delta^{2}}$ which follows from subsection \ref{box}, we get $|R_{slice}| \leq |K| + \frac{1}{\delta^{2}} $. Thus total number of containers can not exceed $ (|K| + \frac{3}{\delta^{3}}) $$\times$$M2^{M}$.
	\begin{remark}
	Guillotine Separability: While slicing and rearranging of rectangles during the fractional packing, the slided rectangles  remain within the initial stripe created by extending the top and bottom edges of pseudoitems and some wide items and therefore vertical boxes as pseudoitems and some wide items which are never sliced are guillotine separable. Also the containers obtained at the end of this stage are guillotine separable as the container's height is equal to the height of the configuration\cite{GalvezGHI0W17} and no container is part of 2 different configurations.
	\end{remark}
	\subsubsection { Integral Packing } 
	We follow a greedy recursive approach to convert the fraction packing obtained in section \ref{fracpack} to an integral packing without loosing much profit from optimal packing same as in \cite{GalvezGHI0W17}. The greedy algorithm $Alg_{G}$ at each iteration proceeds as follows. Given a set of items obtained after fractional packing, pick items with same widths and try packing them in the same container. Either all items can be packed or single item can be left outside, we just discard the item, close our current container and start packing in the other containers. We now propose the recursive formulation for integral packing. Consider a  set of items $S_{0}$ s.t. $S_{0}= \phi$. Run $Alg_{frac}$ on $S_{0}$, followed by $Alg_{G}$ on the output of fractional packing. Analyze the profit of set of items discarded, say $S_{1}$, in greedy algorithm and it is clear that the number is bounded by the number of compartments. Thus the number of items $S_{1}$ i.e. $|(S_{1})|$ $\leq$$ (|S_{0}| + \frac{3}{\delta^{3}}) $$\times$$M2^{M}$. Check if $\profit(S_{1})$ $\leq$ $\eps'_{ra}$$\profit(OPT)$. If yes we have our set $S = S_{0}$ and an integral packing with the set $S_{0} \cup W \setminus S_{1}$. If not we recursively run $Alg_{frac}$ on $S = S_{0} \cup S_{1}$, followed by $Alg_{G}$ on the output of fractional packing. Same as before. we consider the set of discarded items as $S_{3}$ which is bounded by number of compartments $\leq (|S| + \frac{3}{\delta^{3}}) \times M2^{M}$. Proceeding recursively every set $S_{k}$ is bounded by $|S_{k}|$ $\leq$ $ (|S_{0}+S_{1}+S_{2}+...+S_{k-2}+S_{k-1}|$ + $\frac{3}{\delta^{3}}) \times M2^{M}$, which implies $|S_{k}| \leq (\frac{3}{\delta^{3}} M2^{M}\times M2^{M} )^{k}$. Thus $k$ can at most be $\frac{1}{\eps'_{ra}} - 1$. Therefore the number of containers is at most $|S_{\frac{1}{\eps'_{ra}} - 1}| = O_{\epsau',\delta}(1)$. Thus we are packing subset $I'$ of given set of items $I$ in $O_{\epsau',\delta}(1)$ containers without losing no more than constant fraction of profit, i.e., $\profit(I') \leq \eps'_{ra}$$\profit(I)$.
	
	 \begin{remark}
	Guillotine Separability: During integral packing, the rectangles of same width are packed on top of each other in the guillotine separable containers. So after the integral packing the rectangles are still guillotine separable. 
	\end{remark}
	\subsection{Rounding down horizontal and vertical containers}
	\begin{lem}\cite{GalvezGHI0W17}
	Let $C$ be a horizontal (resp. vertical) container defined above, and let $\R_{C}$ be the set of rectangles packed in $C$. Then it is possible to pack a set $\R_{C}'\subseteq \R_{C}$ of profit at least $(1-3\epsau')\profit(\R_{C})$ in a set of at most $\lceil \log_{1+\epsau'}(1/\delta) \rceil /\epsau'^{2}$ horizontal (resp. vertical) containers that can be packed inside $C$ and such that their total area is at most $\area(I_{C})$.
	\end{lem}
	We begin with processing horizontal rectangles first (similar proof holds for vertical rectangles). Define a group of horizontal rectangles $\R_{1},\R_{2},...,$ in a single horizontal container $C$ where the width of the widest rectangle $\R_{j}$ is bigger than the smallest one by a factor of 1+$\epsau'$ and define container $C_{j}$ for each such group. Then in each such new container $C_{j}$  the operations done are either removing a set of low profit rectangles or making one container for a certain set of rectangles and shrinking the container with remaining rectangles. 
\begin{remark}
Guillotine Separability: Operations performed in the processing in this section involve deletion of items and shrinking of containers. It is trivial that deleting items doesn't disturb the guillotine property of the remaining set of items. All containers when considered as pseudorectangles can be separated by guillotine cuts. Shrinking of any such pseudo-rectangle doesn't disturb this guillotine structure. In fact guillotine separability inside such a pseudoitem isn't affected as shrinking of piece (shrinking along axis parallel directions doesn't affect the existing guillotine cuts inside it). Therefore the packing obtained after the whole processing follows guillotine separability.
	\end{remark}
	\subsection{Packing small rectangles}
	\begin{lem}\cite{GalvezGHI0W17}
	In a $1\times(1+\epsau')$  knapsack, there exists a packing of small rectangles into at most $O_{\epsau'}(1)$ area containers such that the property (3) of Lemma \ref{lemma1_old} is satisfied.
	\end{lem}
	We process the knapsack by creating a non-uniform grid by extending each side of the containers until they hit the boundary or the other containers. Plus a uniform grid is created by adding vertical and horizontal rectangles which are spaced by a distance of $\epsau'$. Now cells  which do not overlap with any other container are defined to be free cell. Then the small rectangles are packed into free cells with both dimensions at least $8\mu/\epsau'$ and possibly an extra container $C_{S}$ with height and width $\epsau'/2$ using NFDH. Then items $R'_{D}\subseteq R_{D}$ in an area container $D$ is taken and repacked into a container $D'$ such that $\profit(R'_{D})\geq(1-4\epsau')\profit(R_{D})$ and width of $D'$ is $\width(D)$ and height is $(1-\epsau')\height(D)$. The previous procedure is not applied to $C_{S}$.
	\begin{remark}
	Guillotine Separability: During construction only we get the sequence of guillotine cuts along the grid which separated out the containers individually. Since the grid is formed by extending the boundaries of the containers, we can claim that they form guillotine cuts because containers so obtained from previous subsection were guillotine separable. Moreover the packing of small rectangles inside the grid cells can be isolated from same sequence of cuts. Since items within such a packing are packed using NFDH the arrangement is guillotine separable. Thus we conclude that the obtained arrangement follows guillotine property.
	\end{remark}
	\subsection{Rounding containers to a polynomial set of sizes}
	\begin{definition}
	For a set $\R$ of rectangles, we define $WIDTHS(\R)=\{\width_{j}|R_{j}\in \R\}$ and $HEIGHTS(\R)=\{\height_{j}|R_{j}\in \R\}$.
	\end{definition}
	\begin{definition}
	Given a finite set $P$ of real numbers and a fixed natural number $k$, we define the set $P^{(k)}=\{(p_{1}+p_{2}+...+p_{l}\}+ip_{l+1}|p_{j}\in P \forall j,l\leq k,0\leq i\leq n, i  \in I \}$.
	\end{definition}
	\begin{lem}
	\label{lemma2}\cite{GalvezGHI0W17}
	Let $\eps > 0$, and let $\R$ be set of rectangles packed in a horizontal or vertical container $C$. Then for any $k\geq 1/\eps$, there is a set $\R' \subseteq \R$ with profit $\profit(\R')\geq(1-\eps)\profit(\R)$ that can be packed in a container $C'$ smaller than $C$ such that $\width(C)\in WIDTHS(\R)^{(K)}$ and $\height(C)\in HEIGHTS(\R)^{(K)}$. 
	\end{lem} 
	\begin{proof} We prove the argument for vertical containers and the similar construction follows for horizontal case. The basic sketch revolves around linear grouping of items according to widths and further rounding to a given set of choices. Consider set of items $R$ in the container in sorted order of widths. If $|R|$ is sufficiently small, i.e., $\leq \frac{1}{\eps}$ then we already meet the condition $\sum_{r \in R}w_{r} \in WIDTHS(\R)^{(K)}$. If not, we consider the set of items, say $R'$, with largest width from the set of sorted items such that $|R'| \leq \frac{1}{\eps}$. Using averaging argument, there must exist an item $r \in R'$ such that $\profit(r) \leq \eps\profit(R')$. We can safely discard $r$ without losing no more than constant $O(\eps)$ profit. Consider the set $R \setminus R'$, every item in the set has width almost $\width(r)$, sum of widths of all items in $R\setminus \{r\}$ is at most $(n - \frac{1}{\eps})\width(r)$. Thus we can pack all the items in $R \backslash {r}  $ in a new container. The sum of widths of items in the container is $\width( R \backslash {r})$ $=$ $\width(R') + (\lceil \frac{\width(R\backslash R')}{\width(r)} \rceil ) \width(r) $ $\leq$ $\width(R') + n$$\width(r) $. Thus width of the new container belongs to the set $WIDTHS(R)^{(K)}$. The given choice of factor $(\lceil \frac{\width(R\backslash R')}{\width(r)} \rceil )$ $\leq$ $n$ implies that the width of new container formed is at most width of original container considered.
	\end{proof}
	\begin{remark}
	Guillotine Separability: Applying processing done in Lemma \ref{lemma2} to a vertical container maintains guillotine separability as items are stacked beside each other, i.e., items are packed side by side sharing at most one common vertical edge and can thus be separated by single stage vertical guillotine cuts. Similar argument follows for proving guillotine separability of horizontal containers. Thus Lemma \ref{lemma2} maintains guillotine separability.
	\end{remark}
	\begin{lem}
	\label{lemma3}\cite{GalvezGHI0W17}
	Let $\eps >0$, and let $\R$ be a set of rectangles that are assigned to an area container $C$. Then there exists a subset $\R'\subseteq\R$ with profit $\profit(\R')\geq(1-3\eps)\profit(\R')$ and a container $C'$ smaller $C$ such that: $\area(\R')\leq\area(C)$,$\width(C')\in WIDTHS(\R)^{(0)}$, $\height(C')\in HEIGHTS(\R)^{(0)}$, and each $R_{j}\in\R'$ is $\frac{\eps}{1-\eps}$-small for $C'$.
	\end{lem}
	\begin{proof}
It follows from the initial assumption that items in any container $C$ are $\eps$-small for $C$, i.e., $\forall R \in \R$ which are packed inside $C$ $\width(R)$ $\leq$ $\eps\width(C)$ and $\height(R)$ $\leq$ $\eps\height(C)$. Thus $a(R) \leq \eps^{2} a(C)$. We first shrink the container C such that boundaries of C coincide with extreme edges of items in the set of packed items $\R$. Let $\width_{max}(\R)$ and $\height_{max}(\R)$ be the maximum height and width of items in $\R$, respectively. We further reduce the dimension of C by rounding off to maximum multiple of maximum item width (resp.~height) less than the original dimension. The new container $C_{1-\eps}$ formed is such that the $\width(C_{1-\eps})$ $=$ $\width_{max}(I)$ $ \lfloor \frac{\width(C)}{\width_{max}(I)} \rfloor$ and  $\height(C_{1-\eps})$ $=$ $\height_{max}(I)$ $ \lfloor \frac{\height(C)}{\height_{max}(I)} \rfloor$.
We then follow a greedy approach to pack a subset $\R'$ $\subset $ $\R$ in container $C_{1-\eps}$. Pack the item in order of non-increasing area-density (profit/area) using NFDH. 
	Now we analyze the process followed above and prove the bounds. We get $a(C') = ( 1 - \eps )^{2} a(C)$ $\geq$ $a(C)$. Since we pack as many items as possible in the area and the area of each item in $\R$ is at most $\eps^{2}$, the area of items $I'$ packed in $C'$  is $a(\R')$ $\geq$ $(1-\eps - \eps^{2} ) a(C) $. Thus we get $\profit(\R') \geq (1 - 3$$\eps) \profit(\R)$. The items in $\R'$ are $\frac{\eps}{1-\eps}$-small for the new container. Also by rounding the dimensions of container to integral multiple of maximum dimensions of items packed inside, we get $\width(C')$ $\in$ $WIDTHS(I)^{(0)}$ and $\height(C')$ $\in$ $HEIGHTS(I)^{(0)}$. This concludes the proof of lemma.
	\end{proof}
	\begin{remark}
	Guillotine Separability : All containers when considered as pseudo-rectangles can be separated by guillotine cuts. Shrinking of any such pseudo-rectangle doesn't disturb this guillotine structure. In fact, guillotine inside such a pseudoitem isn't affected as shrinking of piece (shrinking along axis parallel directions doesn't affect the existing guillotine cuts inside it). Since items within such a packing are packed using NFDH the arrangement is guillotine separable. Thus we conclude that the arrangement so obtained follows guillotine property.
	\end{remark}
	By considering $\eps=\epsau'$ and applying Lemma \ref{lemma2} and Lemma \ref{lemma3} to all containers, we have completed the proof of Lemma \ref{lemma1_old} as $\frac{\epsau'}{1-\epsau'}\leq\epsau$. Also we have proved that Lemma \ref{lemma1_old} can be extended to the case of guillotine separable rectangles.

\section{Tools}

\subsection{Generalized Assignment Problem}\label{sec:GAP}
\emph{Maximum Generalized Assignment Problem (GAP)} states as follows:
Given $m$ bins with capacities $C_{j}$, $\forall j\in [m]$, a set of $n$ items is to be packed in bins with different sizes and profits for different bins. Thus each item $i \in [n]$ is given with an associated profit $\profit_{ij}$ and size $s_{ij}$ corresponding to bin $j$, $\forall j\in [m]$. The goal is to find a packing of a subset $I\subseteq [n]$ into the bins maximizing the total profit of items packed.
In our case we will be using a specific variant when there is only constant number of bins, i.e., $m=O(1)$. We adapt the following lemma from \cite{GalvezGHI0W17} which gives a PTAS for the variant:
\begin{lem}\label{lem:GAP}
\cite{GalvezGHI0W17}
There is an algorithm for maximum generalized assignment problem with $k$ bins that runs in time $O\big( (\frac{1+\eps}{\eps})^kn^{k/\eps^2+k+1})\big)$ and returns a solution that has profit at least $(1-3\eps)\profit(OPT)$, for any fixed $\eps > 0$. 
\end{lem}

\subsection{NFDH}

\begin{figure}[]
\centering
\resizebox{!}{4cm}{
\begin{tikzpicture}
\draw[thick] (0,0) rectangle (4,3.75);
\draw (0,2)--(4,2);
\draw (0,3)--(4,3);
\draw[solid,fill=lightgray] (0,0) rectangle (2,2);
\draw[solid,fill=lightgray] (2,0) rectangle (3.5,1.2);
\draw[solid,fill=lightgray] (0,2) rectangle (1.5,3);
\draw[solid,fill=lightgray] (1.5,2) rectangle (3,2.9);
\draw[solid,fill=lightgray] (3,2) rectangle (3.8,2.8);
\draw[solid,fill=lightgray] (0,3) rectangle (2,3.75);
\draw[solid,fill=lightgray] (0,3) rectangle (2,3.75);
\draw[solid,fill=lightgray] (2,3) rectangle (3,3.6);
\draw[solid,fill=lightgray] (3,3) rectangle (3.5,3.5);
\end{tikzpicture}}
\caption{2 Stage Packing using NFDH}
\label{NFDH:fig}
\end{figure}
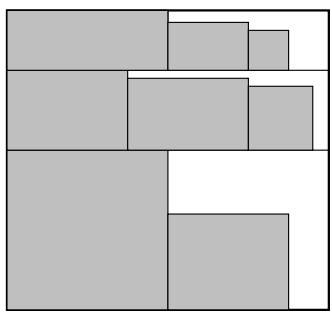

\begin{lem}
\label{lem:NFDH}
[NFDH\cite{coffman1980performance}]Given a $N\times N$ knapsack and set of small items $\R_{small}$ ( $\forall i \in \R_{small}$, $\width(i)\leq \eps N$ and $\height(i)\leq \eps N)$, we can pack a subset of items $\R' \subseteq \R$ such that $a(\R') \geq min\{ a(\R), (1-2\eps)N^{2}\}$ and the packing is $2$-stage guillotine separable. 
\end{lem}
\begin{proof}
For complete proof we refer the reader to \cite{coffman1980performance}. We now prove that resulting packing from procedures in the proof forms a $2$-stage guillotine separable packing. The processing in the proof packs items in horizontal strips each of width $N$. Further in each strip, items are packed side by side sharing vertical edge with the next item. This gives us a $2$-stage packing, with first stage cuts separating horizontal strips from each other and items within second stage cuts separating items separating items in individual strips as shown in Figure \ref{NFDH:fig}.
\end{proof}

\end{document}